\documentclass[prb,showpacs,twocolumn,amsmath,superscriptaddress,amssymb,amsfonts,aps,floatfix]{revtex4}

\usepackage{graphicx, color,rotating} 
\usepackage{dcolumn} 
\usepackage{bm} 
\usepackage{epsfig}
\usepackage{bbm}

\begin{document}
 
\title{ The effect of spin fluctuations on the electronic structure in
iron based superconductors }

\author{Andreas~Heimes}
\affiliation{SEPnet and Hubbard Theory Consortium, Department of Physics, Royal Holloway, University of London, Egham, Surrey TW20 0EX, United Kingdom}
\affiliation{Institut f\"ur Theoretische Festk\"orperphysik and DFG-Center for Functional Nanostructures, Karlsruhe Institute of Technology, D-76128 Karlsruhe, Germany}

\author{Roland~Grein}
\affiliation{SEPnet and Hubbard Theory Consortium, Department of Physics, Royal Holloway, University of London, Egham, Surrey TW20 0EX, United Kingdom}
\affiliation{Institut f\"ur Theoretische Festk\"orperphysik and DFG-Center for Functional Nanostructures, Karlsruhe Institute of Technology, D-76128 Karlsruhe, Germany}

\author{Matthias~Eschrig}
\affiliation{SEPnet and Hubbard Theory Consortium, Department of Physics, Royal Holloway, University of London, Egham, Surrey TW20 0EX, United Kingdom}

\date{\today}

\begin{abstract}
Magnetic inelastic neutron scattering studies of iron-based superconductors 
reveal a strongly temperature-dependent spin-fluctuation spectrum in the normal conducting state,
which develops a prominent low-energy resonance feature when entering the superconducting state. 
Angle-resolved photoemission spectroscopy (ARPES) 
and
scanning tunneling spectroscopy (STS) 
allow us to study the fingerprints of fluctuation modes via their
interactions with electronic quasiparticles.
We calculate such fingerprints in 122 iron pnictides using an experimentally motivated spin-fluctuation spectrum 
and make a number of predictions that can be tested in ARPES and 
STS experiments.
This includes discussions of the quasiparticle scattering rate and the superconducting order parameter.
In quantitative agreement with experiment we reproduce the quasiparticle dispersions obtained from 
momentum distribution curves as well as energy distribution curves. We discuss the relevance of
the coupling between spin fluctuations and electronic excitations for the superconducting mechanism.
\end{abstract}

\maketitle

\section{Introduction}

Shortly after the discovery of high-temperature superconductivity in Fe-based pnictide compounds by Hosono and collaborators \cite{Hosono2008} a magnetic Cooper-pairing mechanism was proposed. \cite{Mazin08, Zlatko09} A pure electron-phonon interaction as the primary pairing mechanism was, however, found to be unlikely. \cite{Boeri08,Yang2009} It is believed that the proximity of antiferromagnetic spin density waves and superconductivity in the phase diagram \cite{Zhao2008,Nandi2010} is likely to support a pairing scenario where superconductivity is driven dominantly by an electron-electron interaction mediated by spin fluctuations. \cite{Hirschfeld2011, Heimes2011}
The spin excitation spectrum of the iron pnictides (and iron chalcogenides) shows pronounced similarities with other superconductors where a magnetic pairing mechanism is under debate. 
This includes for instance a strong temperature dependence of the spectrum in the normal state, 
which was studied in iron pnictides by Inosov {\it et al.} \cite{Inosov2009},
as well as the presence of a spin resonance feature below a gapped continuum in the superconducting state, \cite{Inosov2009,Christianson2008,Korshunov2008}  which also appears in cuprates and in some heavy fermion superconductors. 

Angle-resolved photoemission spectroscopy (ARPES) experiments on iron pnictides\cite{richard11} reveal a sharp Fermi surface consisting of multiple electron and hole pockets (see e.g. Ref. \onlinecite{Fink2009} and references herein). These exhibit comparable superconducting order parameter amplitudes at Fermi surface sheets which are nearly nested by the antiferromagnetic wave vector, $\bm Q$. \cite{Ding2008,Nakayama2009} 
It thus seems likely that magnetic and electronic order are closely related. \cite{MazinSchmalian2009} In the case of antiferromagnetic spin fluctuations as the origin for pairing, the order parameter must have different sign at the electron and hole pockets. \cite{Parker2008,Chubukov2008,Maier2008} In this context, we recently endeavored to answer the question of how strongly electrons couple to spin fluctuations in these compounds. \cite{Heimes2011} To this end, we
investigated low energy dispersion anomalies, whose position and shape can be traced back to a coupling of bosonic modes - a method which has been proven to be a powerful tool previously in the case of cuprate superconductors. \cite{Eschrig2006}
In iron-based superconductors such anomalies have been observed in the hole-doped 122 compound Ba$_{1-x}$K$_x$Fe$_2$As$_2$ by Wray {\it et al.} \cite{Hasan2008}, Koitzsch {\it et al.}, \cite{Koitzsch09} and Richard {\it et al.} \cite{Richard2008} These experiments reveal that the electronic dispersion features a self-energy effect, which is most pronounced at an energy $\epsilon_{0}\approx 25\,$meV. 

Shortly before that, inelastic neutron scattering studies by Christianson {\it et al.} \cite{Christianson2008} of the same compound revealed the development of a spin resonance in the superconducting state. This resonance appears at an energy $\Omega_{\rm res}\approx 14\,$meV, is situated at the in-plane antiferromagnetic wavevector $\bm Q=(\pi,\pi)$ and is weakly dispersive in the $q_z$ direction. \cite{Lumsden2009, Park_et_al_2010, Zhang2011}  
Guided by similar studies in cuprate superconductors, we look for corresponding signatures in the available ARPES data.\cite{Hasan2008,Koitzsch09,Richard2008}
There, it was observed that (i) the superconducting excitation gap at the Fermi surface pockets nested by $\bm Q$ has the absolute value $ \Delta\approx 12\,$meV, 
(ii) the self-energy effect occurs at an energy $\epsilon_0\approx \Omega_{\rm res}+ \Delta$, and (iii) $\epsilon_0$ follows an order-parameter-like evolution in 
temperature as the resonance energy does. 
Thus, it is intuitively plausible to conjecture that the magnetic resonance is responsible for the observed self-energy effects. 
By numerical calculation we show that this conjecture is theoretically well founded.

Motivated by the experimental procedures \cite{Richard2008} to quantify self-energy effects in angle-resolved photoemission experiments,
we concentrated in a previous Letter on the so-called {\it effective self-energy} as extracted directly from the electronic spectral function.\cite{Heimes2011} Such an approach allowed us to make an immediate comparison with ARPES experiments. In the present work we are taking a more detailed look into our model,
introducing the formalism used for our calculations, discussing the renormalization effects entering the electronic structure as well as the Fermi surface, and applying our model to photoemission and tunneling experiments.

\section{The spin-fluctuation spectrum and the resonance mode}
The appearance of a low-energy resonance mode in the dynamic magnetic susceptibility upon entering the superconducting state is a feature well known from unconventional superconductors (for a review concerning cuprates see Ref.~\onlinecite{Eschrig2006}). As it results from scattering between Fermi surface sheets with opposite order-parameter sign its detailed experimental and theoretical investigation is a powerful tool to study the symmetry of the superconducting gap. \cite{Korshunov2008, Maier2009} 
Various microscopic theoretical treatments are able to reproduce the resonance feature, 
such as, for example, in methods employing the random phase approximation (RPA) or in the fluctuation exchange (FLEX) approximation. 
However in all known treatments specific assumptions and approximations are necessary.
Strong correlations usually lead to vertex corrections in a Feynman diagrammatic technique, which, in most cases, are either neglected or, in the absence of any good expansion parameter, introduced 
by {\it ad hoc} methods or educated guesses (as in conserving approximations such as FLEX).
As our results do not depend on the specific microscopic model for the resonance excitation,
we prefer here a semiphenomenological approach.
Most of the magnetic correlations are included automatically within the semiphenomenological approach used here by relying on experimental results.

In the following sections we introduce an effective model for the spin susceptibility and establish the connection to recent neutron scattering experiments.

\subsection{Spin susceptibility}
In this section
we summarize the semiphenomenological approach we employ in this paper, which has already been successfully used to describe optimally doped and overdoped cuprate superconductors. \cite{Eschrig2000,Abanov2000, Abanov2000-2, Eschrig2006, EschrigNormanApr2003}
To motivate its application to iron-based superconductors we recall one possible interpretation of the resonance mode in the spin susceptibility. 
We underline, however, that this interpretation is not necessary for our predictions to hold, as we base the spin excitation spectrum on the experimental observations.

Let us divide the phase space of electronic excitations into high-energy and low-energy regions, the latter being located around the Fermi surfaces and populated by the electronic quasiparticles (see e.g. Ref.~\onlinecite{Eschrig99}). Incoherent spin fluctuations are dominated by high-energy electronic excitations; however, fine features such as the resonance mode result from modifications in the low-energy electronic spectrum.
In the case of Fermi surface nesting in the vicinity of certain``nesting points''
a common approximation is given by the RPA enhanced susceptibility

\begin{eqnarray}
 \label{eq1} 
  \hat \chi(\omega,\bm q)=\left\{\hat \chi_{\rm high}^{-1}(\bm q)-\hat \Gamma^{(1)}_{\rm high}(\bm q) \, \hat \Pi(\omega,\bm q)  \,\hat \Gamma^{(2)}_{\rm high}(\bm q) \right\}^{-1},
\end{eqnarray}

where $\hat \Gamma^{(1,2)}_{\rm high}(\bm q)$ are vertex functions involving only high-energy processes, $\hat \Pi (\omega,\bm q)$ describes the polarization due to low-energy processes, and a hat denotes a matrix structure due to orbital degrees of freedom. \cite{Eschrig2006} (note that the vertex functions occur twice here as they contain no low-energy inclusions). We neglect for simplicity in the following a possible $\bm q$-dependence of the vertex functions.
The high-energy part $\hat \chi_{\rm high}$ accounts for intermediate or long-range antiferromagnetic correlations and thus can be well approximated by the phenomenological Ornstein-Zernike form $\hat \chi_{\rm high}(\bm q)=\hat \chi_{\bm Q}/(1+\xi^2\,|\bm q - \bm Q|^2)$ with $\xi \equiv \xi ({\bm Q}) $ a correlation length, and where $\hat \chi_{\bm Q}$ is nonzero only for the relevant orbitals. 

In order to find a suitable analytic form that fits the experimental data near the antiferromagnetic wave vector $\bm Q$, it is a common procedure to start from an approximation for $\hat \Pi(\omega, \bm Q)$ in the energy region of interest. To this end, let us consider a simple two band model consisting of holelike and electronlike bands, $\zeta_{\rm h}$ and $\zeta_{\rm e}$ respectively, nearly nested by the antiferromagnetic wave vector $\bm Q$, and let us neglect for a moment the orbital structure.\\
 In the normal state interband scattering can lead to a relaxation process exciting a particle-hole pair around the Fermi surface. Such excitations most likely appear near the nesting points at the electron (e) and hole (h) pockets, i.e. where $\zeta_{\rm h}(\bm k)\approx \zeta_{\rm e}(\bm k - \bm Q)+\omega $.
Linearizing the dispersions around the nesting points, $[\zeta_{\rm h}(\bm k),\zeta_{\rm e}(\bm k - \bm Q)]=[k_x,k_y] \hat {\bm V}^T$, will lead to ${\rm Im}\Pi (\omega, \bm Q) \propto \Gamma^{(1)}_{\rm high} \Gamma^{(2)}_{\rm high}\int d\zeta_{\rm h} \int d\zeta_{\rm e} \, {\rm det }(\hat{\bm V}^{-1}) [f(\zeta_{\rm h})-f(\zeta_{\rm e}) ] \, \delta(\omega-[\zeta_{\rm h} - \zeta_{\rm e}])$, where $f$ is the Fermi distribution function, and $ \hat {\bm V}$ is a matrix with components proportional to the Fermi velocity components. For small frequencies $\omega\rightarrow 0$ the imaginary part is linear in energy (and the real part negligibly small), leading to an approximation for the susceptibility

\begin{eqnarray}
 \label{eq2}
  \chi^{\rm c}(\omega,\bm q)=\frac{\chi_{\bm Q}}{1+\xi_{\rm c}^2\,|\bm q - \bm Q|^2- \imath (\omega /\Omega_{\rm max})}.
\end{eqnarray}

A detailed investigation of the spin dynamics in optimally doped BaFe$_{\rm 1.85}$Co$_{\rm 0.15}$As$_{\rm 2}$ was performed by Inosov {\it et al.} \cite{Inosov2009} By including a temperature dependence in the parameters in Eq.~\eqref{eq2}, that is replacing $\chi_{\bm Q} \rightarrow \chi_T$, $\xi_{\rm c} \rightarrow \xi_T$ and $\Omega_{\rm max}\rightarrow \Omega_{\rm max}^T$, they were able to show that the energy and momentum dependence as well as the temperature behavior of Eq.~\eqref{eq2} fit well with the normal state behavior, which validates theoretical models based on an itinerant description of magnetic excitations. 

In the superconducting state a low energy resonance appears with a weight that follows the temperature dependence of the superconducting gap. This again can be understood in a similar manner. Particle-hole excitations in the superconducting state that result from scattering between the two bands are suppressed below the two-particle excitation gap 

\begin{equation}
2 \breve \Delta_{\bm q} \equiv {\rm min}_{\left\{ \bm k\in FS \right\}}\left( |\Delta_{\rm h}(\bm k)| + |\Delta_{\rm e}(\bm k - \bm q)|\right) ,
\end{equation}

where $ \Delta_{\rm e/h}$
is the superconducting gap at the electron/hole pocket. Excitations in the continuum set in above this threshold, i.e., ${\rm Im} \Pi(\omega,\bm q)\propto 2 \breve \Delta_{\bm q} \,{\rm sign}(\omega)\, \theta(|\omega| - 2 \breve \Delta_{\bm q})$. 
For different signs between $\Delta_{\rm h}(\bm k) $ and $\Delta_{\rm e}(\bm k-\bm q)$, the coherence factors appearing in $\Pi (\omega, \bm q)$ are such that
the real part has a logarithmic divergence at $2\breve \Delta_{\bm q}$. Furthermore, it can be expanded for small $\omega$, i.e., ${\rm Re}\Pi(\omega,\bm q)\propto 
\omega^2/2 \breve \Delta_{\bm q}$. We define the resonance energy $\Omega_{{\rm res},\bm q}$ via $\chi_{\rm high}(\bm q)\Gamma_{\rm high}^2\,\Pi(\omega, \bm q)\approx \omega^2/\Omega^2_{{\rm res},\bm q}$ and insert this expression into Eq.~\eqref{eq1} to get

\begin{eqnarray}
  \label{eq3}
  \chi^{\rm r}(\omega,\bm q)=\frac{\chi_{\bm Q}}{1+\xi_{\rm r}^2\,|\bm q - \bm Q|^2- (\omega +\imath \Gamma_{\rm res})^2/\Omega_{{\rm res},\bm q}^2}.
\end{eqnarray}

Here we have introduced a small broadening $\Gamma_{\rm res}\ll \Omega_{{\rm res},\bm q}$ of the resonance mode, which accounts for its experimentally observed finite width. 
The resonance mode can be identified with the pole in Eq. \eqref{eq3}, which is moved by the amount $\Gamma_{\rm res}$into the lower 
complex half plane. If the correlation length $\xi_{\rm r}$ is sufficiently large, the
susceptibility is sharply peaked around $\bm Q$, in the region 
$|\bm q - \bm Q| \ll \xi_{\rm r}^{-1}$.  

We rewrite Eq.~\eqref{eq3} as
\begin{eqnarray}
  \label{eq4}
  \chi^{\rm r}(\omega,\bm q)=\frac{w_{\bm Q}/\pi}{1+\xi_{\rm r}^2|\bm q - \bm Q|^2} \; \cdot \; \frac{2 \Omega_{\rm res}(\bm q)}{\Omega_{\rm res}(\bm q)^2 - (\omega + \imath \Gamma_{\rm res})^2},
\end{eqnarray}
with $\Omega_{\rm res}(\bm q)=\Omega_{{\rm res},\bm q} (1+\xi_{\rm r}^2 |\bm q- \bm Q|^2)^{1/2}$ and
$w_{\bm Q} = \pi \chi_{\bm Q}\Omega_{{\rm res},\bm Q}/2$.
In the following we adopt the approximation to
neglect the dispersion of the resonance energy with momentum. 
This is, we replace Eq.~\eqref{eq4} by
\begin{eqnarray}
  \label{eq4a}
  \chi^{\rm r}(\omega,\bm q)=\frac{w_{\bm Q}/\pi}{1+\xi_{\rm r}^2|\bm q - \bm Q|^2} \; \cdot \; \frac{2 \Omega_{\rm res}}{\Omega_{\rm res}^2 - (\omega + \imath \Gamma_{\rm res})^2}.
\end{eqnarray}
This form has the advantage of fitting the energy and momentum width independently.
From experiment is it known that the dispersive features have much less weight then the resonance itself, and thus this will lead to a strongly reduced influence on the electronic single particle dispersions. There have been experimental and theoretical investigations of the doping dependence of the bosonic dispersion.\cite{Castellan2011, Maiti2011} They show that the incommensurability of the resonance and the dispersion of the resonance mode depend on the doping level. Below and up to optimal doping the resonance energy is centered around the antiferromagnetic wave vector, whereas at higher dopings the spectral weight becomes large around two wave vectors and is therefore incommensurate.\cite{Castellan2011} Within the accuracy of the known parameters, neglecting the dispersive features will be sufficient for our purpose.

\subsection{Ba$_{\rm 1-x}$K$_{\rm x}$Fe$_{\rm 2}$As$_{\rm 2}$} \label{sec1b}

In this section we outline our phenomenological approach for the case of iron based superconductors. We shall focus on hole-doped Ba$_{\rm 1-x}$K$_{\rm x}$Fe$_{\rm 2}$As$_{\rm 2}$, since we will compare our predictions with the available experimental angle-resolved photoemission spectra for this compound. 
 
Spin fluctuations that originate from the spin of the conduction electrons are closely connected to the Fermi surface topology of the electronic structure. In our model we employ a tight-binding fit in an orbital basis that was obtained from the density functional theory (DFT) band structure of BaFe$_{\rm 2}$As$_{\rm 2}$ by Graser {\it et al.} \cite{Graser2010} The dominant contribution to the density of states in the energy range $\pm $1-2 eV around the chemical potential originates from the five Fe 3d orbitals (with some hybridization with the As 4p orbitals mainly for energies above the chemical potential). This energy range will be sufficient for our purposes. The resulting reduction of the Hamiltonian to a five-orbital basis reads
\begin{eqnarray}
  \label{eq5}
  H_0=\sum_{\bm k \sigma}\sum_{mn}\,d^\dagger_m(\bm k\sigma)[\zeta_{mn}(\bm k) + \delta_{mn} \epsilon_n] \,d_n(\bm k\sigma).
\end{eqnarray}
Here $d_m^\dagger(\bm k\sigma)$ creates an electron with momentum $\bm k$ and spin $\sigma$ in the orbital $m$, where $m=1,2 \, \cdots \,5$ corresponds to the five orbitals $d_{xz},\,d_{yz},\,d_{x^2-y^2},\,d_{xy},\,d_{3z^2-r^2}$. The parameters $\zeta_{mn}$ and $\epsilon_n$ are listed in Ref.~\onlinecite{Graser2010}. 
The canonical transformation 
\begin{equation}
d^\dagger_\mu(\bm k \sigma)=\sum_m \,a^m_\mu(\bm k)d^\dagger_m(\bm k \sigma)
\end{equation}
diagonalizes the Hamiltonian, leading to eigenvalues $\zeta'_\mu(\bm k)$ and eigenvectors $a_\mu^m(\bm k)=\langle m \bm k|\mu \bm k \rangle$, where $\mu$ represents the band index. Because $\zeta_{mn}(-\bm k)=\zeta_{mn}(\bm k)^*$ the eigenvectors can be choosen such that $a_\nu^n(-\bm k)=a_\nu^n(\bm k)^*$ holds. The set of eigenvectors for each $\bm k$ is orthonormal (or can be chosen so), i.e. $\sum_m a_\mu^m(\bm k)^*a_\nu^m(\bm k)=\delta_{\mu \nu}$ and $\sum_\mu a_\mu^m(\bm k)^* a_\mu^n(\bm k)=\delta_{mn}$. 

In order to simulate hole-doping we apply a rigid shift of the chemical potential, i.e. we define the new band structure $\zeta_\mu(\bm k)= \zeta'_\mu(\bm k)-\mu_0$ with $\mu_0=-50 \,{\rm meV}$. This was shown to be at least applicable for materials in which doping occurs via substitution in regions outside of the conducting Fe-As planes and subsequent charge transfer, as in the case of Ba$_{\rm 1-x}$K$_{\rm x}$Fe$_{\rm 2}$As$_{\rm 2}$. \cite{ChangLui2008,Neupane2011} 
This assumption is consistent with our results for weak to moderate coupling, which show that 
the renormalization of the chemical potential due to the coupling between electronic excitations and spin fluctuations is in this case well approximated by 
a linear dependence between the chemical potential and the 
charge carrier concentration (for details see below).

The dominant processes for spin fluctuations are the interactions between electrons at the various Fe orbitals of the same atom (inter- and intra-orbital Coulomb interaction, Hund's coupling, and intra-orbital pair hopping). Thus, spin fluctuations are described naturally in an orbital basis, whereas electronic excitations are easier to discuss within a band picture. Because we are interested in the influence of spin fluctuations on electronic excitations, it is useful to recall the orbital characters at each Fermi surface point.

\begin{figure}[tbhp]
 \begin{center}
  $\begin{array}{cc}
	\includegraphics[width=1\columnwidth]{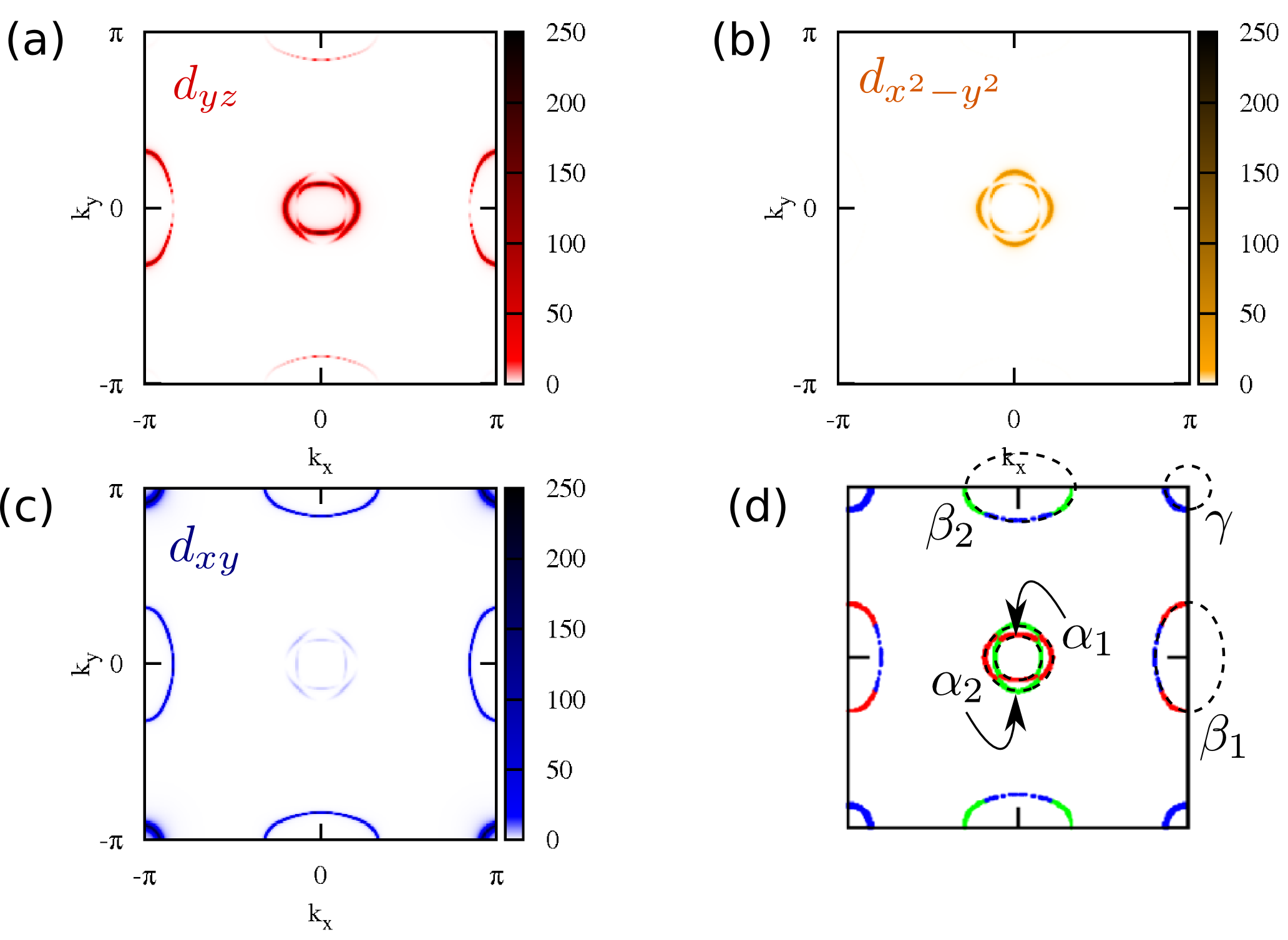}\\
	\includegraphics[width=1\columnwidth]{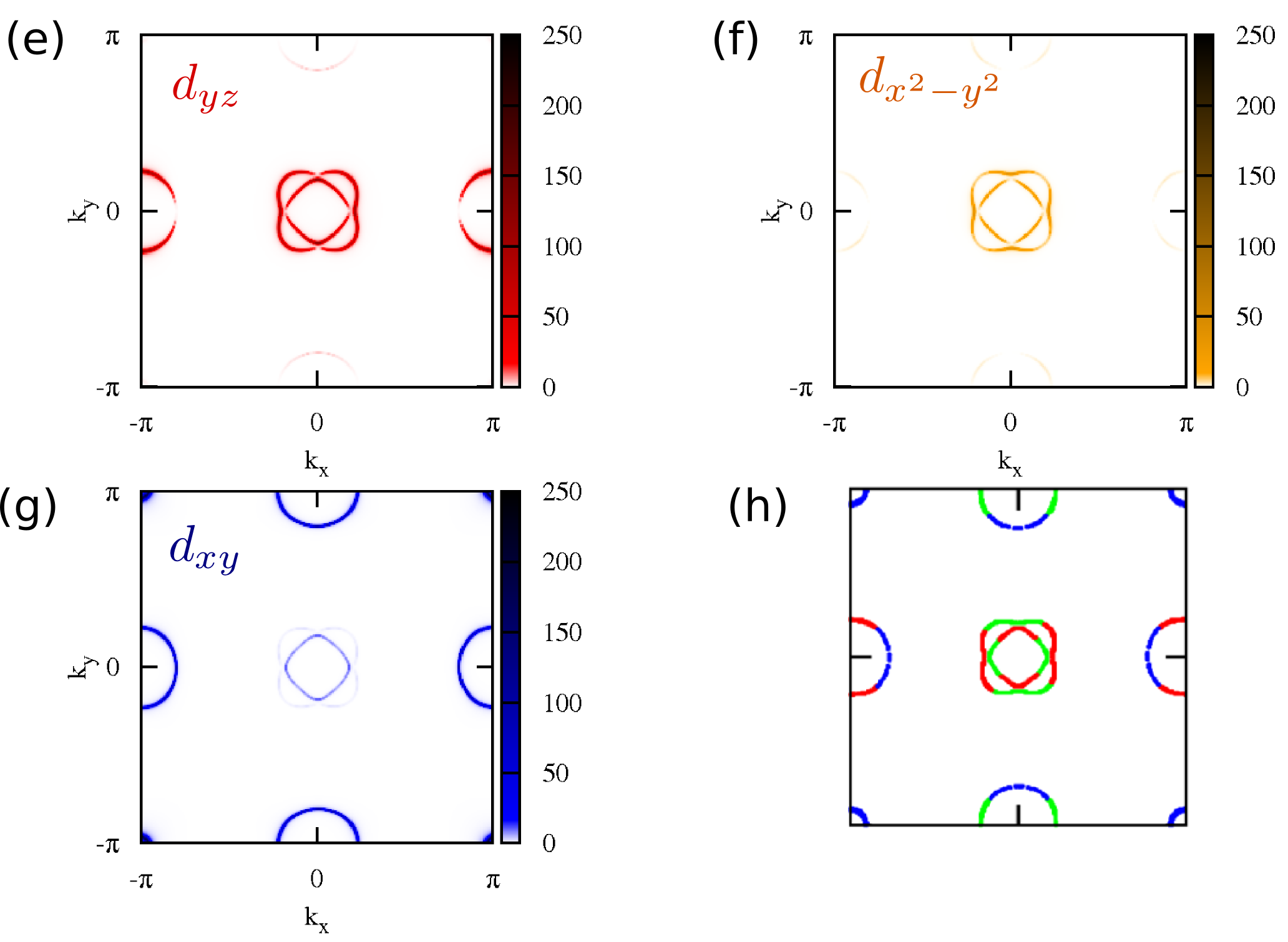}\\
	\includegraphics[width=1\columnwidth]{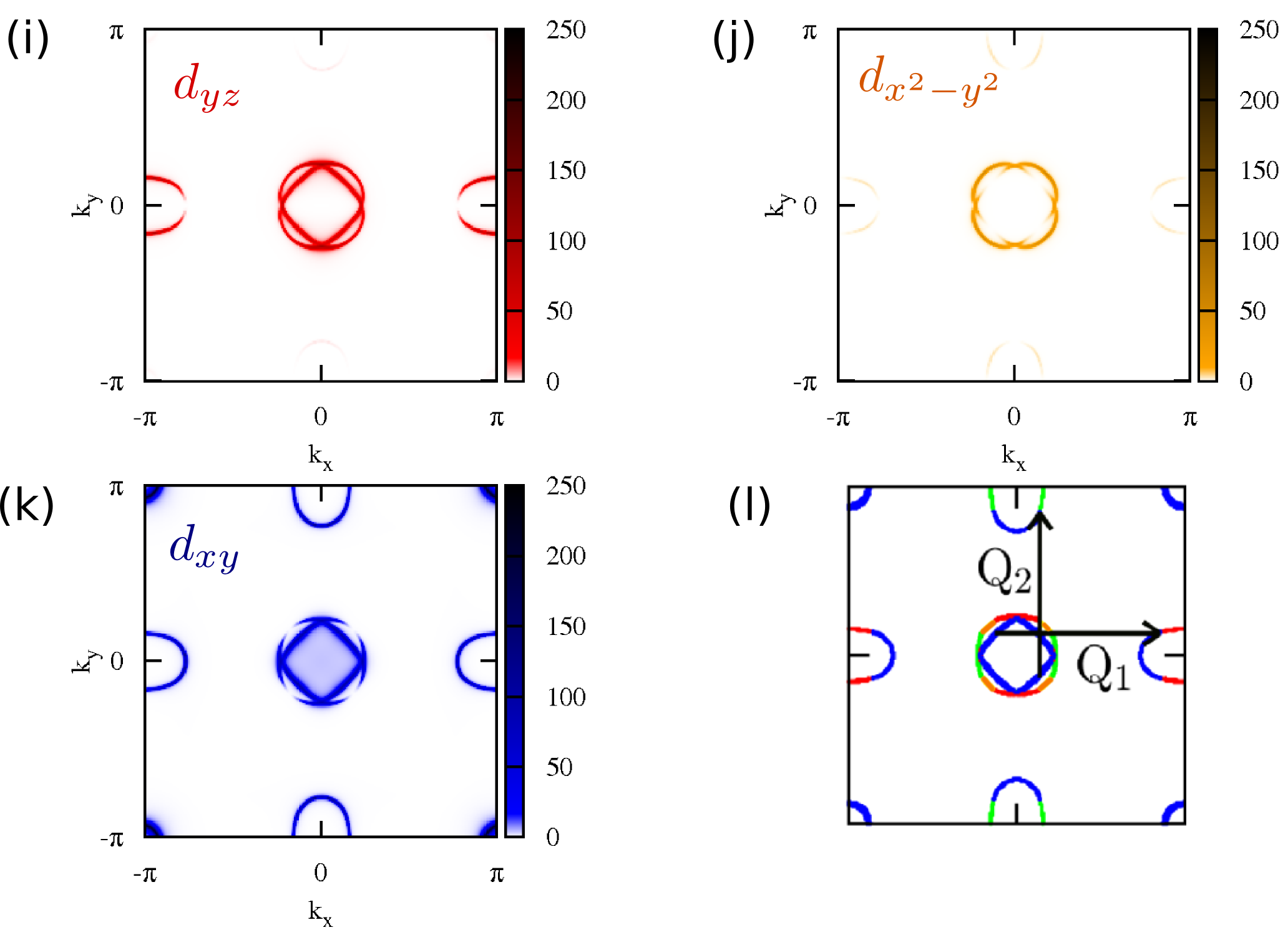} 
  \end{array}$
   \caption{Orbital character at the Fermi surface for Ba$_{1-x}$K$_x$Fe$_2$As$_2$  
($x $ = 0.4): 
(a)-(c) Partial spectral function $A_n(\bm k)=\sum_\mu |a_\mu^n(\bm k)|^2 \,\delta/(\delta^2+\zeta_\mu(\bm k)^2)$ at $k_z=0$ for the orbitals $n=d_{yz},\, d_{x^2-y^2}$ and $d_{xy}$, respectively. 
The corresponding figure for the $d_{xz}$ orbital is obtained from (a) by a 90$^o$ rotation.
(d) Dominant orbital contributions to the Fermi surface at $k_z=0$ (with the
$d_{xz}$ orbital in green). 
The various Fermi surface sheets are named $\alpha_{1,2}$ (hole sheets), $\beta_{1,2}$ (electron sheets), and $\gamma $ (hole sheet).
(e)-(h)
the same as in (a)-(d) for $k_z=\pi/2$.
(i)-(l)
The same as in (a)-(d) for $k_z=\pi$.
The two nesting vectors $\bm Q_1$ and $\bm Q_2$ are shown in (l).}
   \label{fig1}
 \end{center}
\end{figure}

Throughout this paper, we adopt the notation of Ref.~\onlinecite{Graser2010}
corresponding to 1 Fe/unit cell. 
Fig.~\ref{fig1} summarizes the orbital contributions to the Fermi surface for three values of $k_z$: $0$, $\pi/2$, and $\pi$.

Starting with $k_z=0$, 
Figs.~\ref{fig1}(a)-\ref{fig1}(c) show the partial spectral function for the bare bands,
\begin{equation}
A_n(\bm k)=\sum_{\mu}a_\mu^n(\bm k)a_\mu^n(\bm k)^\ast \,\frac{\delta}{\delta^2+\zeta_\mu(\bm k)^2}
\end{equation}
for the orbitals $n=d_{yz},\,d_{x^2-y^2}$ and $d_{xy}$, respectively (where $\delta =4\,$meV is an broadening factor), whereas Fig.~\ref{fig1}(d) shows the dominant orbital contributions to the Fermi surface only. 
Corresponding orbital characters for $k_z=\pi/2$ and $k_z=\pi$ are shown in
Figs.~\ref{fig1}(e)-\ref{fig1}(h) and \ref{fig1}(i)-\ref{fig1}(l), respectively.

Magnetic excitations can be described in terms of the dynamic spin susceptibility $\chi^{mn}_{pq}$, where the measured susceptibility $\chi$ is the sum over all orbital contributions, $\chi=\sum_{mn} \chi^{mm}_{nn}$. 
As can be seen from Fig.~\ref{fig1},
the hole pockets at $(0,0,k_z)$ and $(\pi,\pi,k_z)$ are nearly nested (taking into account the finite momentum width $\xi_{\rm r}^{-1}$ of the spin excitation) to the electron pockets at $(0,\pi,k_z)$ and $(\pi,0,k_z)$ by the wave vectors $\bm Q_1\equiv (\pi,0,q_z)$ and $\bm Q_2 \equiv (0,\pi,q_z)$, 
which connect Fermi surface sheets with same orbital character and correspond to antiferromagnetic correlations in the respective direction. 
Fluctuation exchange approaches have shown that the magnetic mode predominantly scatters between orbitals with the same character. \cite{Stanescu2008, Kemper2010, Graser2009}
According to that we neglect all inter-orbital contributions in our model, i.e. we only couple the part 
$\chi_n\equiv\chi_{nn}^{nn}$ to the electronic excitations, and we assume
$\chi\approx \sum_n \chi_n$.

As can be seen from Fig.~\ref{fig1}, the $d_{x^2-y^2}$ orbital has negligible
intra-orbital contributions to nesting, because the electron pockets have very little of its character. Thus, we only take into account the remaining orbitals
($n=d_{xz},\,d_{yz},\,d_{xy}\equiv 1,\,2,\,4$). To simplify notation, we introduce the parameters 
\begin{eqnarray}
  \label{eq6}
  b_{n,\alpha}=\left\{ \begin{array}{cc} 1/2 & {\rm for}\, (n\alpha)=(21),(12),(41),(42) \\
                                         0  & {\rm else}
                     \end{array}\right\}
\end{eqnarray}
to account for the orbital selective mode coupling via the wave vectors $\bm Q_{\alpha=1,2}$ (see Fig.~\ref{fig1}(l)). 
With this, the normal state susceptibility will be 
modelled by an equivalent of the Ornstein-Zernike form in Eq.~\eqref{eq2}, 
\begin{eqnarray}
  \label{eq7}
  \chi_n^{\rm c}(\omega,\bm q)=\sum_{\alpha=1,2} \frac{b_{n,\alpha}\,\chi^n_T}{1+\xi_T^2|\bm q - \bm Q_\alpha|^2 - \imath (\omega/\Omega_{\rm max}^T)},
\end{eqnarray}
with temperature-dependent parameters $\chi^n_T=\chi^n_0/(T+\theta)$, $\xi_T=\xi_0/\sqrt{T+\theta}$, and $\Omega_{\rm max}^T=\Omega_0(T+\theta)$, as motivated by Ref. \onlinecite{Inosov2009}. Here $\theta$ is the Curie-Weiss temperature.

In the superconducting state a particle-hole excitation gap opens up in the spin excitation spectrum, a resonance peak appears within this gap,
and spectral weight is shifted into this magnetic resonance. Accordingly, the magnetic spectrum now consists of two parts, the low-energy resonance and the particle-hole continuum.

Particle-hole excitations appear above a temperature dependent threshold of 
\begin{equation}
2\breve \Delta(T) \approx {\rm min}_{\left\{ |\bm k_{\rm F}^\mu- \bm k_{\rm F}^\nu|\approx \bm Q_\alpha \right\}}\left(|\Delta^T_\mu(\bm k^\mu_{\rm F})| + |\Delta^T_\nu(\bm k_{\rm F}^\nu)| \right) ,
\end{equation}
where $\Delta^T_{\mu}$ and $\Delta^T_{\nu}$ are the superconducting gaps at the nested pockets, and $\bm k_{\rm F}^\mu$ and $\bm k^\nu_{\rm F}$ are the corresponding Fermi wave vectors. For high excitation energies the susceptibility should recover the normal state behavior. 
Motivated by these observations
we approximate the gapped continuum by the same functional form (but with different magnitude; see below) as in the normal state, Eq.~\eqref{eq7}, and write for the imaginary part of the complex dynamical susceptibility 
\begin{equation}
\chi''^{\rm sc}_n(\omega ,\bm q)=\chi''^{c}_n(\omega,\bm q)
\quad \mbox{for} \quad |\omega|>2\breve \Delta(T).
\end{equation}
Neutron scattering experiments have shown that the resonance follows an order parameter like evolution. 
Therefore, we assume the temperature dependence
\begin{equation}
\Omega_{\rm res}^T=\Omega_{\rm r} \sqrt{1-T/T_{\rm c}}, 
\end{equation}
and we apply the Lorentzian form of Eq.~\eqref{eq4} in order to model the resonance below the threshold, i.e. 
\begin{equation}
\chi''^{\rm sc}_n(\omega , \bm q)=\chi''^{r}_n(\omega,\bm q)
\quad \mbox{for} \quad |\omega|<2\breve \Delta(T),
\end{equation}
with
\begin{eqnarray}
  \label{eq8}
  \chi_n^{\rm r}(\omega,\bm q)=\sum_{\alpha=1,2} \frac{2\,w_{n,\alpha}(\bm q)}{\pi} \;\cdot\;\frac{\Omega_{\rm res}^T}{(\Omega_{\rm res}^T)^2-(\omega+\imath \Gamma_{\rm res})^2}.
\end{eqnarray}
Here the momentum distribution enters via the weight function
\begin{eqnarray}
  \label{eq9}
   w_{n,\alpha}(\bm q)=\frac{b_{n,\alpha}w^n_T}{1+\xi_{\rm r}^2|\bm q-\bm Q_{\alpha}|^2}.
\end{eqnarray}
Note that the momentum and orbital dependence of the susceptibility is contained in the factors $b_{n,\alpha}$ (Eq.~\ref{eq6}).
We determine the real part (up to a constant, see below) from ${\chi''}_n^{\rm sc}$ by exploiting Kramers-Kronig relations

\begin{eqnarray}
{\chi'}_n^{\rm sc}(\omega,\bm q)=\frac{1}{\pi}\,\mathcal{P} \int_{-\infty}^{ \infty}
d\omega' \, \frac{{\chi''}_n^{\rm sc}(\omega',\bm q)}{\omega'-\omega}. 
\end{eqnarray}

\begin{figure}[b]
 \begin{center}
	\includegraphics[width=1\columnwidth]{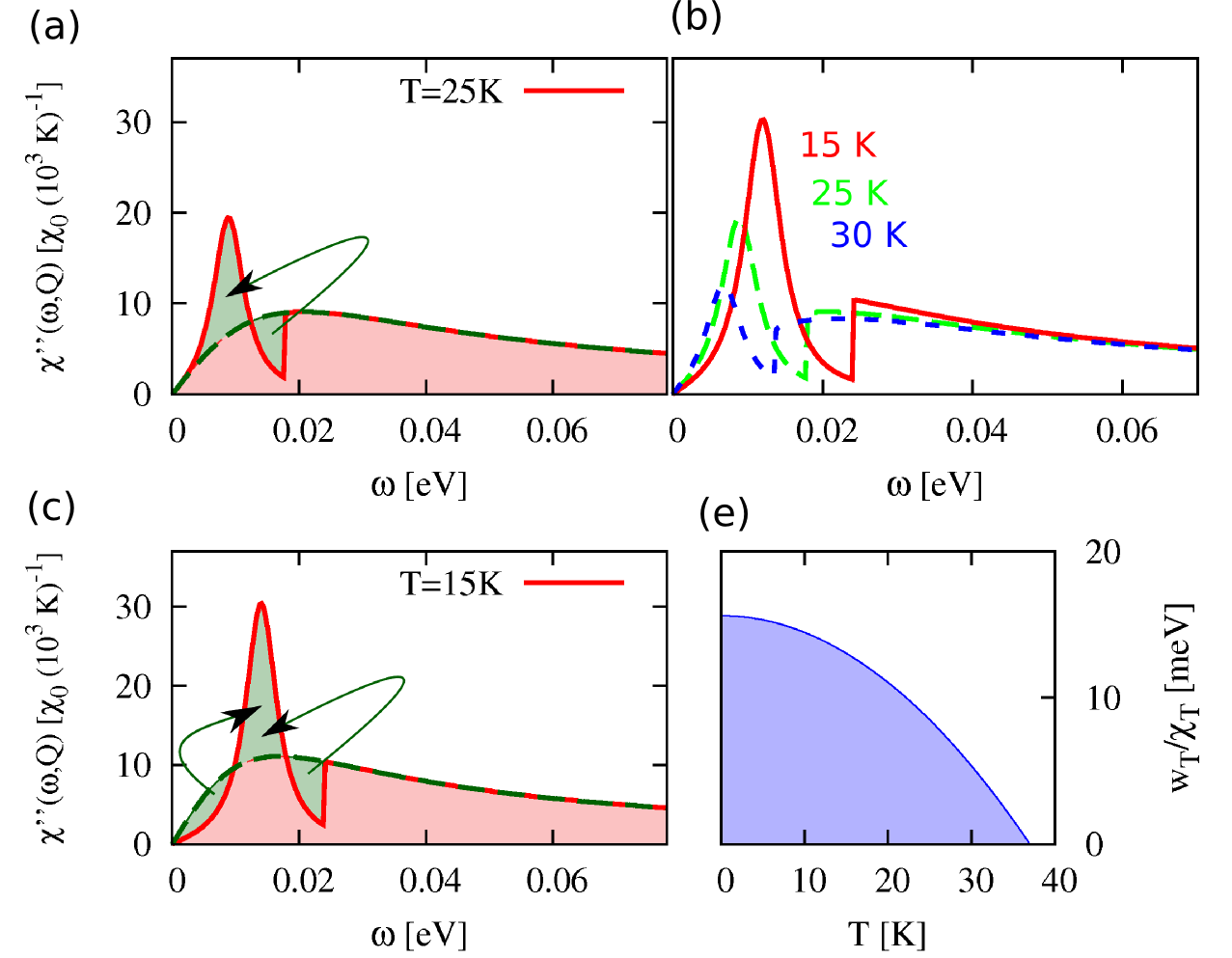}
   \caption{Energy dependence of the dynamic spin susceptibility in units of $\chi_0\,(10^3 {\rm K})^{-1}$: (a) and (c) illustration of the local sum-rule at two different temperatures. (b) Temperature evolution of the magnetic resonance. (c) Temperature dependence of the resonance weight factor $w_T/\chi_T$. }
   \label{fig2}
 \end{center}
\end{figure}

The resulting energy dependence of the dynamical susceptibility and its variation with temperature are shown in Fig.~\ref{fig2}.

The resonance weight factors $w_T^n$ in Eq.~\eqref{eq9} are determined by a 
local sum rule (Appendix \ref{appB}) which fixes the ratio between resonance and 
continuum, $w^n_T/\chi^n_T$. They are chosen so that the integrated spin structure factor, $\int_{-\infty}^{\infty}d\omega \int d\bm q \, S(\omega,\bm q)$, with $S(\omega,\bm q)=2\hbar \sum_n \chi_n''(\omega,\bm q)/(1-e^{-\hbar\omega/k_BT})$, remains temperature independent. In general the weight factors $\chi_T^n$ and $w^n_T$ depend on the orbital character. In order to determine the relative weight between $w_T^{xz}=w_T^{yz}$ ($\chi_T^{xz}=\chi_T^{yz}$) and $w_T^{xy}$ ($\chi_T^{xy}$) we studied the electronic dispersions at the X point of the Brillouin zone and find that experiments are best reproduced when the weight is roughly equal. We thus set $w_T^{xz}=w_T^{yz}=w_T^{xy}\equiv w_T$ and $\chi_T^{xz}=\chi_T^{yz}=\chi_T^{xy}\equiv \chi_T$ and are left with
\begin{eqnarray}
 \int_{-2\bar\Delta(T)}^{2\bar\Delta(T)} d\omega\int d^3\bm q \, \frac{{\chi''}_n^{r}(\omega,\bm q)-{\chi''}_n^{c}(\omega,\bm q)}{1-e^{-\hbar\omega/{k_B T}}}=0
\end{eqnarray}
to fix $w_T/\chi_T$. We illustrate this procedure in Figs.~\ref{fig2}(a) and \ref{fig2}(c): When integrating over energy and momentum the green area above and under the dashed curve has to be (approximately) the same. The ratio is in excellent agreement with the functional form 
$w_T/\chi_T= (1-T^2/T^2_{\rm c}) w_{T=0\,{\rm K}}/\chi_{T=0\,{\rm K}} $ (Fig.~\ref{fig2}(e)).

We apply the model above to Ba$_{\rm 0.6}$K$_{\rm 0.4}$Fe$_{\rm 2}$As$_{\rm 2}$ and use the experimentally motivated parameter set in Table~\ref{tab1}. 
\begin{table}[t]
\begin{center}
\caption{Parameter set appropriate for Ba$_{0.6}$K$_{0.4}$Fe$_2$As$_2$}
\label{tab1}
\begin{tabular}{|c|c|c|c|c|c|}
\hline 
\hline
$\Omega_0$ &
$\xi_0$ &
$\theta$ &
$\xi_{\rm r}$ &
$\Omega_{\rm r}$ &
$\Gamma_{\rm res}$ 
\\
\hline
&&&&&\\
$\quad 0.375 \, \frac{\mbox{meV}}{\mbox{K}} \quad $ &
$\quad 5.84 $ K$^{1/2}\quad $ &
$30 $ K &
$2$ &
$15.5 $ meV &
$3$  meV 
\\ 
&&&&&\\
\hline
\hline
\end{tabular}
\end{center}
\end{table}
The only remaining parameter to be fixed is the quantity $g^2\chi_0$, where the overall weight $\chi_0$ can in principle be extracted from experiment but is not known for this compound, and $g$ is the coupling constant between electrons and spin fluctuations.
To account for the periodicity in momentum space we replace the factors 
$|\bm q - \bm Q_\alpha|^2$ in Eq.~\eqref{eq7} and \eqref{eq9} by 
\begin{equation}
|\bm q - \bm Q_\alpha|^2 \to
4\left[\sin^2\left(\frac{q_x-Q_{\alpha x}}{2}\right)+\sin^2\left(\frac{q_y-Q_{\alpha y}}{2}\right)\right]
\end{equation}
and neglect the $q_z$-dependence for our purpose as it varies weakly. \cite{Lumsden2009, Park_et_al_2010} We use the in-plane lattice constant $a$ and the out-of-plane lattice constant $c$ as units of length. 

The resulting energy dependence of the spin-excitation spectrum and its evolution with temperature is presented in Fig.~\ref{fig2} (b).
Experimentally the resonance appears at an energy of $\Omega_{\rm res}^{T=7\,{\rm K}}\approx 14 {\rm meV}\,$ \cite{Christianson2008} below the particle-hole continuum $\omega<2\breve \Delta(T=15 \, {\rm K}) \approx 24\,{\rm meV}$. \cite{Ding2008} In momentum space the mode is peaked around the wave vectors $\bm Q_{\alpha=1,2 }$ with a correlation length $\xi_{\rm r}$ of nearly twice the lattice constant. When the temperature is decreased below $T_{\rm c}$, the resonance gains weight and eventually becomes the dominating part of the spectrum at low energies. Note that the gain in weight is due to two parallel processes: a suppression of high-energy weight due to an increase in correlations (already present in the normal state), and an increase of the superconducting gap (and thus the spin excitation gap) with decreasing temperature.

As we will see later, the characteristic temperature dependence of the resonance mode imprints itself onto the electronic spectrum, where the resonance leads to a significant effect below $30\,$K.

\section{Coupling to spin fluctuations}
We are ultimately interested in the renormalization of the electronic dispersion 
and in the electronic lifetime 
as a result of the coupling of conduction electrons to spin fluctuations both in the normal and in the superconducting state. We model this coupling by an effective electron-electron interaction mediated by the exchange of spin fluctuations. We concentrate on the leading (quadratic) contribution in the coupling constant. This approximation can be justified in part by a phase-space consideration. Small parameters are introduced by the restricted phase-space areas available for electronic quasiparticle excitations as well as for the low-lying spin-fluctuation modes. This introduces stringent kinematic restrictions. \cite{Abanov2002}

\subsection{Formalism}
In this section we summarize the formalism we use to study these effects.
We use a perturbative approach based on Gor'kov Green's functions with the coupling between the
conduction electrons and the spin fluctuations as the expansion parameter.
The unperturbed Green's functions are diagonal in band index, with normal (diagonal) and anomalous (off-diagonal) components $G^{(0)}_\mu(\epsilon,\bm k)$ and $F^{(0)}_\mu(\epsilon, \bm k)$. The renormalized Green's functions, $G_{\mu\nu}$ and $F_{\mu\nu}$ are not diagonal in band index due to the interactions induced by spin-fluctuation exchange. The Green's functions in the orbital basis, $G_{mn}$ and $F_{mn}$, are related to those in band representation, $G_{\mu \nu}$ and $F_{\mu \nu }$, by
\begin{eqnarray}
  \label{eq11}
  G_{mn}(\epsilon, \bm k)&=&\sum_{\mu\nu}\,a_\mu^m(\bm k)a_\nu^n(\bm k)^* \, G_{\mu\nu}(\epsilon,\bm k), \\
  \label{eq12}
   F_{mn}(\epsilon, \bm k)&=&\sum_{\mu\nu}\,a_\mu^m(\bm k)a_\nu^n(-\bm k) \, F_{\mu\nu}(\epsilon,\bm k). 
\end{eqnarray}

We couple electrons to the spin-fluctuation spectrum with an energy and momentum independent coupling constant $g$ (instantaneous and local coupling). The retarded diagonal and off-diagonal self-energies are then given by
\begin{eqnarray}
  \label{eq13}
  \Sigma_{mn}^{\rm R}=\delta_{mn}\Sigma^{\rm R}_{n}, \quad \Phi_{mn}^{\rm R}=\delta_{mn}\Phi_n^{\rm R},
\end{eqnarray}
written in terms of the retarded ($\rm R$) and Keldysh ($\rm K$) Green's functions 
\begin{eqnarray}
  \label{eq14}
  \Sigma_n^{\rm R} &=& -\frac{\imath}{2}g^2\,\left(G_{nn}^{\rm K} \ast \chi^{\rm R}_n + G_{nn}^{\rm R} \ast \chi_n^{\rm K}\right), \\
 \label{eq14b}
  \Phi_n^{\rm R} &=& -\frac{\imath}{2}g^2\,\left(F_{nn}^{\rm K} \ast \chi^{\rm R}_n + F_{nn}^{\rm R} \ast \chi_n^{\rm K}\right),
\end{eqnarray}
with $(A\ast B)(\epsilon,\bm k)=
\int \frac{d\omega}{2\pi} \int \frac{d^3\bm q }{(2\pi)^3}
A(\epsilon-\omega,\bm k - \bm q)\,B(\omega,\bm q)$ as explained in Ref.~\onlinecite{Eschrig2006}. 
In equilibrium the Keldysh components are given by
\begin{eqnarray}
 \label{eq15a}
  \chi_n^{\rm K}(\omega,\bm q) &=& 2 \imath \,{\rm Im} \chi_n^{\rm R}(\omega,\bm q)\,[1+2\,b(\omega)],\nonumber \\
  G_{nn}^{\rm K}(\epsilon,\bm k) &=& 2 \imath \,{\rm Im} G^{\rm R}_{nn}(\epsilon,\bm k)\, [1-2f(\epsilon)], \nonumber \\
  F_{nn}^{\rm K}(\epsilon,\bm k) &=& [F^{\rm R}_{nn}(\epsilon,\bm k)-F_{nn}^{\rm R}(-\epsilon,-\bm k)]\, [1-2f(\epsilon)],
\quad \nonumber
\end{eqnarray}
where $f$ and $b$ are the fermionic and bosonic distribution functions.
The normal and superconducting state dispersion relations are given by $\chi_n=\chi_n^{\rm c}$ and $\chi_n=\chi_n^{\rm sc}$ respectively, where $n=1,2,\,4$. The self-energies then enter the Dyson equation in terms of Nambu-Gor'kov Green's functions
\begin{eqnarray}
  \hat G_{mn}^{\rm R^{-1}} &=&  \hat G_{mn}^{(0)\rm R^{-1}}-\hat \Sigma_{mn}^{\rm R} \nonumber \\
  \label{eq16}
                   &=& \sum_\mu a_\mu^m a_\mu^{n*}\,(\hat G_{\mu }^{(0)\rm R^{-1}}-\hat \Sigma_n^{\rm R}),
\end{eqnarray}
where we used the completeness relation $\sum_\mu a_\mu^m a_\mu^{n*}=\delta_{mn}$ and
\begin{eqnarray}
  \label{eq17}
  \hat G_{mn}^{\rm R} &=&\left(\begin{array}{cc} G^{\rm R}_{mn} & F^{\rm R}_{mn} \\
                                      \tilde F^{\rm R}_{mn} & -\tilde G^{\rm R}_{mn}
                           \end{array} \right), \,\, \\
  \label{eq18}
   \hat \Sigma_{mn}^{\rm R} &=&\left(\begin{array}{cc} \Sigma^{\rm R}_{mn} & \Phi^{\rm R}_{mn} \\
                                      \tilde \Phi^{\rm R}_{mn} & -\tilde \Sigma^{\rm R}_{mn}
                           \end{array} \right),
\end{eqnarray}
with $\tilde A(\epsilon,k)=A^*(-\epsilon,-k)$. From Eq.~\eqref{eq16} we obtain 
\begin{eqnarray}
  \label{eq19}
  &&\hat G_{mn}^{\rm R^{-1}}(\epsilon, \bm k)=\sum_\mu a_\mu^m(\bm k)a_\mu^n(\bm k)^*   \\
  \nonumber
  &\times& \left[Z_n(\epsilon,\bm k)(\epsilon+\imath \delta)\hat{\mathbbm 1} - \zeta_\mu^n(\epsilon,\bm k) \hat \tau_3 - \Delta^n(\epsilon,\bm k) \hat \tau_1 \right],
\end{eqnarray}
where $\hat {\mathbbm 1}$ is a $\rm 2 \times 2$ unit matrix 
and $\hat \tau_i$ ($i=1,2,3$) are 
the Pauli matrices in Nambu space. The renormalized dispersion and order parameter as well as the renormalization function are given by
\begin{eqnarray}
  \label{eq20}
  \zeta_\mu^n(\epsilon,\bm k)&=&\zeta_\mu(\bm k) + \frac{\Sigma_n^{\rm R}(\epsilon,\bm k) + \tilde \Sigma^{\rm R}_n(\epsilon,\bm k)}{2}, \\
  \label{eq21}
  \Delta^n(\epsilon,\bm k)&=&\Delta^n_{\bm k} + \Phi_n^{\rm R}(\epsilon,\bm k), \\
  \label{eq22}
  Z_n(\epsilon,\bm k)&=& 1-\frac{\Sigma_n^{\rm R}(\epsilon,\bm k) - \tilde \Sigma^{\rm R}_n(\epsilon,\bm k)}{2(\epsilon+\imath \delta)},
\end{eqnarray}
using the fact that in our case $\Phi_n^{\rm R}=\tilde \Phi_n^{\rm R}$. 
Note that the real part of the dynamical susceptibility is determined by a Kramers-Kronig analysis only up to a constant. This would lead to energy independent contributions to $\zeta_\mu^{n} $ and $\Delta^n$.
However, any such contribution can be thought about as absorbed into the band structure and $\Delta^n_{\bm k}$, which enter in our approach as phenomenological parameters. Finally,
by inverting Eq.~\eqref{eq19} numerically the spectral function is obtained from the first diagonal element of Eq.~\eqref{eq17},
\begin{eqnarray}
  \label{eq23}
  A(\epsilon,\bm k)=
-\frac{1}{\pi} {\rm Im} \sum_m  G_{mm}^{\rm R}(\epsilon, \bm k)
.
 \end{eqnarray}
For analytic properties of the self-energies and a sum rule  we refer to Appendix \ref{appD}.
The density of electrons is given  in terms of the spectral function by
\begin{equation}
\label{eq28-2}
\rho =\frac{2 }{V}\int d\epsilon \int \frac{d^3\bm k}{(2\pi)^3}\, f(\epsilon )A (\epsilon,\bm k), 
\end{equation}
When switching on the interactions, it must be ensured that the chemical potential is adjusted such that the electronic density stays constant.

 We calculate the convolutions in Eqs.~\eqref{eq14} and \eqref{eq14b} numerically by a fast Fourier transform using bare Green's functions (a procedure supported by the numerical studies in Ref.~\onlinecite{Vilk1997}) with a broadening parameter of $\delta=4\,$meV. For this we use a very fine $512 \times 512 \times 8$ $\bm k$-mesh and $128$ points in energy space. In addition a high-energy cutoff of $\omega_{\rm c}=200\,$meV was introduced in the spectrum of spin excitations. The exact value of this is, however, less important, as any change in the high-energy part contributes only to an additional renormalization factor $Z^{\rm HE}$ which we address next.

\subsection{High-energy cutoff and high-energy renormalization}

As already mentioned, we use a high-energy cutoff $\omega_{\rm c}=200\,$meV in the spin excitation spectrum. The precise value of this cutoff is, however, not essential. To see this, let us assume that we change the high-energy cutoff from
the value $\omega_{\rm c} $ to $\omega_{\rm c}+\Delta \omega $.
Accordingly, the self-energy will have two terms, which we call the
low-energy and the high-energy part, i.e. $\Sigma^{\rm R}_{\epsilon,\bm k}=\Sigma^{\rm LE}_{\epsilon,\bm k} + \Sigma_{\epsilon,\bm k}^{\rm HE}$. 
For sufficiently large cutoff $\omega_{\rm c}$,
the high-energy part contributes at low energies (well below $\omega_{\rm c}$)
mostly to the real part and is linear in energy, i.e. $\Sigma_{\epsilon,\bm k}^{\rm HE}\approx {\Sigma'}_{\epsilon,\bm k}^{\rm HE}\approx -b_n\,\epsilon\quad(b_n \in \mathbb R)$ (see Appendix \ref{appC}). We exploit this fact to define an energy independent high-energy renormalization factor $Z^{\rm HE}_n=1-{\Sigma'}_{\rm HE}/\epsilon=1+b_n$.

The new Green's functions and self-energies corresponding to 
the cutoff $\omega_{\rm c}+\Delta \omega $ are determined by the assignments
$\zeta_\mu^n \to \zeta_\mu^n/Z_n^{\rm HE}$,
$\Delta^n \to \Delta^n/Z_n^{\rm HE}$, and
$Z_n \to 1+(Z_n-1)/Z_n^{\rm HE} $ in the expression for $[\hat G_{mn}^{\rm R}]^{-1}$,
and $[\hat G_{mn}^{\rm R}]^{-1} \to Z_n^{\rm HE}[\hat G_{mn}^{\rm R}]^{-1}$.
These relations simplify further for equal coupling constants for all orbitals, in which case $Z_n^{\rm HE} \equiv Z_{\rm HE}$ is independent of the orbital index $n$,
and $\hat G_{mn}^{\rm R} \to \hat G_{mn}^{\rm R}/Z_{\rm HE}$.

We treat $Z_{\rm HE}$ as a free parameter, which is of order 1 and modestly temperature dependent. This temperature dependence might be at first view a bit surprising, as from Fermi liquid theory one is used to temperature variations of the high-energy vertices that are negligible. However, one should remember that in our case the susceptibility is strongly temperature dependent in the normal state, which includes temperature-dependent shifts of spectral weight between low and high-energy. This is manifestly non-Fermi-liquid behavior.

\begin{figure}[t]
 \begin{center}
    \includegraphics[width=0.7\columnwidth]{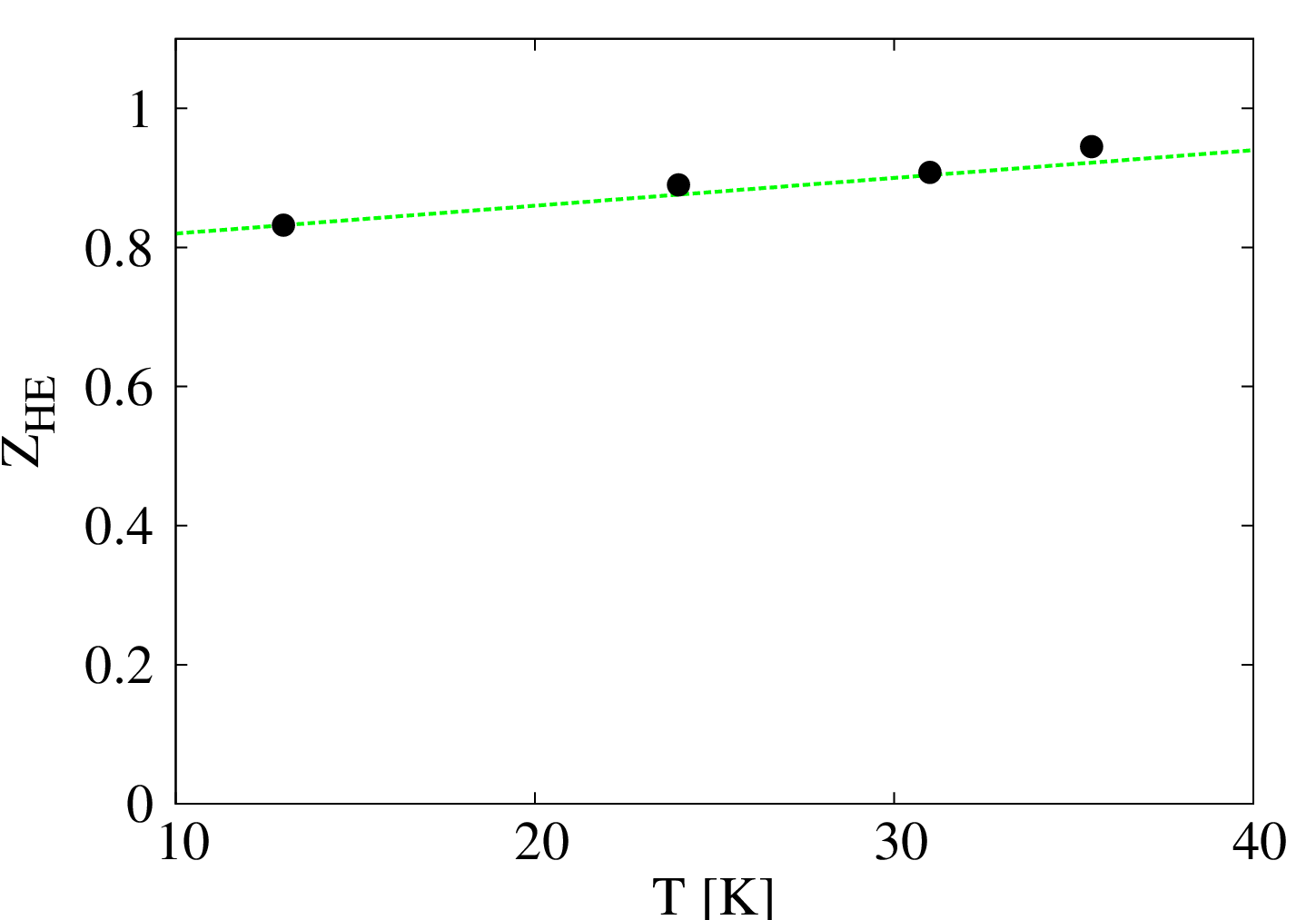} 
   \caption{The high-energy renormalization factor $Z_{\rm HE}$ shows a linear dependence in temperature.}
   \label{fig14}
 \end{center}
\end{figure}

We fix the high-energy renormalization factor above a certain reference temperature $T_{\rm ref}$ (which we chose as 50K) in the normal state $Z_{\rm HE}^{T_{\rm ref}}=1$, and we determine $Z^T_{\rm HE}$ for lower temperatures so that the superconducting and the normal state dispersion merge for high energies, as observed in experiment. \cite{Richard2008, Hasan2008} 
Our numerical solutions show that $Z_{\rm HE}^T$ varies slowly 

in the temperature range $T=10-50\,K$ 
(Fig.~\ref{fig14} shows an example for $g^2\chi_0=1.17\times 10^3 \mu_{\rm B}^2$eV K). We underline that this additional high-energy renormalization applies only for energies well below the spin-fluctuation cutoff $\omega_{\rm c}$.

We note that the thus determined $Z_{\rm HE}$ is lower than one below a temperature of 50 K. 
In general, the renormalization function due to the entire spin-fluctuation spectrum must be larger than one for zero energy.\cite{Dolgov94} However, this refers to the sum of our model spectrum and the correction due to the high-energy cutoff. 
Our model susceptibility overestimates the high-energy contributions, so that a negative correction at high energies is in place, leading to negative values for $b_n$; the sum $Z_n+b_n$ at zero energy is, however, always larger than one.

\section{Self energy effects}
In this section we discuss effects resulting from the self-energies,
Eqs.~\eqref{eq14} and \eqref{eq14b}. We start with the diagonal self-energies, which determine the quasiparticle scattering rate and the quasiparticle band renormalization, and proceed then with the off-diagonal self-energies, which renormalize the superconducting order parameter.

All results in this section are for 
$g^2\chi_0=1.17\times 10^3 \mu_{\rm B}^2$eV K.

\subsection{Scattering rate and band renormalization}

\begin{figure}[b]
 \begin{center}
	\includegraphics[width=1.\columnwidth]{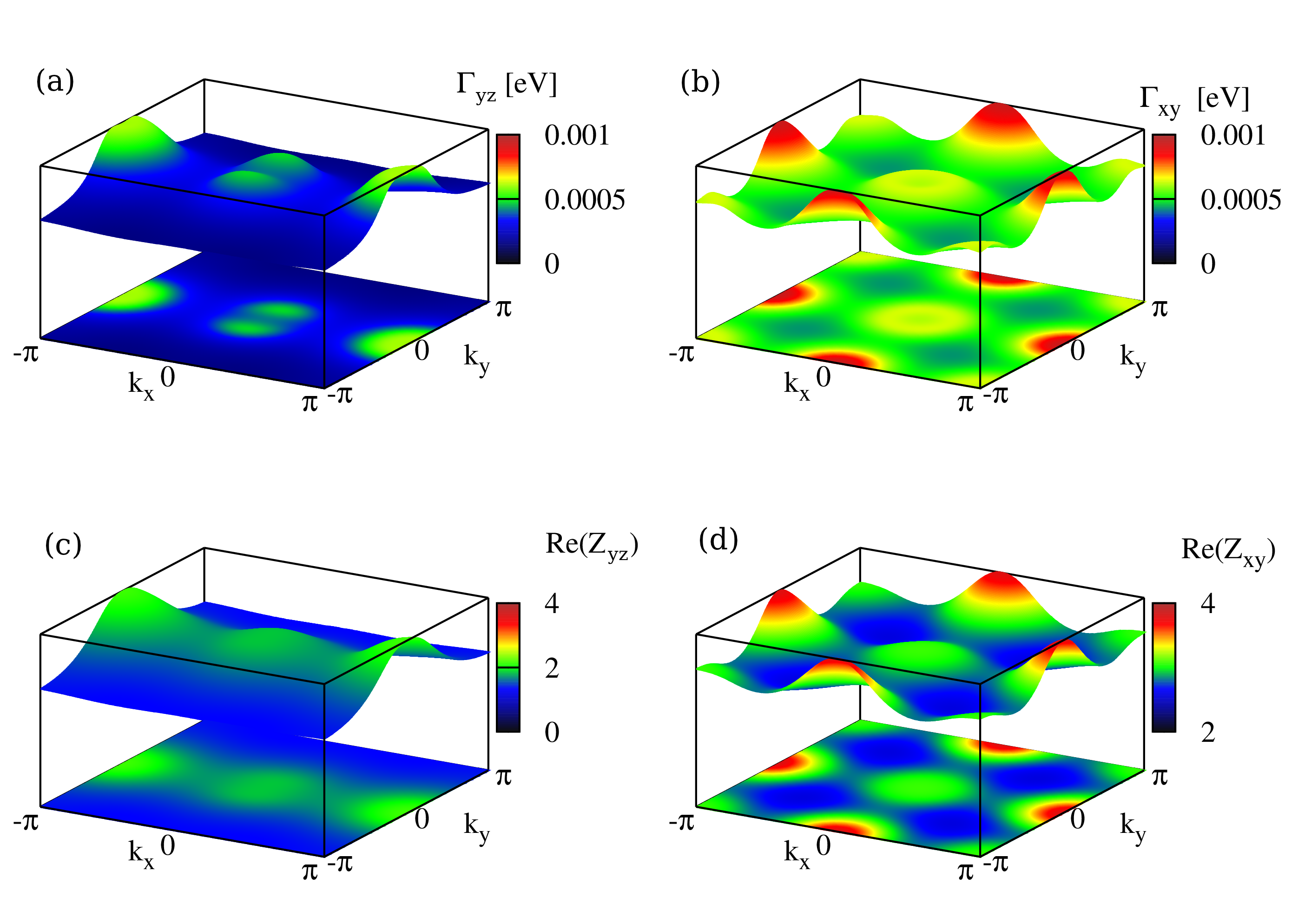}
   \caption{Momentum dependence of the low-energy scattering rate $\Gamma_n(\epsilon\rightarrow 0,\bm k)$ in the $d_{yz}$ ($n=yz$) and the $d_{xy}$ ($n=xy$) channel at $T=31 $K [(a) and (b)], as well as the momentum dependence of the renormalization factor $Z_n(\epsilon \rightarrow 0,\bm k)$ for the respective orbitals $n$ [(c) and (d)].  }
   \label{fig3}
 \end{center}
\end{figure}

\begin{figure}[b]
 \begin{center}
	\includegraphics[width=0.8\columnwidth]{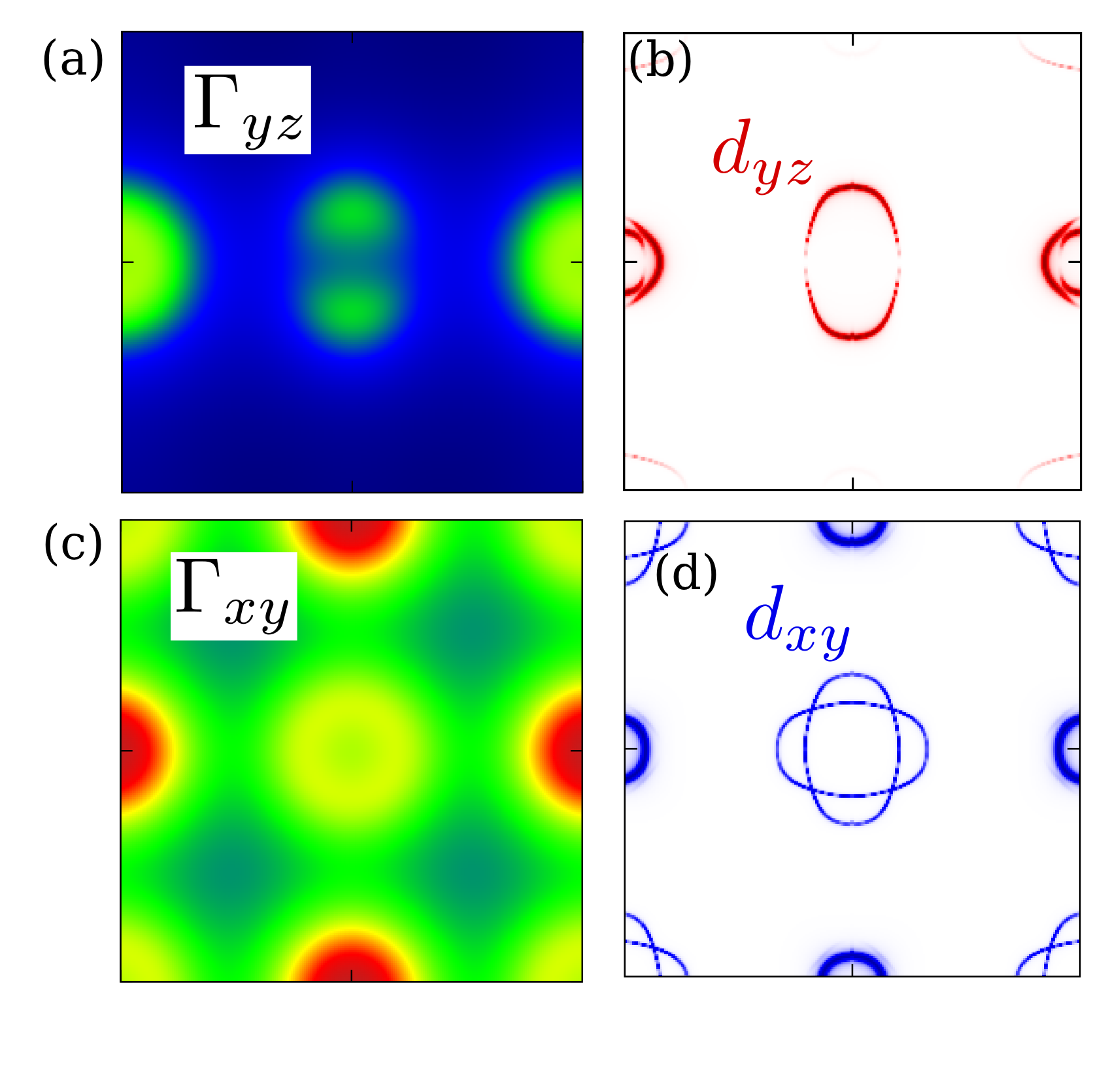}
   \caption{Intra-orbital scattering channels: The significant features in the $d_{yz}$ and $d_{xy}$ scattering rate, shown in (a) and (c) correspond to the different orbital contributions to the Fermi surface. We show in (b) and (d) the contributions of the $d_{yz}$ and the $d_{xy}$ orbitals to the Fermi surface like in Fig.~\ref{fig1} but shifted by the scattering vector $\bm Q_1$ in the case of $d_{yz}$ and $\bm Q_1$ and $\bm Q_2$ in the case of $d_{xy}$. The scattering rate reflects the underlying structure of the Fermi surface shifted by the antiferromagnetic wave vectors.   }
   \label{fig3-2}
 \end{center}
\end{figure}

The imaginary part of the diagonal self-energy, $\Sigma_n^{\rm R}(\epsilon,\bm k)$, determines the scattering rate of electronic quasiparticles. In the presence of a renormalization factor $Z_n(\epsilon , \bm k)$, Eq.~\eqref{eq22}, the scattering rate is renormalized and is given in the normal state for
the respective orbital $n=d_{xz},\,d_{yz},\,d_{xy}$ by
\begin{eqnarray}
  \label{eq24}
  \Gamma_n(\epsilon,\bm k)=-\frac{{\rm Im}\left\{\Sigma_n^{\rm R}(\epsilon, \bm k)\right\}}{{\rm Re}\left\{Z_n(\epsilon, \bm k)\right\}}.
\end{eqnarray}
The influence of the spin resonance mode on electronic quasiparticles is most pronounced in the vicinity of the Fermi surface.
Therefore the scattering rate should exhibit characteristics of the Fermi surface topology. In Figs.~\ref{fig3}(a) and \ref{fig3}(b) we show the $k_x$ and $k_y$ dependence of the low-energy $d_{yz,xy}$ scattering rate in the $k_z=0$ plane and we show the quasiparticle renormalization factor ${\rm Re}\{Z_n\}$ in Figs.~\ref{fig3}(c) and \ref{fig3}(d).

In the case of the $d_{xz,yz}$ orbitals [Fig.~\ref{fig3}(a) and \ref{fig3}(b)], both feature a two-hump structure in the middle of the Brillouin zone and a even larger peak at $\bm K_{yz}=(\pm \pi,0)$ and $\bm K_{xz}=(0,\pm \pi)$. As we demonstrate in Fig.~\ref{fig3-2}, the hump structure is a clear signature of the respective $d_{xz,yz}$-orbital contribution on the Fermi surface. In Figs.~\ref{fig3-2}(b) and \ref{fig3-2}(d), we have shifted the partial spectral functions shown in Figs.~\ref{fig1}(a) and \ref{fig1}(c) by the antiferromagnetic wave vector $\bm Q_1$ in the case of the $d_{yz}$ orbital and by $\bm Q_1$ and $\bm Q_2$ in the case of the $d_{xy}$ orbital. It can be seen that the two-hump-structure in Fig.~\ref{fig3-2}(a) corresponds to the $d_{yz}$ contributions of the $\beta_1$ pockets which lie opposite and well separated. On the other hand, the peak feature at $\bm K_{yz}=(\pm \pi,0)$ occurs due to the hole-like Fermi surface sheets at the $\alpha$ pockets where the $d_{yz}$ contributions enclose the $\Gamma=(0,0)$ point. The scattering rate is broadened compared to the Fermi surface contributions simply because the susceptibility is broadened in momentum space.
Looking at the $d_{xy}$ orbital in Figs.~\ref{fig3-2}(c) and \ref{fig3-2}(d) we essentially observe the same characteristics. The broader peak at the $\Gamma$ point results from the $\beta$ pockets and the sharp ones at the $\bm K_{xy}^1=(\pm \pi,0)$ and $\bm K_{xy}^2=(0,\pm \pi)$ points originate from the $\gamma $ Fermi surface sheets at $(\pm \pi, \, \pm \pi)$.

\begin{figure}[]
 \begin{center}
	\includegraphics[width=1.\columnwidth]{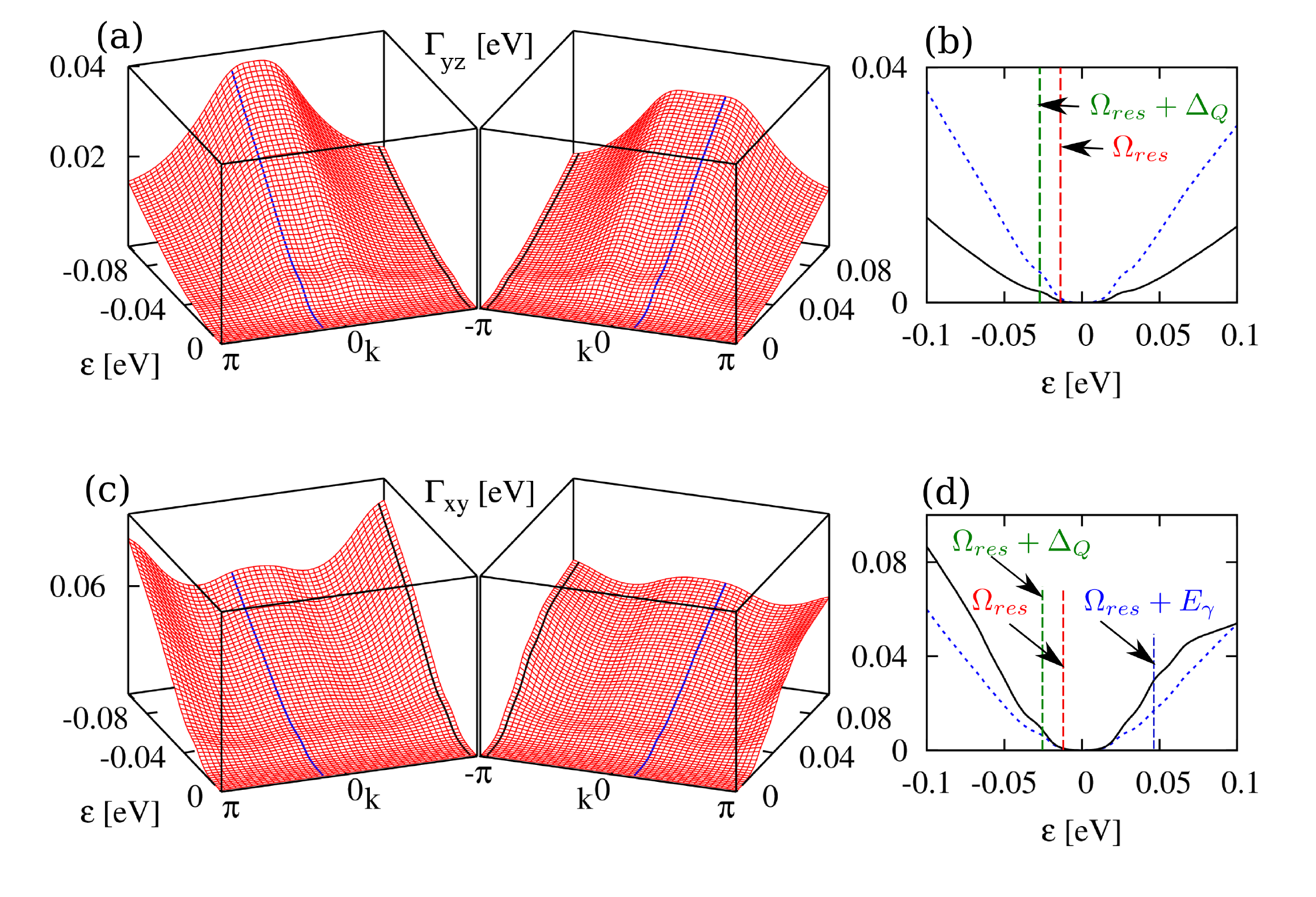} 
   \caption{Scattering rate $\Gamma_n$ at $T=15\,$K: Energy-momentum dependence along the cut $\{k_y\in[-\pi:\pi),\,k_x=k_z=0\}$ in the $d_{yz}$ and $d_{xy}$ channel [(a) and (c)]. Energy dependence of the scattering rate at fixed momentum as indicated at the left hand side by the dotted blue and full black line, respectively [(b) and (d)]. Scattering sets in at the energy $\Omega_{\rm res}$. Peaks appear at the energies $\Omega_{\rm res}+ \Delta_{\bm Q}$ and $\Omega_{\rm res}+E_\gamma $ which are due to the coupling of the resonance to excitations at the Fermi surface and the van Hove singularities at the hole pockets centered at $(\pi,\pi, z)$, respectively.}
   \label{fig4b}
 \end{center}
\end{figure}

In the superconducting state the opening of a particle-hole excitation gap leads to a suppression of the particle decay below $\epsilon<2\breve \Delta(T)$. However, the appearance of the resonance allows for scattering processes below the continuum threshold. As a consequence, inelastic scattering sets in above the resonance energy, i.e., $\epsilon > \Omega^T_{\rm res}$, as depicted in Fig.~\ref{fig4b}, where the scattering rates $\Gamma_{d_{yz}}$ and $\Gamma_{d_{xy}}$ are shown for $T=15\,$K. The largest contributions to scattering arise from states near the Fermi surfaces. This means that the resonance imprint is most significant at an energy of $\Omega_{res} + \Delta_{\bm Q} $, where for each considered Fermi surface wave vector, $\bm{k}$, $\Delta_{\bm Q}$ denotes the gap averaged over a region of diameter $\xi_{\rm r}^{-1}$ around each of the points $\bm k \pm \bm Q_{1/2} $ (which are close to another Fermi surface nested to the original one) [Fig.~\ref{fig4b}(b)]. An additional peak appears in the $d_{xy}$-scattering rate at an energy of $\Omega_{\rm res}+E_\gamma $ which is due to the van Hove singularity at the $\gamma$ pockets [Fig.~\ref{fig4b}(d)]. In the following, we will refer to $\Omega_{res} + \Delta_{\bm Q} $ as the kink-energy. In Appendix~\ref{appA} we present similar investigations of the diagonal self-energies in Eqs.~(\ref{eq20}) and (\ref{eq22}).

\subsection{The superconducting order parameter}
The origin of the pairing instability may well be related to the spin-fluctuation continuum, as demonstrated by recent FLEX calculations. \cite{Schmalian2010} Our model is restricted to the low-energy region in the spin excitation spectrum. Consequently, we have to include the high-energy incoherent part, which considerably contributes to pairing, separately. The energy range of interest ($|\omega|<\omega_{\rm c}=200\,{\rm meV}$) gives only a partial contribution (about 40-50\%) to the value of the superconducting order parameter. Thus, we must add an additional contribution $\Delta_{\bm k}^n$ resulting from the high-energy incoherent spin fluctuations. 

According to theoretical calculations the order parameter dominantly has a $s^\pm$ symmetry \cite{Hirschfeld2011, Thomale2009, Graser2009, Graser2010}, which follows approximately the form 
\begin{eqnarray}
  \label{eq25}
  \Delta^\pm_{\bm k}=\Delta_0 \cos(k_x)\cos(k_y).
\end{eqnarray}
For our model we have chosen as a first initial guess $ \Delta_{\bm k}^n=\Delta^\pm_{\bm k}$, which is independent of the orbital character. In the absence of detailed experimental data, we resort to the temperature dependence $\Delta_0(T)=\Delta_0(0)\sqrt{1-T/T_{\rm c}}$, which is true in the BCS-limit in the vicinity of $T_{\rm c}$, but deviates from the true value at much lower temperatures. ARPES measurements resolve a superconducting gap of $\Delta(15{\rm K})\approx 12 \,{\rm meV}$ at the inner hole-like $\alpha$-pocket as well as at the nested electron pockets. \cite{Ding2008} We choose $\Delta_0(0)=18.1 {\rm meV}$ so that the renormalized gap matches the experimental one at the particular points in the Brillouin zone.

\begin{figure}[t]
 \begin{center}
	\includegraphics[width=1.\columnwidth]{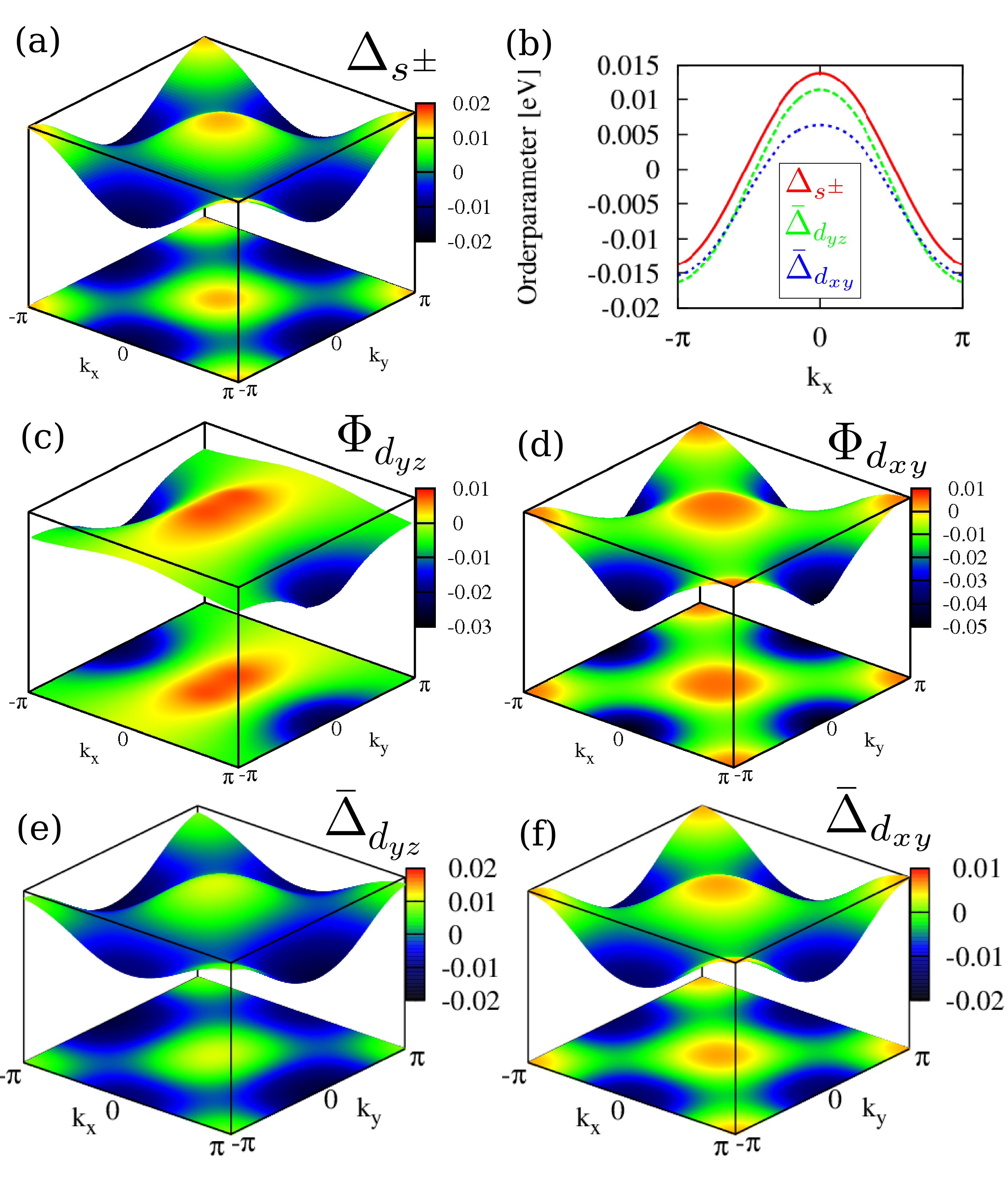} 
   \caption{Momentum dependence of the superconducting gap at $\epsilon=0\,$eV and $T=15\,$K as calculated with the ansatz in Eq.~\eqref{eq25} for the high-energy contributions: (a) $s^\pm$-gap. (c) and (d) show the off-diagonal self-energies in the $d_{yz}$ and $d_{xy}$ channel. (b) compares $\Delta^\pm$ and the renormalized order parameter $\bar \Delta_n=(\Delta^{\pm}+\Phi_n^{\rm R})/Z_n$ along the cut $\{k_x\in[-\pi:\pi),\,k_y=k_z=0\}$. (e) and (f) show the renormalized order parameter $\bar \Delta_n$ in the $d_{yz}$ and $d_{xy}$ channel. }
   \label{fig5a}
 \end{center}
\end{figure}

The low-energy spin fluctuations then lead to an orbital-dependent part, $\Phi_n$ according to Eq.~\eqref{eq14b}. We plot these and the $s^\pm$ gap in Fig.~\ref{fig5a}. 
From Fig.~\ref{fig5a}(c) it can be seen that
in the $d_{xz,yz}$ channel the coupling via the $\bm Q_{1,2}=(\pm \pi,0),\,(0,\pm \pi)$ wave vector favors an order parameter with opposite sign on the $\alpha$ and the $\beta_{1,2}$ pockets. 
The corresponding analysis for the $d_{xy}$ channel in Fig.~\ref{fig5a}(d) shows that
the presence of the holelike pockets at $(\pi, \pi, z)$ 
strongly support pairing
with opposite signs between the holelike $\gamma $ pockets and the electronlike $\beta_{1,2} $ pockets and with equal signs between the $\alpha_{1,2}$ and $\gamma $ pockets. Taken together, all three orbitals 
promote an order parameter of the form of Eq.~\eqref{eq25}. 

According to Eq.~\eqref{eq19}, the renormalized gap is given by 

\begin{equation}
\label{eq37-1}
\bar \Delta^{\rm R}_n(\epsilon,\bm k)=\frac{\Delta^\pm_{\bm k} + \Phi^{\rm R}_n(\epsilon,\bm k)}{Z_n(\epsilon,\bm k)}. 
\end{equation}

In Fig.~\ref{fig5a} we compare the high-energy contribution $\Delta_{\bm k}^\pm$, the off-diagonal self-energy $\Phi^n_{\bm k}$, and the renormalized gap given by Eq.~(\ref{eq37-1}). We find that the spin-fluctuation spectrum in the energy range $|\omega|\leq 200 {\rm meV}$ contributes nearly 40\%-50\% to the observed order parameter. To see this, note that at the $\Gamma$-point $Z_{d_{yz}}$ equals approximately 2 [see Fig.  \ref{fig3}(c)], thus the low-energy contribution $\Phi_{d_{yz}}$ must be of the same order as $\Delta^{\pm}$ to arrive at the $\bar{\Delta}_{d_{yz}}$ shown in Figs. \ref{fig5a}(b) and  \ref{fig5a}(e). Furthermore the superconducting gap originating from the low-energy part of the spin-fluctuation spectrum supports an $s^\pm$ state. We performed additional calculations for an extended $s$-wave contribution [$\cos(k_x)+\cos(k_y)$] to the order parameter ansatz in Eq.~(\ref{eq25}). According to our findings, a large extended $s$-wave contribution would imply a nodal structure in the partial density of states (for its definition see Sec.~\ref{EDC}). ARPES\cite{richard11} and quasiparticle interference (QPI)  experiments\cite{Allan2012, Haenke2012} however show a fully gapped order parameter. For our purpose it is sufficient to use the ansatz in Eq.~(\ref{eq25}).

In Appendix~\ref{appA} we also discuss the energy and momentum dependence of the off-diagonal selfenergy, Eq.~(\ref{eq21}).

\section{The electronic Spectral Function}
\begin{figure*}[]
 \begin{center}
	\includegraphics[width=1.0\textwidth]{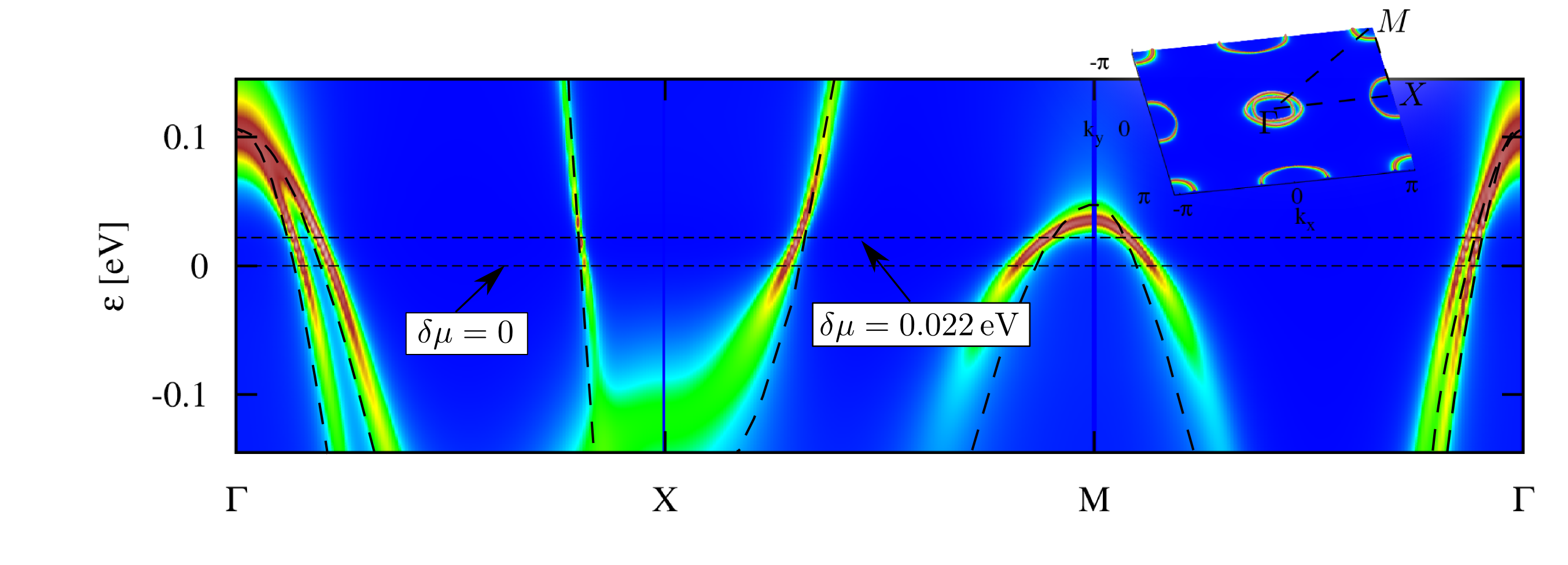} 
   \caption{Spectral function along the 
cut in the first Brillouin zone as indicated in the inset.
Here, $T=50\,$K (which corresponds to the normal state reference temperature we have used for analyzing the spectral functions in the superconducting states as discussed in the text), and $g^2\chi_0=1.17\times 10^3 \mu_{\rm B}^2$eVK.
The black wide-dashed lines present the bare band structure. Zero energy corresponds to the unrenormalized chemical potential ($\mu_0$, for the bare band structure), and the shift of the chemical potential, $\delta \mu$, for the renormalized band structure is indicated as a second dashed line.}
   \label{fig7}
 \end{center}
\end{figure*}

Self-energies are measurable via their effect on the line shape of the spectral function and the associated dispersive features. The spectral function can be measured by ARPES techniques. 
In this section we will apply procedures to our theoretically obtained spectral function
that routinely have been applied in ARPES experiments.
This allows us to compare our results directly with
experimentally observed quantities, such as effective self-energies extracted from ARPES measurements. 
In addition we will present results that can be compared with other spectroscopic experiments such as $c$-axis tunneling through superconductor-insulator-normal metal and superconductor-insulator-superconductor junctions.

\subsection{Normal state dispersion, Fermi surfaces, and chemical potential}

\begin{figure}[b]
 \begin{center}
	\includegraphics[width=0.9\columnwidth]{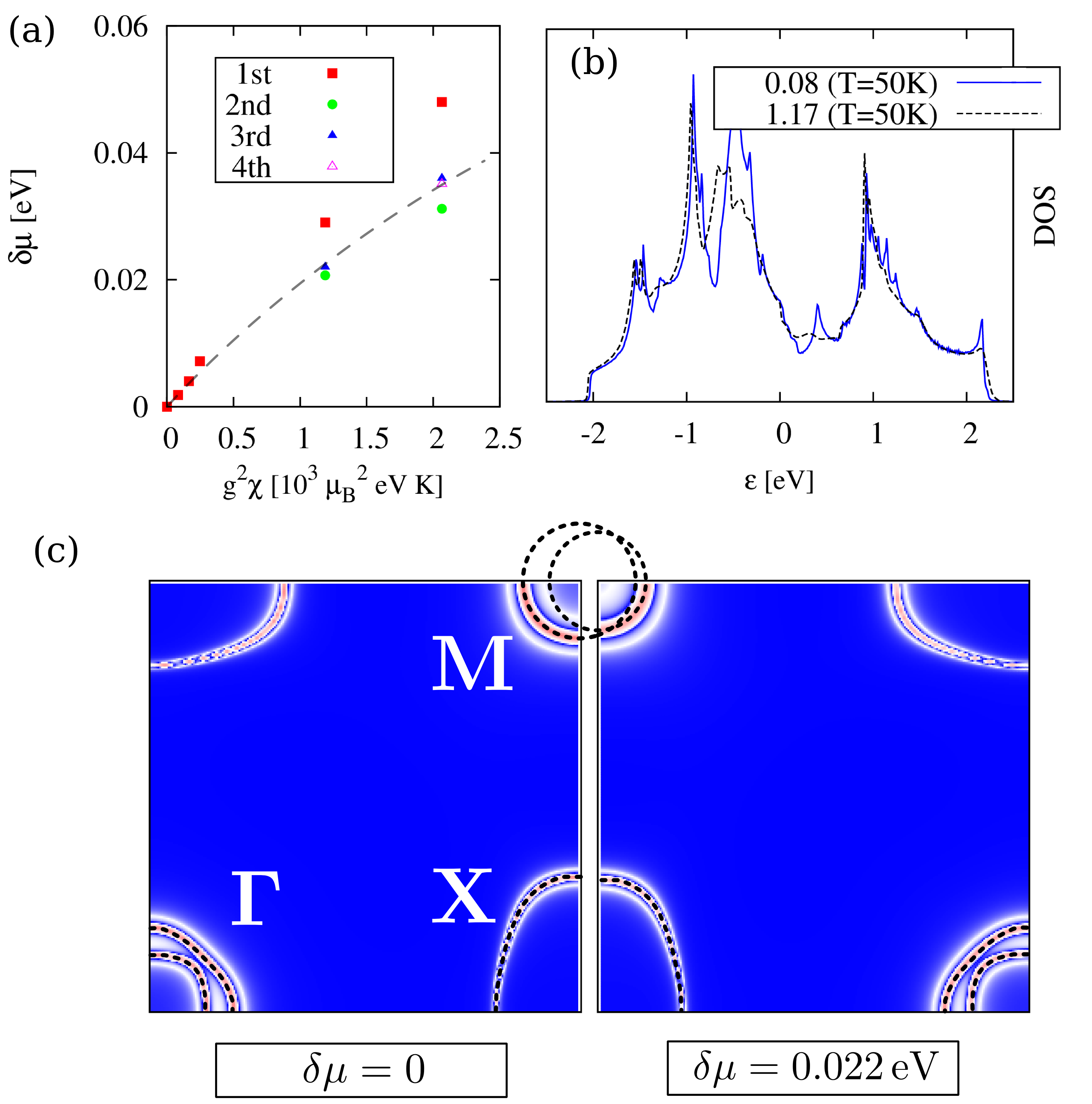} 
   \caption{
(a) Dependence of the chemical potential $\delta \mu$ on the coupling constant $g^2\chi_0$. Shown is the result after the first, second, third, and fourth iteration at $T=50\,$K. (b) The electronic density of states for different coupling parameters $g^2\chi_0=0.08\,\times 10^3 \mu_B^2\,eV\,$K and $1.17\, 
\times 10^3 \mu_B^2\,eV\,$K at $T=50\,$K. (c) Fermi surfaces in a quarter of the Brillouin zone at $k_z=0$ for $g^2\chi_0=0$ and $\delta \mu=0$ and for $g^2\chi_0=1.17\times 10^3 \mu_B^2\,eV\,$K and $\delta \mu=22\,$meV, at $T=50\,$K.}
   \label{fig8}
 \end{center}
\end{figure}

Dispersions obtained by ARPES experiments usually differ from density functional calculations by a renormalization factor of roughly 1.5-2.\cite{Ding2008_2, Lu2009, Sekiba2009,Kordyuk2011} 
We show that this renormalization can be well accounted for by the incoherent high-energy spin-fluctuation spectrum coupling to the electronic excitations.
In Fig.~\ref{fig7} we compare the spectral function obtained from the bare electronic structure with the one obtained after coupling to the spin-fluctuation continuum at $T=50\,$K. 
As can be seen, there is a pronounced energy broadening of the spectral function
at high energies, whereas it remains sharp in the vicinity of the Fermi surface. 
The high-energy broadening increases with excitation energy.
Furthermore, the figure shows that the energy bands are strongly
renormalized, in particular near the X and M points.
By choosing the coupling constant $g$ to be the same for all orbitals we are able to reproduce the experimentally observed shallow electron pocket near the X point.\cite{Ding2008_2, ChangLui2008} 

The Luttinger theorem requires that the volume of the Fermi surfaces must stay constant as a consequence of particle conservation. 
In Fig.~\ref{fig8}~(a) the chemical potential $\delta \mu$ necessary to maintain a constant particle number is shown. By shifting the bands, $\zeta(\bm k,\mu_0)=\zeta' (\bm k) - \mu_0 \rightarrow \zeta (\bm k,\mu) =\zeta' (\bm k) - \mu $, the density of electrons
$\rho $, given by Eq.~\eqref{eq28-2},
is fixed. In the following, we denote by $A_\mu$ the spectral function calculated for 
chemical potential $\mu $ and for self-energies $\hat \Sigma [A_\mu ]$. 
For the unperturbed band structure 
$\mu_0=-50\,$meV as chosen in Sec.~\ref{sec1b} in order to simulate hole doping.

To obtain the correct chemical potential we need to find 
$\mu $ such that $\rho[A_\mu ]=\rho_0$ with $A_\mu=A[\mu, \Sigma_0(\mu )]$, where
$ \Sigma_0(\mu )$ is calculated from the unperturbed spectral functions,
$A^{(0)}_{\mu }$, with chemical potential $\mu $. 
In a first step we calculate self energies $ \Sigma_0(\mu_0)=\Sigma_0[A^{(0)}_{\mu_0}]$. We then,
in a next step, calculate $A_{\mu_0}=A[\mu_0, \Sigma_0(\mu_0)]$ and from this $\rho_1=\rho[A_{\mu_0 }]$.
In order to obtain a first approximation to the chemical potential,
we linearize its 
functional form as function of the charge carrier density, defining
\begin{eqnarray}
  \label{eq28-1}
   \mu_1=\mu_0 + \frac{\rho_1-\rho_0}{\partial \rho /\partial \mu}.
\end{eqnarray}
The partial derivative is obtained by making a second calculation for a shifted potential $\mu=\mu_0+\Delta\mu$ with $\Delta\mu=5\,$meV and repeating the procedure above leading to $\rho_1^*=\rho[A_{\mu}]$ and
\begin{eqnarray}
 \frac{\partial \rho}{\partial \mu}\approx \frac{\rho_1^*-\rho_1}{\Delta\mu}.
\end{eqnarray}
We then repeat these steps: $\Sigma_0(\mu_1)=\Sigma_0[A^{(0)}_{\mu_1}]\to A_{\mu_1}=A[\mu_1,\Sigma_0(\mu_1)] \to \rho_2=\rho[A_{\mu_1}] \to \mu_2=\mu_1+\frac{\rho_2-\rho_1}{\partial \rho/\partial \mu}$, and so forth until convergence. 

In Fig.~\ref{fig8}(a) we show $\delta \mu_j= \mu_j-\mu_0$ for $j=1,2,3$
as a function of $g^2\chi_0$. 
In Fig.~\ref{fig8}(b) we compare the density of states for two different coupling strengths for $T=50\,$K.

Finally, in Fig.~\ref{fig8}(c) we show the Fermi surface for $g^2\chi_0= 0$ and for $g^2\chi_0= 1.17\times 10^3 \mu_B^2\,eV\,$K as obtained from the above procedure.
The main effect is that all Fermi surface sheets shrink with increasing interaction,
whereby a redistribution of populated states from the electronlike band to the holelike bands takes place near the Fermi surfaces.
This has the most pronounced effect on the small hole pockets at the M points.
We speculate that for sufficiently strong spin fluctuations a Lifshitz transition is triggered with an interaction-induced collapse of the entire hole pocket.

Experimentally, it was shown that a shift of the chemical potential corresponds to a linear change in doping, \cite{ChangLui2008,Neupane2011} meaning that in the weak-coupling regime the coupling strength is linear to the doping level. This observation would correspond in our calculations to solutions that are already to a good approximation reached after the first iteration, $\mu_1 \approx \mu $. 
Our results support this picture for the weak-coupling regime,
$g^2\chi_0<1\times 10^3 \mu_B^2\,eV\,$K; however, they show pronounced deviations at stronger coupling.

\subsection{Angle-resolved photoemission}

In photoemission experiments the intensity of photoelectrons is proportional to $f(\epsilon)A(\epsilon,\bm k)$, where $f$ is the Fermi distribution function and $A$ is the spectral function. In this section we discuss 
our results for the
spectral function $A(\epsilon,\bm k)$ obtained from Eq. \eqref{eq23}. 
For this purpose we follow a procedure commonly used to analyze ARPES data: The superconducting state dispersion is compared to that of a reference dispersion in the normal state (at a temperature well above $T_{\rm c}$) and renormalization effects are directly inferred by quantifying the differences. In addition, a linewidth analysis gives information about the imaginary parts of the diagonal self-energies. These procedures give naturally not as precise information as an orbitally resolved measurement. The reason is that the spectral function is obtained from the orbital-dependent self-energies via a matrix inversion [see Eqs.~\eqref{eq19} and \eqref{eq23}], and all orbital contributions to the self-energy mix with each other when considering the spectral function. 
In other words, the self-energies are not diagonal in the band index, whereas the renormalized tight binding bands are not diagonal in the orbital index.
This refers also to orbital-resolved photoemission experiments, as each orbital component of the spectral function is obtained by the same matrix inversion, Eq.~\eqref{eq19}, thus mixing orbital components of the self-energies. Nevertheless, it is possible to compare theoretical and experimental data when applying identical procedures to the theoretically and experimentally obtained spectral functions. We call the related quantities {\it effective self-energies}.
All results in this section are for $g^2\chi_0=1.17\times 10^3 \mu_{\rm B}^2$eV~K.

\subsubsection{Effective self-energies: dispersions and linewidth}

Motivated by experimental findings, we use a fitting procedure for our theoretically obtained spectral functions as discussed in the following.
For fixed energy the momentum dependence of the spectral function (MDC: momentum distribution curve)\cite{Norman01} is peaked and often well approximated by a Lorentzian. 
In addition to MDCs,
also the energy dependence for fixed momentum (EDC: energy distribution curve) can be useful to analyze the dispersion anomalies of superconductors that arise from coupling to bosonic modes. Whereas in the cuprates the EDCs exhibit a pronounced dip-hump structure, \cite{Eschrig2006, EschrigNormanApr2003} there is no such pronounced behavior for iron-based superconductors. \cite{Richard2008} 
Here the self-energy effects are best extracted by examining the MDC-derived dispersions. 

When concentrating on such a Lorentzian peak centered at momentum $\bm k$ and energy $\epsilon_{\bm k}$ for a certain band well separated from all other bands, we consider the immediate region around this point in energy-momentum space. The energy dispersion $\epsilon_{\bm k}$ of the considered band in relation to the bare band dispersion defines an effective $\Sigma'$ via the relation
\begin{eqnarray}
  \label{resigma}
\varepsilon_{\bm k}=\zeta_{\bm k}+\Sigma'_{\rm LE}(\varepsilon_{\bm k},\bm k). 
\end{eqnarray}
This equation is motivated by the analogous equation for a single band system,
where $\Sigma'_{\rm LE}$ describes the true low-energy part of the self-energy, and where the above equation represents an implicit equation for the renormalized dispersion $\varepsilon_{\bm k}$.
In the present case, we instead calculate the renormalized dispersion using the true self-energies, and then we define the effective self-energy $\Sigma'_{\rm LE}$ via Eq.~\eqref{resigma}.

By assuming that ${\Sigma'}_{\rm LE}$ is nearly constant in the 
momentum region over which the Lorentzian describing the MDC spectral function is spread,
the MDC spectral function can be written in the form
\begin{eqnarray}
  \label{eq30}
    A(\epsilon,\bm k)=\frac{1}{\pi} \frac{\tilde \Sigma''_{\rm LE}(\epsilon,\bm k)}{\{\tilde {\bm v}_\epsilon(\bm k - \bm k_\epsilon)\}^2 + \{\tilde \Sigma''_{\rm LE}(\epsilon,\bm k)\}^2},
\end{eqnarray}
where $\bm k_{\epsilon}$ is obtained from $\epsilon=\tilde \varepsilon_{\bm k_\epsilon}\equiv\varepsilon_{\bm k_\epsilon}/Z_{\rm HE}$ and 
\begin{eqnarray}
  \label{eq31}
  \tilde \Sigma''_{\rm LE}&=&\Sigma''_{\rm LE}/Z_{\rm HE}, \\
  \label{eq32}
  (\tilde\varepsilon_{\bm k}-\epsilon)|_{\bm k \approx \bm k_\epsilon}&\approx &\frac{d \tilde\varepsilon_{\bm k}}{d \bm k}\Big |_{\bm k=\bm k_\epsilon} (\bm k - \bm k_\epsilon)   
\equiv \tilde{\bm v}_\epsilon\,(\bm k-\bm k_\epsilon), 
\qquad
\end{eqnarray}
and the group velocity is related to the effective self-energy by
\begin{eqnarray}
  \tilde{\bm v}_\epsilon
  &=&
\frac{1}{Z_{\rm HE}}
\left\{\frac{ \partial \zeta_{\bm k}/\partial \bm k + \partial \Sigma'_{\rm LE}/\partial \bm k}{1-\partial \Sigma'_{\rm LE}/\partial \varepsilon_{\bm k}}\right\}_{\bm k=\bm k_\epsilon} 
  \label{eq33}
\end{eqnarray}

The {\it bare} dispersion $\zeta_{\bm k}$ cannot be measured. However, it is temperature independent, such that the quantity
\begin{eqnarray}	
\label{eq35}
\Sigma'(\epsilon) \equiv
\tilde \varepsilon_{\bm k_\epsilon}^T-\tilde \varepsilon_{\bm k_\epsilon}^{T_{\rm ref}}
\end{eqnarray}

gives a measure of the self-energy effects only,
$\Sigma'(\epsilon) =
\tilde\Sigma'_{\rm LE}(\epsilon,\bm k_\epsilon)^T -
\tilde\Sigma'_{\rm LE}(\epsilon,\bm k_\epsilon)^{T_{\rm ref}} $.
It can be extracted experimentally by taking the difference 
between two MDC dispersions at fixed $\bm k_\epsilon$. Here $T_{\rm ref}$ is a reference temperature that should be chosen deep in the normal state. 

Another interesting quantity, which determines the quasiparticle lifetime and is accessible by Eq.~\eqref{eq30}, is the linewidth function $
  \tilde \Sigma''(\epsilon)=
\Sigma''_{\rm LE}(\epsilon,\bm k_\epsilon)/Z_{\rm HE}$, which we obtain by determining the full width at half maximum (FWHM) of the Lorentzian in reciprocal space, $|\delta \bm  k_\epsilon|$, and the group velocity $\tilde {\bm v}_\epsilon$ 
in the direction of $\delta \bm k_\epsilon$,
\begin{eqnarray}
  \label{eq34}
  \tilde \Sigma''(\epsilon)=\tilde{\bm v}_\epsilon \delta \bm k_\epsilon .
\end{eqnarray}

\begin{figure}[t]
 \begin{center}
	\includegraphics[width=1.0\columnwidth]{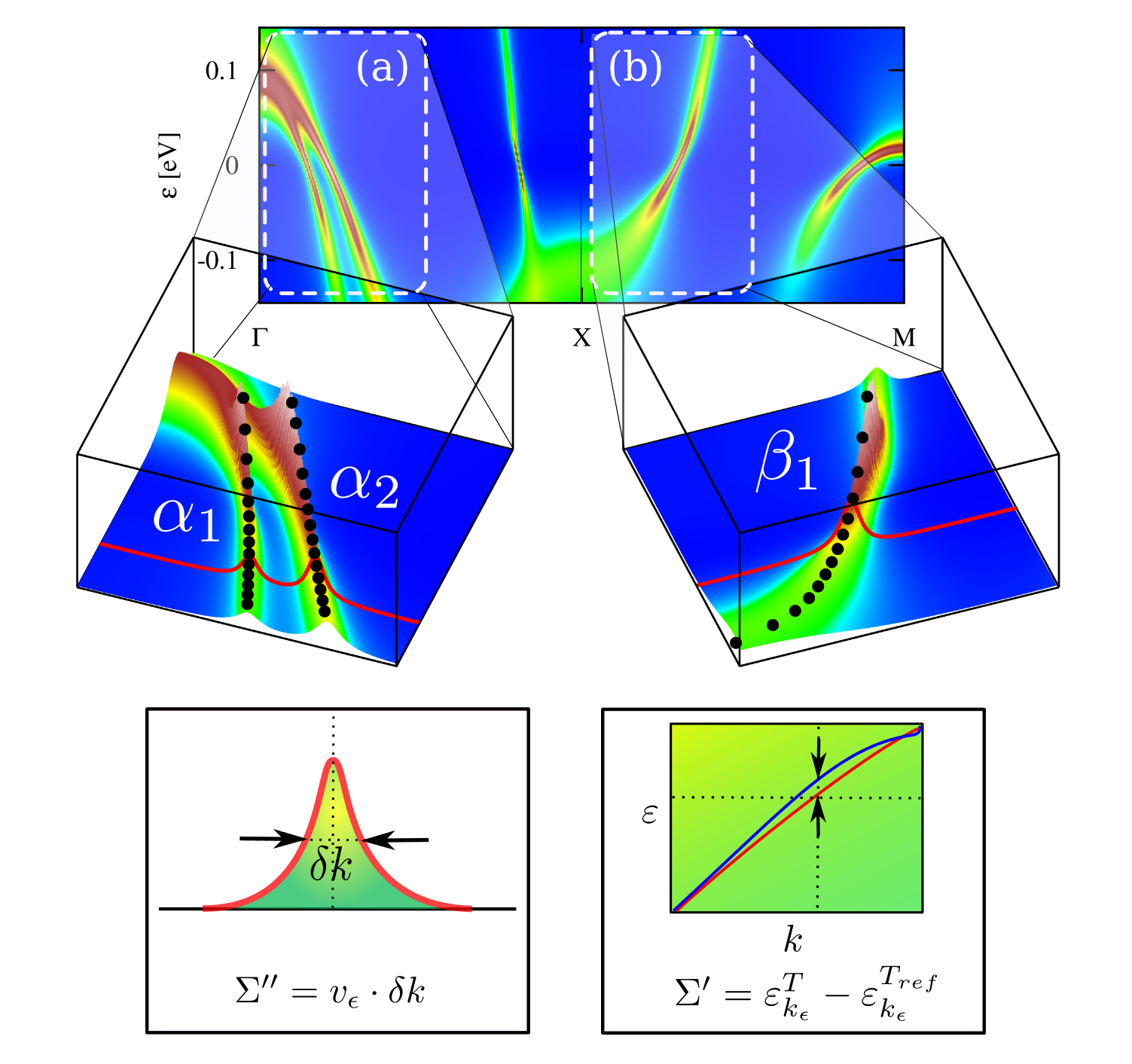} 
   \caption{Effective Self-energies: Illustration of the procedure used to extract self-energy effects from the electronic spectral function $A(\epsilon,\bm k)$ at $T_{\rm ref}=50\,$K. The width of $A(\epsilon,\bm k)$ for fixed energy determines the scattering rate $\Sigma''$, whereas the difference of two dispersion relations at different temperatures gives the real part of the effective self-energy, $\Sigma'$.    }
   \label{fig9}
 \end{center}
\end{figure}

The procedure explained above is illustrated in Fig.~\ref{fig9}. 
We choose $T_{\rm ref}=50\,$K. 
Because the exact value of $Z_{\rm HE}$ is unknown, we fix it at this temperature, $Z_{\rm HE}^{T_{\rm ref}}=1$, and determine $Z_{\rm HE}$ in the superconducting state so that the dispersions merge with the normal state reference at high energies. In Fig.~\ref{fig9} the reference dispersion is shown along a cut in the first Brillouin zone for $k_z=0$. In the extracts (a) and (b) we present the dispersion branches of the $\alpha_{1,2}$ and the $\beta_1$ bands. Below the Fermi energy we fitted the MDC dispersion to the spectral functions, which are shown by the black solid dots, and calculated the linewidth by the procedure explained above. 

\subsubsection{Quasiparticle lifetime in the normal state}

\begin{figure}[t]
 \begin{center}
    \includegraphics[width=1\columnwidth]{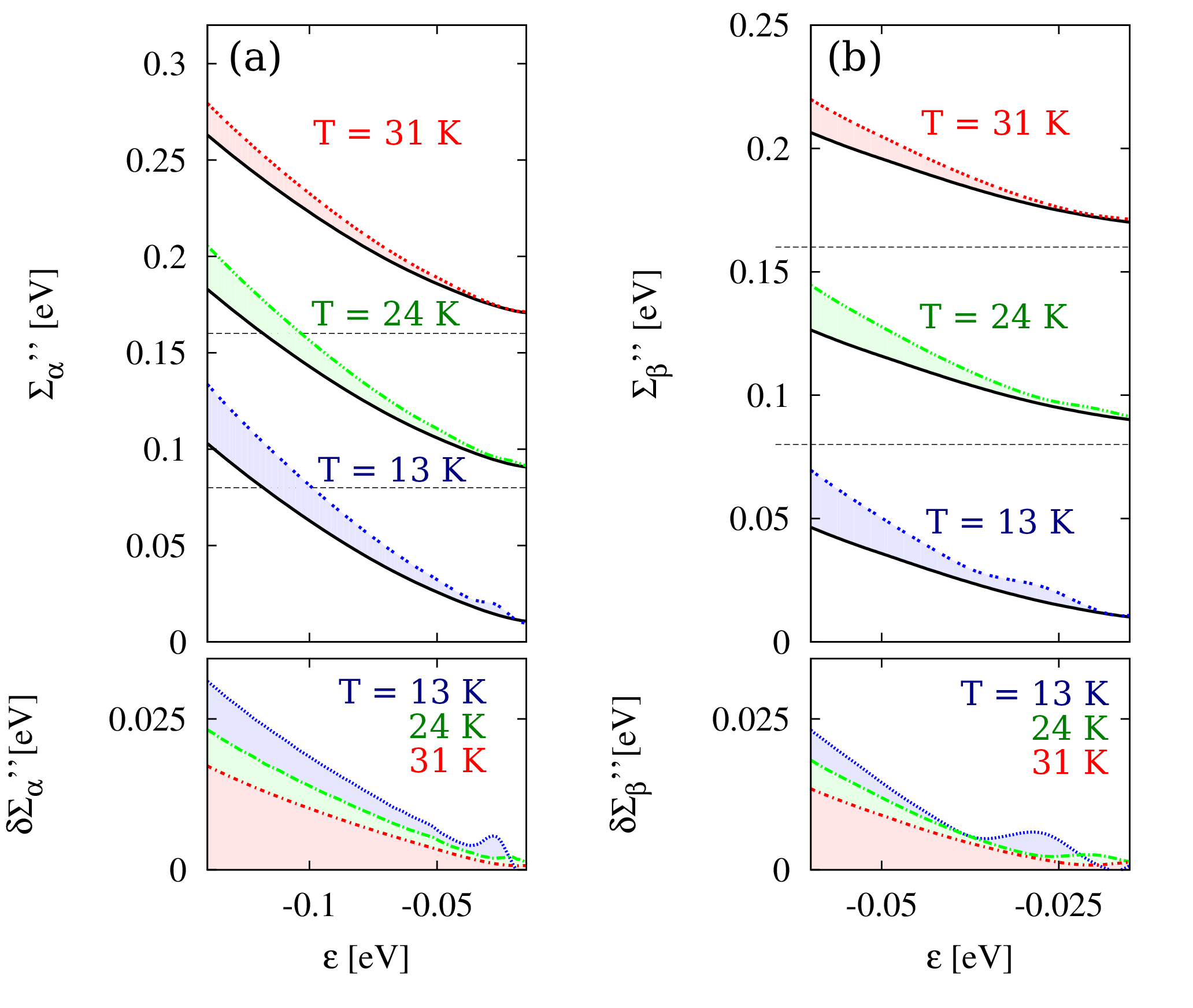} 
   \caption{(a) Temperature dependence of the imaginary part $\Sigma''$ of the effective self-energy at the $\alpha_1$ band extracted from the cut in Fig.~\ref{fig9}(a). The black solid curve corresponds to the normal state at $T=50\,$K. The dashed blue, green, and red curves correspond to $T=13,\,24,\,31\,$K respectively. The difference of the imaginary parts of the effective self-energy, $\Sigma''(T)-\Sigma''(50 {\rm K})$, at $T=13,\,24,\,31\,$K is shown at the bottom. (b) The same as in (a) along the cut at the $\beta_1$ band which is depicted in Fig.~\ref{fig9}(b). }
   \label{fig13}
 \end{center}
\end{figure}

The imaginary part of the effective self-energy exhibits a linear dependence at high energies, $\Sigma''(\epsilon)\propto \epsilon$ (Fig.~\ref{fig13}), consistent with marginal Fermi liquid theory. \cite{Littlewood&Varma1991} This results from the coupling to the spin-fluctuation continuum, in particular from the slow decay towards high energies. The linear in energy contribution increases in magnitude with decreasing temperature, consistent with the temperature dependence of the spin-fluctuation continuum. It persists at high binding energies in the superconducting state.
As we will see in the following section, in the superconducting state the appearance of the spin resonance leads to an additional feature at low energy, a small bump in the imaginary part of the self-energy seen in Fig.~\ref{fig13} for $T=$ 24K and, more pronounced, for $T=$ 13K.

\subsubsection{The superconducting state and the ``kink'' feature}
Upon entering the superconducting state the low-energy resonance in the dynamic magnetic susceptibility appears below the particle-hole continuum. In this section we want to extract the effect of this resonance feature on the electronic structure. For this we compare the normal state and superconducting state dispersions and extract both, the real part as well as the imaginary part of the effective self-energy from those. The influence of the resonance on the dispersion relation can be quantified using, e.g., the relation Eq.~\eqref{eq35},

where we obtain the renormalized dispersion relation $\tilde \varepsilon_{\bm k}^{T<T_{\rm c}}$ by fitting the Lorentzian in Eq.~\eqref{eq30} to the theoretically obtained spectral function in Eq.~\eqref{eq23}. That is a good approximation even in the superconducting state for energies not too close to the superconducting gap.

\begin{figure}[t]
 \begin{center}
    \includegraphics[width=1\columnwidth]{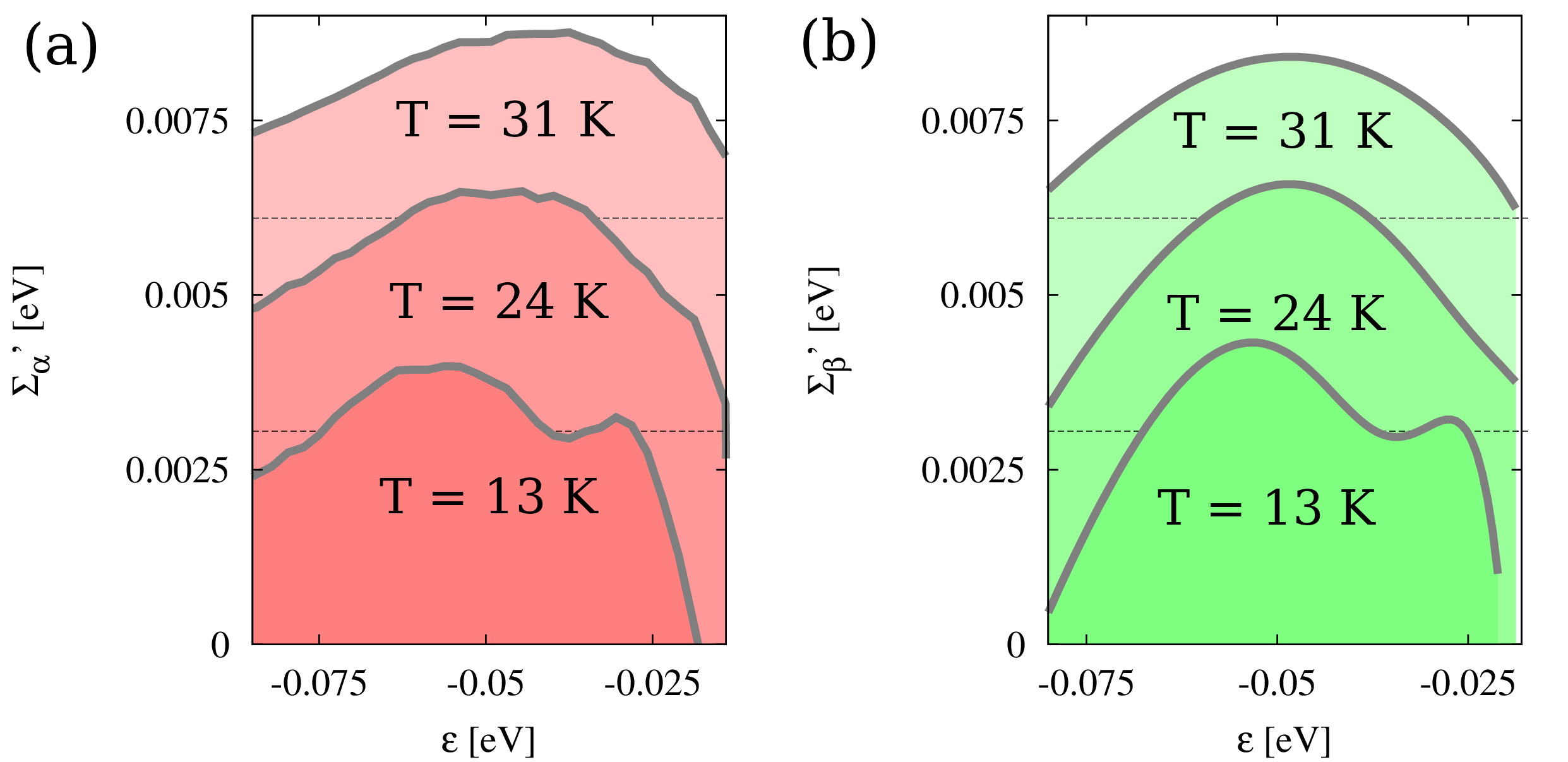} 
   \caption{(a) and (b) The temperature dependence of the real part $\Sigma'$ of the effective self-energy extracted from the dispersion cuts shown in Figs.~\ref{fig9}(a) and \ref{fig9}(b).}
   \label{fig13-b}
 \end{center}
\end{figure}

Now the electronic dispersions $\tilde \varepsilon^T_{\bm k} =\varepsilon^T_{\bm k} /Z_{\rm HE}^T$  at the $\alpha_{1,2}$ bands and the $\beta_1$ band can be determined. In Fig.~\ref{fig13-b} the temperature dependence of the real part of the effective self-energy is shown as obtained in Eq.~\eqref{eq35}. The appearance of the bosonic resonance leads to an effect on the electronic dispersion which is characterized by the development of a peak in the real part of the effective self-energy, $\Sigma'(\epsilon)$ (see Fig.~\ref{fig13-b}), as well as a hump feature in the imaginary part of the effective self-energy $\Sigma''(\epsilon)\equiv \tilde \Sigma''(\epsilon,\bm k_\epsilon)$ (Fig.~\ref{fig13}), which we see even more pronounced by having a look at the difference $\Sigma''(T)-\Sigma''(T_{\rm ref})$ [Fig.~\ref{fig13}(a) and \ref{fig13}(b) at the bottom]. Both effects follow the temperature dependence of the bosonic resonance. The coupling feature is situated at energies $\Delta_{\bm Q}+\Omega_{\rm res}^T$, where $\Delta_{\bm Q}$ is the gap at the corresponding Fermi surfaces that are approximately (i.e. on the scale $\xi_r^{-1}$) nested by an antiferromagnetic wave vector $\bm Q$. The broad maximum in $\Sigma'$ which moves around $\approx \rm 50\, meV$ at all temperature is the result of coupling to the particle-hole continuum.

The fact that the above-discussed kink feature results from coupling to the spin-fluctuation resonance was already predicted in ARPES experiments by Richard {\it et al.}. \cite{Richard2008} There the authors show that the coupling feature follows an order-parameter-like temperature dependence which is consistent with the temperature dependence of the neutron resonance observed by inelastic neutron scattering.\cite{Christianson2008} By comparing the electronic dispersion in the superconducting state with the one in the normal state (see the illustration in Fig. \ref{fig9}) they clearly show that the low-energy kink is emerging in the superconducting phase. Within our calculations we are able to reproduce the shape and the absolute value of the real and imaginary part of the effective self-energy with a coupling strength of $g^2\chi_0=1.17\times 10^3 \,\mathrm{\mu_B^2\,eV\, K}$. \par  
On the other hand, Koitzsch {\it et al.}\cite{Koitzsch09} performed a detailed analysis of the temperature dependence of the electronic dispersion relation deep into the normal state. The authors stress that there exists a bump feature at $\sim$25 meV in the normal state which does not modify at temperatures below $100\,$K, continuously moves to higher energies at higher temperatures and finally disappears above $200\,$K. Although a kink between $25$ and $40\,$meV would be consistent with the phonon spectrum \cite{Christianson2008_2} this would not explain the mentioned temperature dependence. 
However, there is a natural expanation of these observations within our model.
Note that the spin-fluctuation continuum modeled by Eq.~\eqref{eq7} and motivated by neutron scattering experiments \cite{Inosov2009} exhibits a maximum at $\omega=\Omega_{max}^T$. This moves to higher energies with increasing temperature whereas the amplitude of Eq.~\eqref{eq7} continuously vanishes. According to that, the observations of Koitzsch {\it et al.} can be understood as the fingerprint of the spin fluctuation continuum, which persists in the normal state and weakens with increasing temperature. Energy scale, magnitude, and the temperature dependence are all consistent with this interpretation.
The additional change of the kink feature when crossing $T_{\rm c}$, as, e.g., clearly observed in the experiments by Richard {\it et al.}, \cite{Richard2008} was not resolved in the experiments of Koitzsch {\it et al.}

\subsubsection{EDCs in the superconducting state}
\label{EDC}

Recent FLEX calculations have shown that the EDCs show a well-pronounced dip-hump feature. \cite{Schmalian2010} This would essentially lead to a break in the dispersion relation, which is not the case for iron-based superconductors. In this section we want to take a look at the EDCs calculated in our model.

\begin{figure}[b]
 \begin{center}
    \includegraphics[width=0.9\columnwidth]{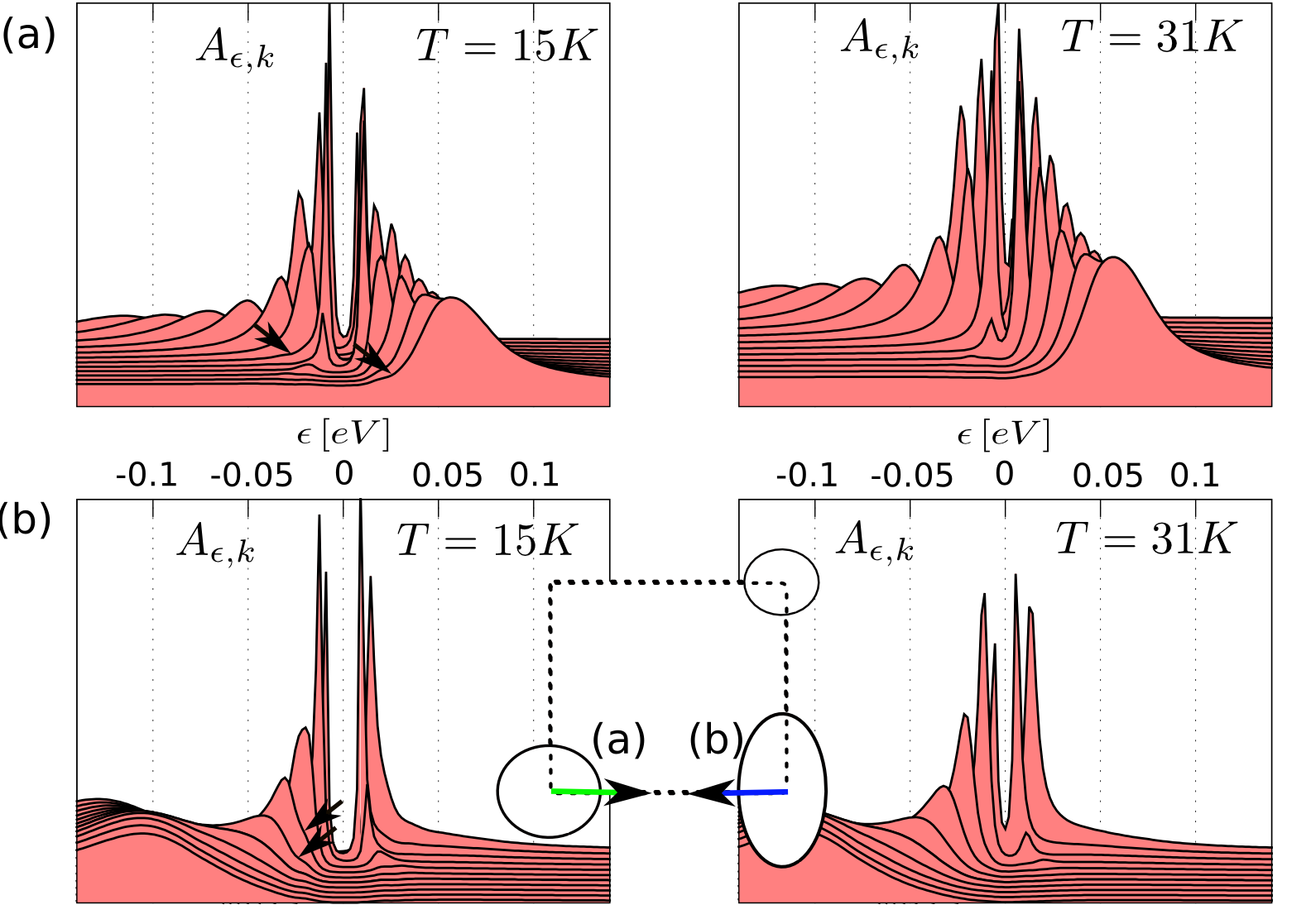} 
   \caption{Energy distribution curves at the $\alpha_{1,2}$ (a) as well the $\beta_1$ pocket (b). We compare the EDC’s at $T = 15$ K (left) and $T = 31$ K (right). The respective cuts are indicated in the inset. }
   \label{fig15}
 \end{center}
\end{figure}
In Fig.~\ref{fig15}(a) and \ref{fig15}(b) we compare the EDC's at different cuts in the first Brillouin zone and at the two temperatures $T=15\,$K and $T=31.0\,$K. As we have seen in the previous chapter, the less pronounced resonance feature at higher temperatures is also less pronounced in the effective self-energies (Fig.~\ref{fig13}). Accordingly, we expect anomalous features, which can be related to the resonance, to appear at $T=15\,$K but to be absent at $T=31\,$K. Actually, for the higher temperature the EDCs on the right-hand side of Fig.~\ref{fig15}(a) and \ref{fig15}(b) show a smooth shape that should be compared to the left-hand side corresponding to the lower temperature. Here the electronic spectral function shows particular variations of this smooth lineshape, namely, small kinks (indicated by arrows). \par

\begin{figure}[t]
 \begin{center}
    \includegraphics[width=0.7\columnwidth]{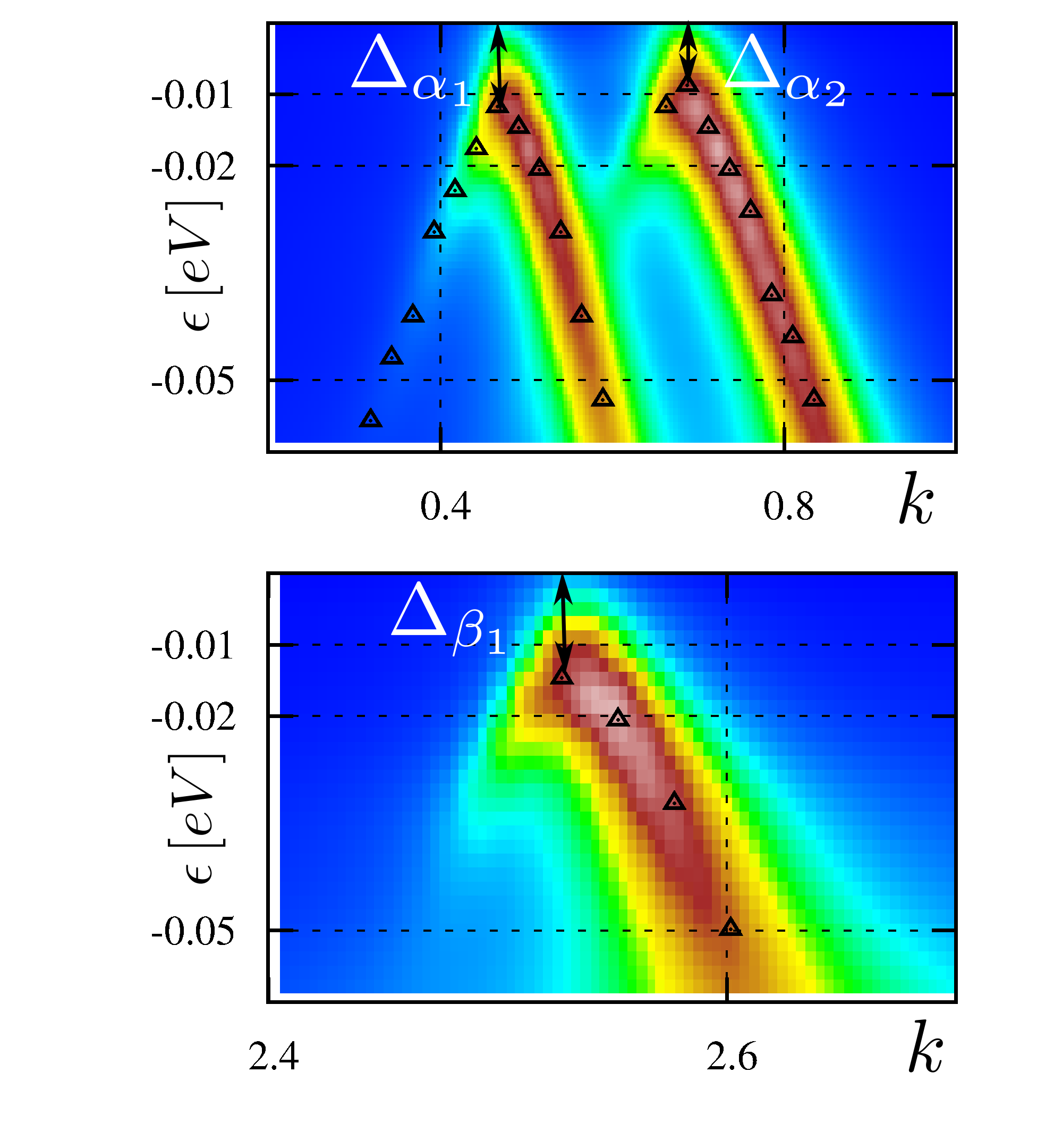} 
\caption{Superconducting state spectral function at $T=15\,$K along the $\Gamma$-$X$ direction as well as the maxima of the EDC curves (black triangles). The superconducting gaps are determined as illustrated in the insets ($\Delta_{\alpha_1}\approx 11.6\,$meV, $\Delta_{\alpha_2}\approx 8\,$meV, and $\Delta_{\beta_1}\approx 11.6\,$meV).   } 
   \label{fig15-1}
 \end{center}
\end{figure}

In Fig.~\ref{fig15-1} the superconducting spectral function along the $\Gamma$-$X$ direction is shown. The maxima of the curves in Fig.~\ref{fig15}(a) and \ref{fig15}(b) give the EDC-derived dispersions indicated by the black triangles. We determine the superconducting gap at the pockets by the maxima of this dispersion, as illustrated in Fig.~\ref{fig15-1}. The superconducting gap is nearly constant along the Fermi surface sheets with $\Delta_{\alpha_1}\approx 11.6\,$meV, $\Delta_{\alpha_2}\approx 8\,$meV and $\Delta_{\beta_1}\approx 12\,$meV.   

\begin{figure}[t]
 \begin{center}
    \includegraphics[width=1.\columnwidth]{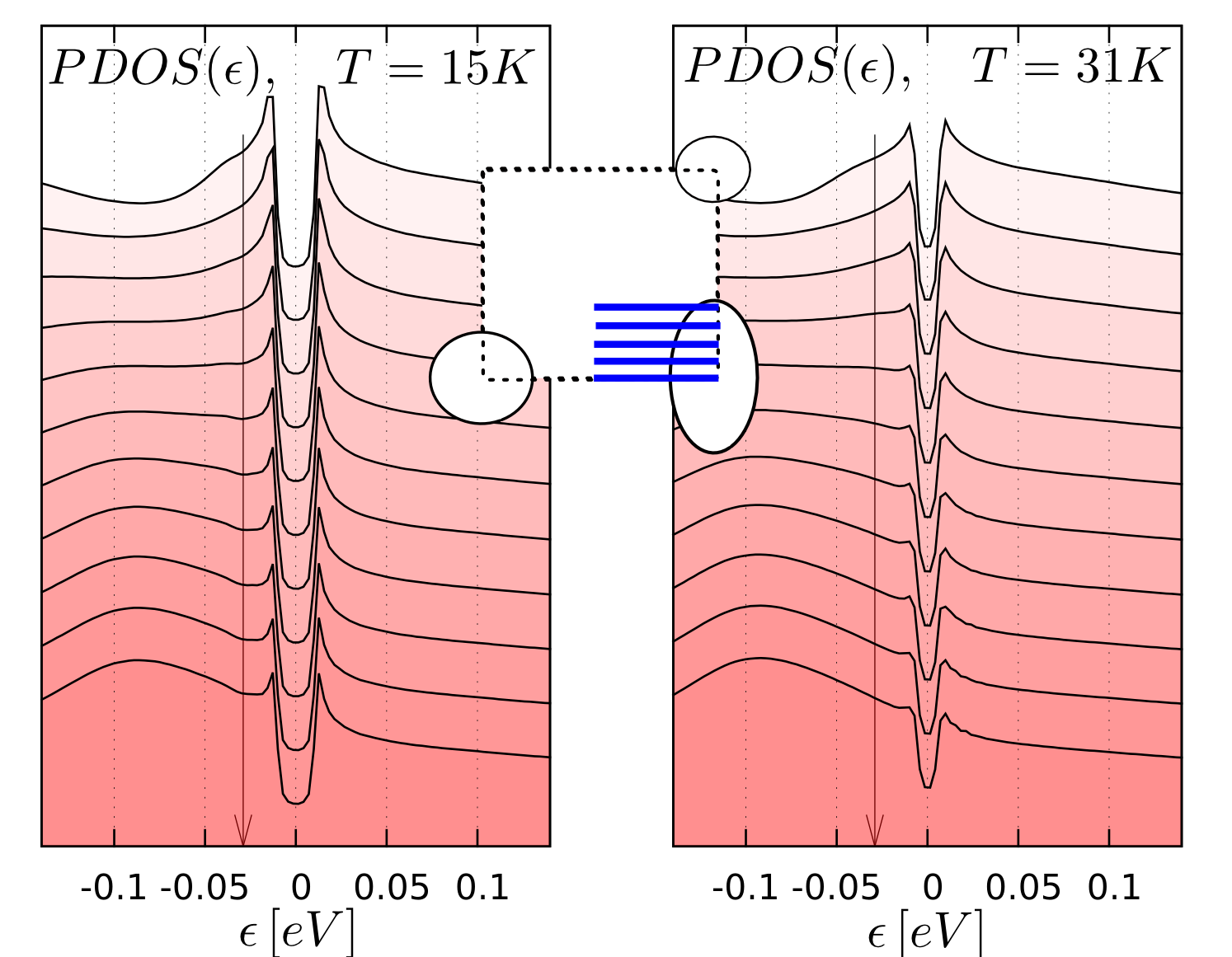} 
   \caption{Partial density of states, i.e., the added spectral function $\sum_{k_x\in[\frac{\pi}{2},\pi]} A_{\bm k}(\epsilon)|_{k_z=0}$. We present the PDOS at different $k_y$-values at the $\beta_1$ band, as illustrated by the inset. The order of the PDOS curves from top to bottom corresponds to the order along the arrow in the inset. At $T=15\,$K the curves exhibit a dip at $\sim 25-30\,$meV which is absent at $T=31\,$K.}
   \label{fig16}
 \end{center}
\end{figure}

\begin{figure}[b]
 \begin{center}
    \includegraphics[width=\columnwidth]{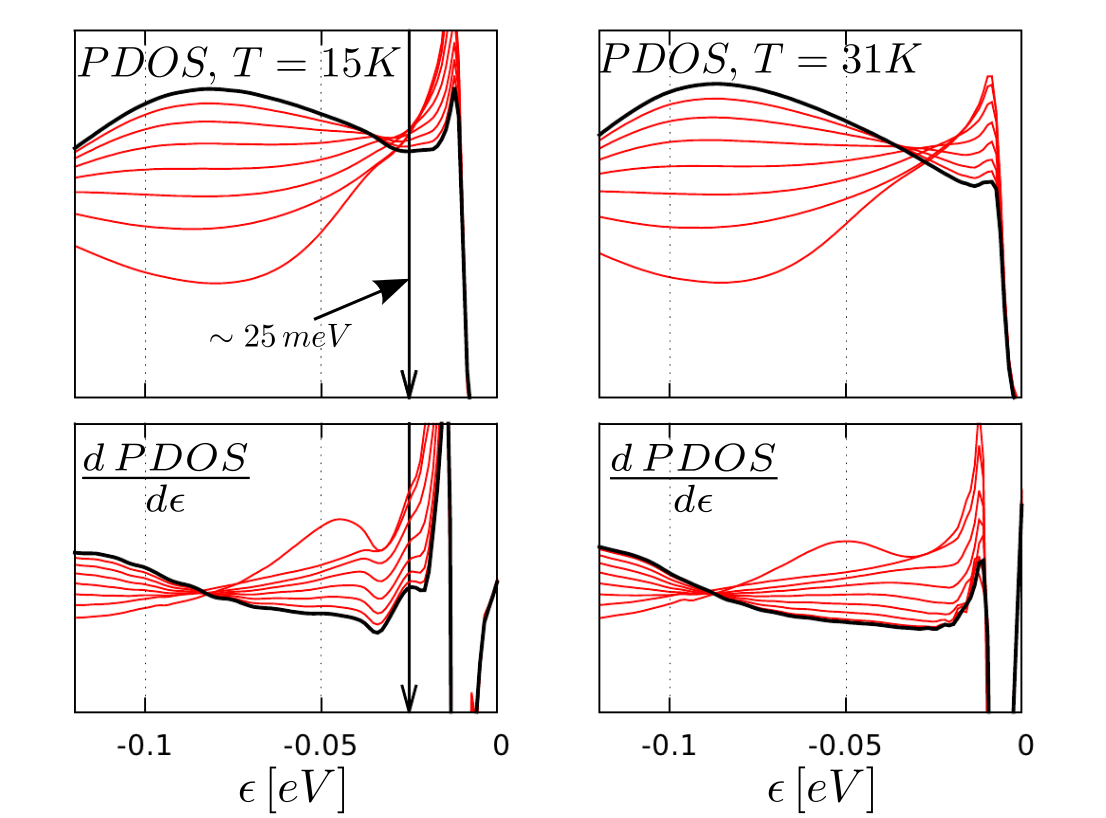} 
   \caption{Partial density of states ${\rm PDOS}(\epsilon)$ curves as in Fig.~\ref{fig16} at $T=15\,$K (left) and at $T=31\,$K (right) and its derivative $\frac{d {\rm PDOS}(\epsilon)}{d \epsilon}$. The dip in the PDOS leads to a peak in its derivative which is absent at $T=31\,$K.}
   \label{fig16-1}
 \end{center}
\end{figure}

Another quantity of interest is the partial density of states (PDOS). Here the spectral function is integrated over a cut that crosses the Fermi surface, as we have illustrated in the inset of Fig.~\ref{fig16}, i.e., $\mathrm{PDOS}(\epsilon)=\sum_{k_x\in[\frac{\pi}{2},\pi]} A_{\bm k}(\epsilon)|_{k_z=0}$. There we present the PDOS curves at different $k_y$ values at the $\beta_1$ pocket. For $T=15\,$K they clearly exhibit a dip feature at an energy of $\Omega_{\rm res} + \Delta_{\bm Q} \sim 25-30 \,$meV, which is absent at the higher temperature $T=31\,$K. \par 

In order to enhance the coupling feature a common procedure is to take the derivative of the partial density of states. This quantity is shown in Fig.~\ref{fig16-1} and we see that $\frac{d {\rm PDOS}(\epsilon)}{d\epsilon}$ exhibits a clear maxima at $T=15\,$K which is absent at $T=31\,$K. The difference of this peak (at $25.5\,$meV) and the quasiparticle peak (here $12.6\,$meV) gives $\approx 12.9\,$meV which corresponds to the resonance energy $\Omega_{\rm res}^{T=15\,{\rm K}}\approx 12\,$meV.
\par 
In conclusion EDC's are not best suited to examine anomalies that arise due to the spin-fluctuation resonance because the features that can be referred to the mode are too small. However, they are essentially more pronounced in the partial density of states [${\rm PDOS}(\epsilon)$], as was already seen in experiment,\cite{Richard2008} and its derivative ($\frac{d { \rm PDOS}(\epsilon)}{d \epsilon}$).

\subsection{Tunneling spectra and density of states}

\begin{figure}[b]
 \begin{center}
    \includegraphics[width=1.\columnwidth]{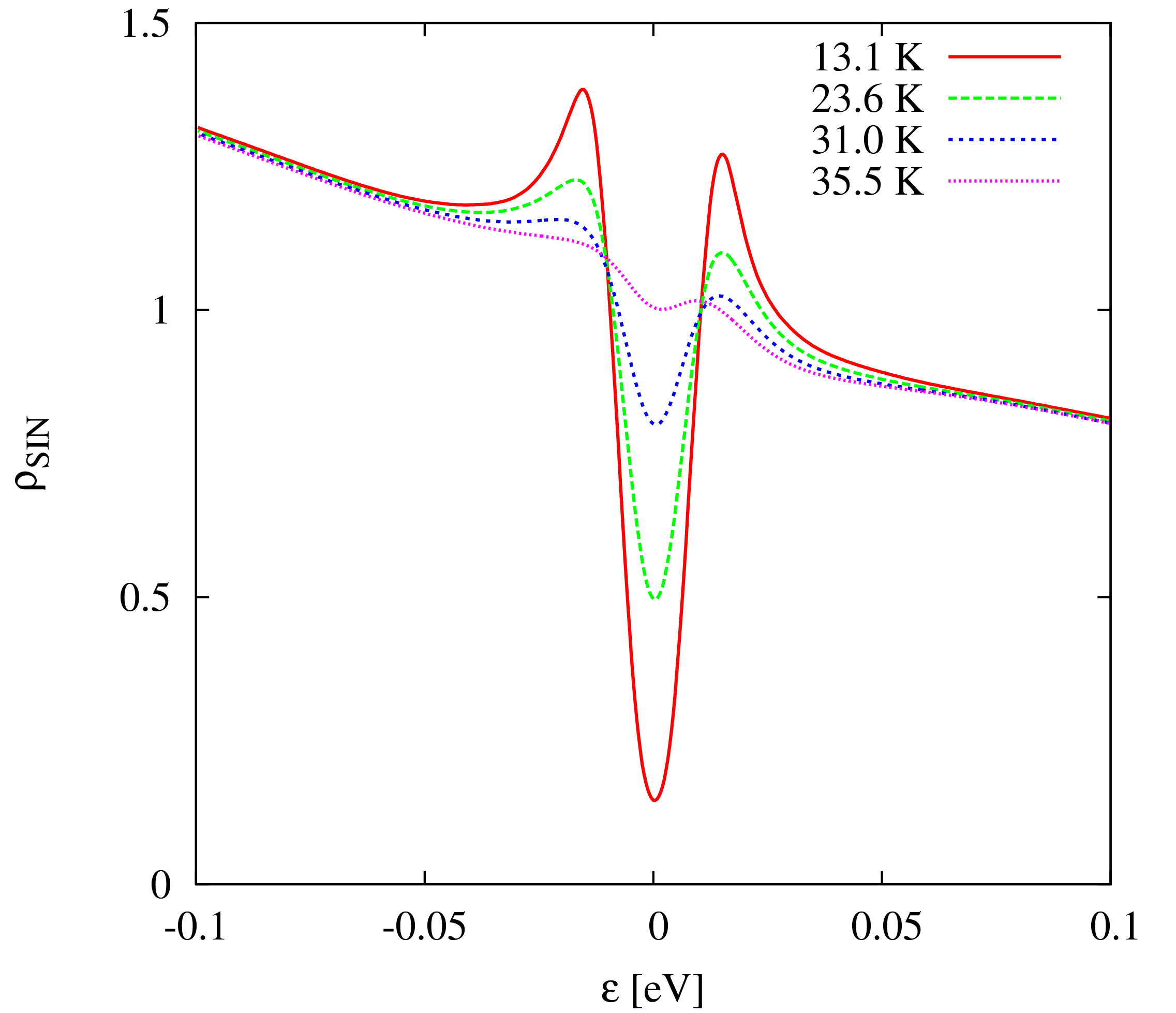} 
   \caption{{SIN tunneling conductance} at different temperatures in the range $13-37\, K$ in the superconducting state for the coupling constant $g^2\chi_0=1.17\times 10^3 \,\mu_B^2\,eV\, K$. The curves show no clear hint of a coupling to spin excitations, such as a dip feature in the case of cuprate superconductors. }
   \label{fig17}
 \end{center}
\end{figure}
Another useful tool to obtain information about the density of states (DOS) is scanning tunneling spectroscopy (STS). 
 
We are interested in the spin conserving tunneling current when applying a voltage $V$. The tunneling current in this case is given by

\begin{equation}
  \label{eq36}
  I(eV)=|T_0|^2
\int_{-\infty}^\infty \frac{d \epsilon }{2\pi}
\nu_1(\epsilon )\nu_2(\epsilon+eV ) \,[f(\epsilon)-f(\epsilon+eV)],
\end{equation}

where $\nu_{1,2}(\epsilon) =\sum_{\bm k} A_{1,2}(\epsilon, \bm k)$ are the densities of states computed from
the spectral functions $A_{1,2}$ of the two materials in contact, and $f$ is the Fermi distribution function. 
We assume here for simplicity an isotropic tunneling matrix element $T_0$.

For a superconductor-insulator-normal metal (SIN) tunneling junction
the energy dependence of the normal metal density of states in the vicinity of the Fermi energy, $\epsilon_F$, is negligible.
In this case one defines the matrix element $|M_{0}|^2=|T_{0}|^2\, \nu_N(\epsilon_F)$, where $\nu_N$ is the 
local density of states in the normal metal. The SIN tunneling conductance is then given by

\begin{eqnarray}
  \label{eq37}
&&  \rho_{SIN}(eV)=\frac{d\,I_{SIN}}{d\,V}\\ 
&&
\qquad\nonumber =e|M_0|^2 \int_{-\infty}^{\infty} \frac{d\epsilon}{2\pi} \, 
\frac{\nu_S(\epsilon )}{4k_B T\cos^2(eV/2 k_B T)}, \qquad
\nonumber
\end{eqnarray}
with
the density of states $\nu_S(\epsilon)=\sum_{\bm k}A_S(\epsilon, \bm k)$, where $A_S$ is the spectral function of the superconductor. For the coupling strength $g^2\chi_0=1.17\times 10^3 \,\mu_B^2\,eV\, K$, which is the value we obtain from our comparison with ARPES experiments, we present the tunneling conductance $\rho_{SIN}(\epsilon)$ in Fig.~\ref{fig17} for different temperatures. 

\begin{figure}[b]
 \begin{center}
    \includegraphics[width=1.0\columnwidth]{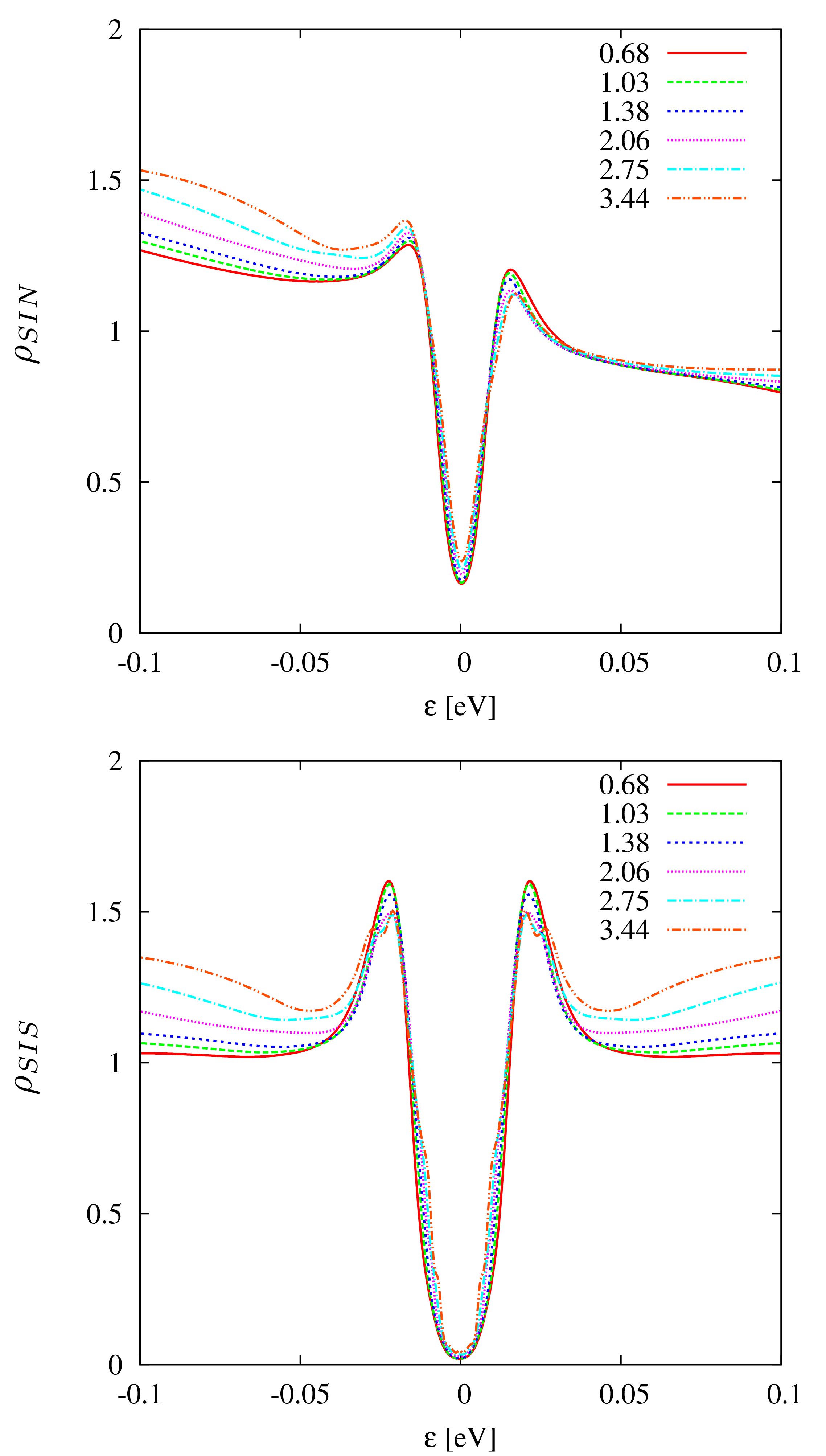} 
   \caption{{SIN (top) and SIS (bottom) tunneling conductance} for different coupling constants $g^2\chi_0=0.68\,...\,3.44 \times 10^3\mu_B^2\,eV\, K$ in units $e|M_0|^2$ and  $e|T_0|^2$, respectively at the temperature $T=15\,K$. }
   \label{fig18}
 \end{center}
\end{figure}

Van Hove singularities at specific points in the Brillouin zone can cause an intensified scattering with spin excitations, leading, e.g., to a well-pronounced break feature in the spectral function and therefore a suppression of the density of states in the affected energy range in cuprate superconductors \cite{Eschrig2000,Kaminski2001,EschrigNormanApr2003,Eschrig2006}. This however is absent in our calculations. 

Our spectra show a strong particle-hole asymmetry that has been also seen in experimental data. \cite{YinMar2009}

Nevertheless, it is interesting to investigate how the tunneling conductance is modified by changes of the coupling strength and when significant features appear at low energies that can be attributed to the effect of spin fluctuations. Therefore we calculated the differential tunneling conductance $dI/dV$ for different coupling strengths $g^2\chi_0=0.68\,...\,2.42\times 10^3\,\mathrm{\mu_B^2\,eV\, K}$ for a SIN as well as a SIS junction. 

Both $\rho_{SIN}$ and $\rho_{SIS}$ are presented in Fig.~\ref{fig18}. 
The SIN tunneling spectra show the development of a dip-hump feature for increasing coupling strength, as in the case for cuprate superconductors. 
As can be seen from Fig.~\ref{fig18}, for SIN tunneling
the coupling predominantly influences the occupied side, whereas the unoccupied side remains nearly unaffected. In contrast,
the SIS spectra are symmetric and therefore the dip feature appears on both sides. 
For a coupling strength of $g^2\chi_0=3.44\times 10^3\,\mathrm{\mu_B^2\,eV\, K}$ we see a clear dip at $\approx 30 \, meV$ that can be traced back to the resonance in the spin-fluctuation spectrum.

\section{Summary and Conclusions}

In this work we presented a model of spin fluctuations coupling to electrons in the normal as well as the superconducting state. 
We find that the high-energy spin fluctuation continuum provides imporant contributions to the pairing interaction, leads to a linear in energy broadening of the quasiparticles, and contributes to the renormalization effects of the dispersion both in the normal and in the superconducting state. In the normal state it leads to a broadened kink feature in the dispersion, similar to that observed in experiments by Koitzsch {\it et al.}\cite{Koitzsch09} In the superconducting state, it leads to a sharper kink feature as experimentally observed by Richard {\it et al.}\cite{Richard2008} and by Wray {\it et al.}\cite{Hasan2008}

The superconducting gap originating from the low-energy ($<$200 meV) part of the dynamic susceptibility supports an $s^\pm$ state, as proposed earlier.\cite{Graser2010} 
The investigation of the scattering rate of quasiparticles and spin fluctuations shows that the renormalization effects are strongest at the approximately nested Fermi surface sheets and exhibit characteristics of these. 

The sharper kink feature at certain energies in the eletronic dispersion in the superconducting state is due to the low-energy resonance in the spin-excitation spectrum. It can be quantified following experimental procedures to extract them directly from the spectral function. 
A comparison with the ARPES experiment by Richard {\it et al.} \cite{Richard2008} enables us to estimate the strength of the coupling between electrons and spin fluctuations. We find an intermediate value, certainly smaller than an analogous analysis in the cuprates yields.

We discuss the renormalization of the Fermi surface and find that coupling to spin fluctuations leads to a shift of the chemical potential to higher values by $\approx 20$ meV, accompanied by a shrinking of all Fermi surfaces. The relation between the chemical potential and the charge carrier density is linear for small to moderate values, and non-linear for higher coupling strengths.

Investigating the energy dependence of the spectral function as well as the partial density of states and the differential tunneling conductance we find that for the coupling strength necessary to reproduce the self-energies of the ARPES experiments, there are no such pronounced coupling features in the tunneling spectra as observed in the case of cuprate superconductors. However, in the partial density of states, obtained when integrating the spectral functions perpendicular to the Fermi surface, coupling features are clearly visible.

We conclude that the coupling between spin fluctuations and electronic excitations provides an important mechanism for superconductivity in iron pnictides.

\section{Acknowledgements}
RG would like to thank the Karlsruhe House of Young Scientists for financial support during his stay at Royal Holloway, University of London.
The work of AH was generously supported by the South East Physics network (SEPnet) during his stay at Royal Holloway, University of London.

\appendix
\section{Energy and momentum dependence of the diagonal and off-diagonal self-energies}
\label{appA}
\begin{figure}[t]
 \begin{center}
	\includegraphics[angle=0, width=1\columnwidth]{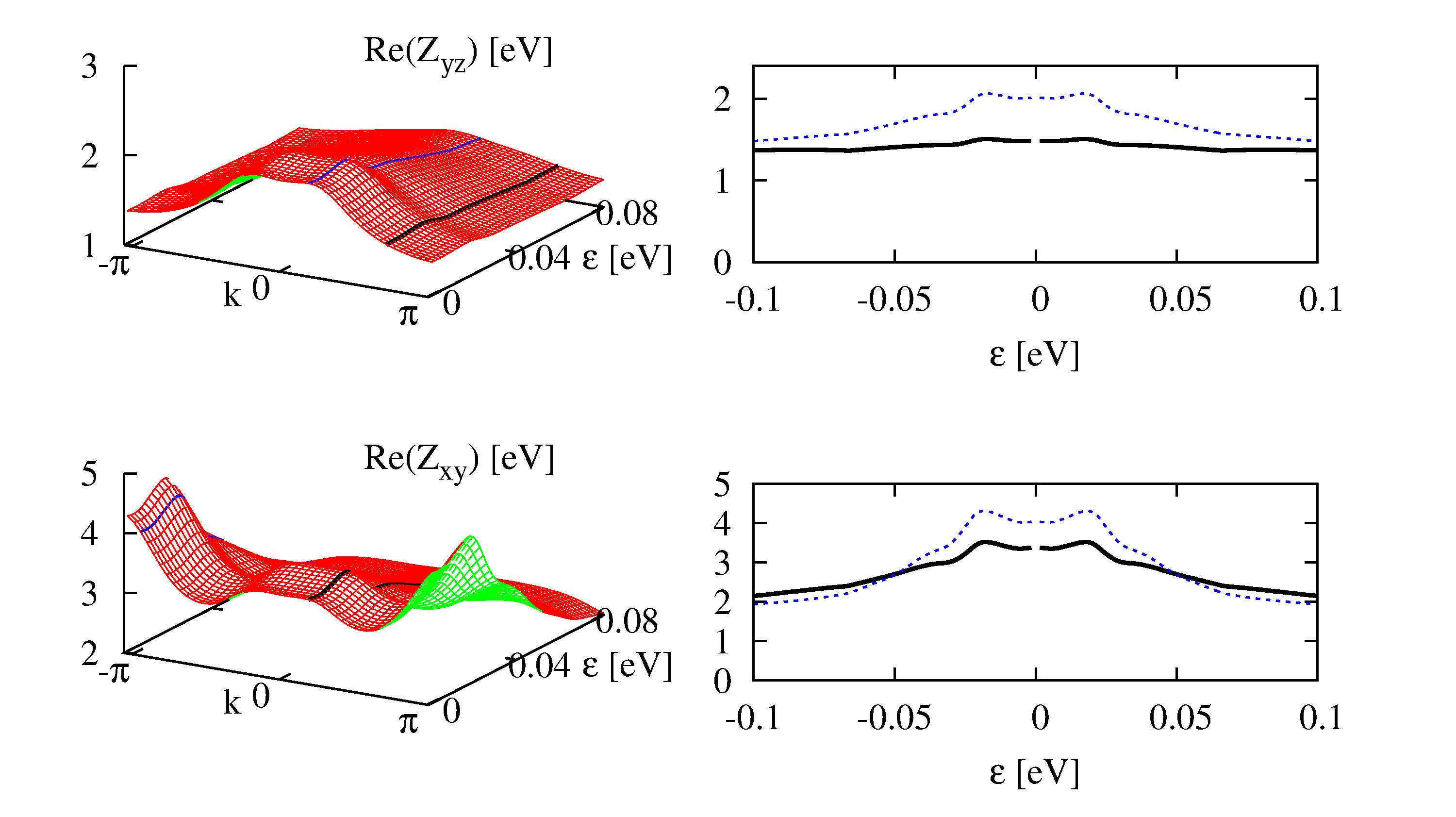} \\
	\includegraphics[angle=0, width=1\columnwidth]{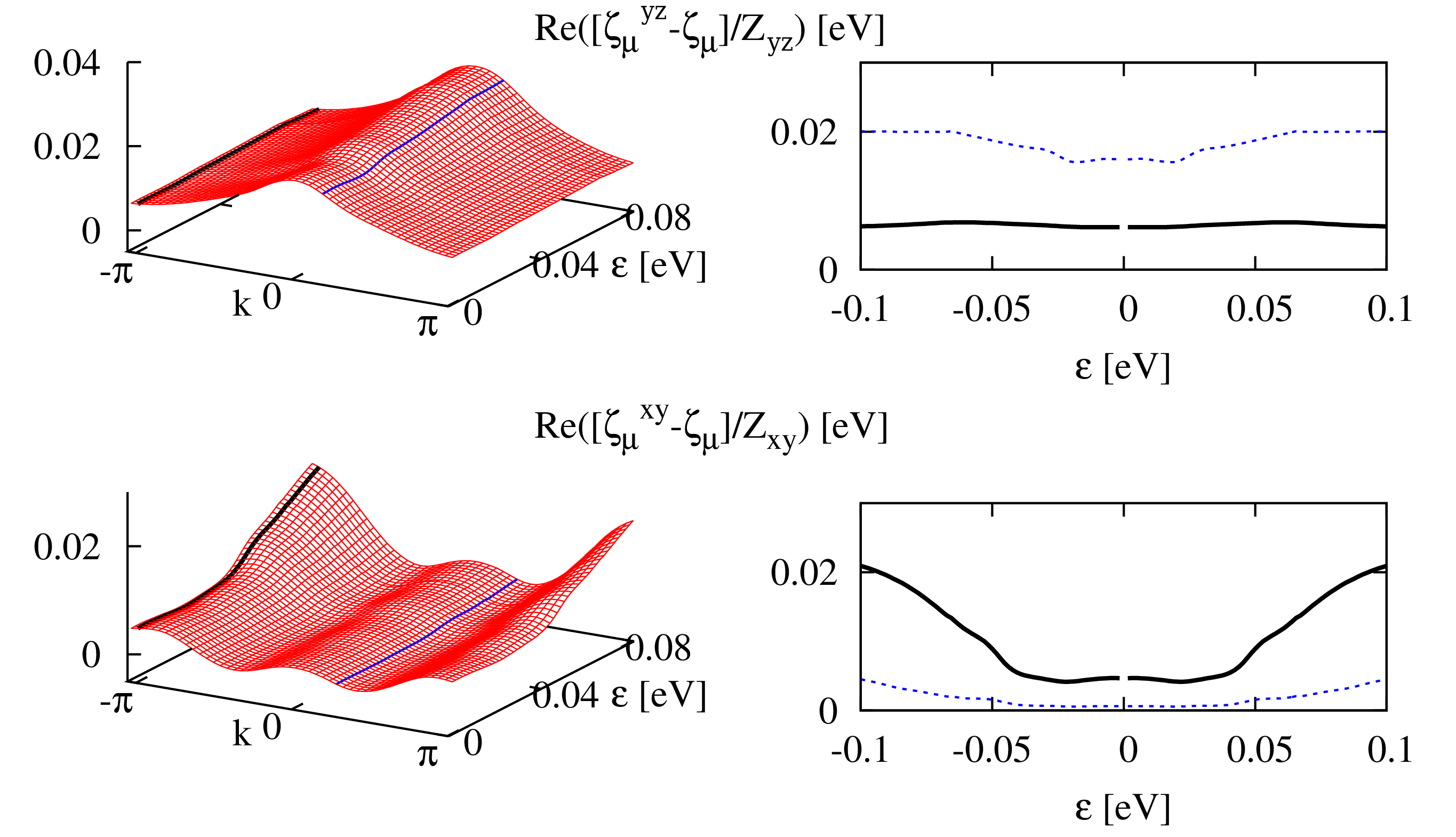} 
   \caption{Energy and momentum dependence of the real part of the renormalization function $Z_n$
and of the real part of the dispersion correction according to Eq.~\eqref{eq20}, $(\zeta_\mu^n-\zeta_\mu)/Z_n= (\Sigma_n^{\rm R}+\tilde \Sigma_n^{\rm R})/2Z_n$,
at $T=15\,$K. 
The energy dependence is shown along the momentum-cut $\{k_y\in[-\pi:\pi),\,k_x=k_z=0\}$.}
   \label{fig4c}
 \end{center}
\end{figure}

In this appendix we discuss the diagonal self-energies entering Eq.~\eqref{eq20} and Eq.~\eqref{eq22}. The energy and momentum dependence of the renormalization function $Z$ is presented in Fig.~\ref{fig4c}. 
At low energies, the renormalization function takes values of about 2 ($yz$ channel) and 4 ($xy$ channel), and at higher energies it takes values of about 1.5 ($yz$ channel) and 2 ($xy$ channel). Results for the $xz$ channel are analogous to the results for the $yz$ channel.

Also shown in Fig.~\ref{fig4c} is
the particle-hole asymmetric part of the diagonal self-energy, corresponding to Eq.~\eqref{eq20}. Looking at the $xy$ channel of the real part we see a drastic change for energies $|\epsilon |> |\Omega_{\rm res}+E_\gamma|$ at $(k_x,\,k_y,\,k_z)\approx (0,\,\pm \pi\,,0)$. This corresponds to the appearance of the shallow electron pocket at the $\beta$ bands.

\begin{figure}[t]
 \begin{center}
	\includegraphics[angle=0, width=1\columnwidth]{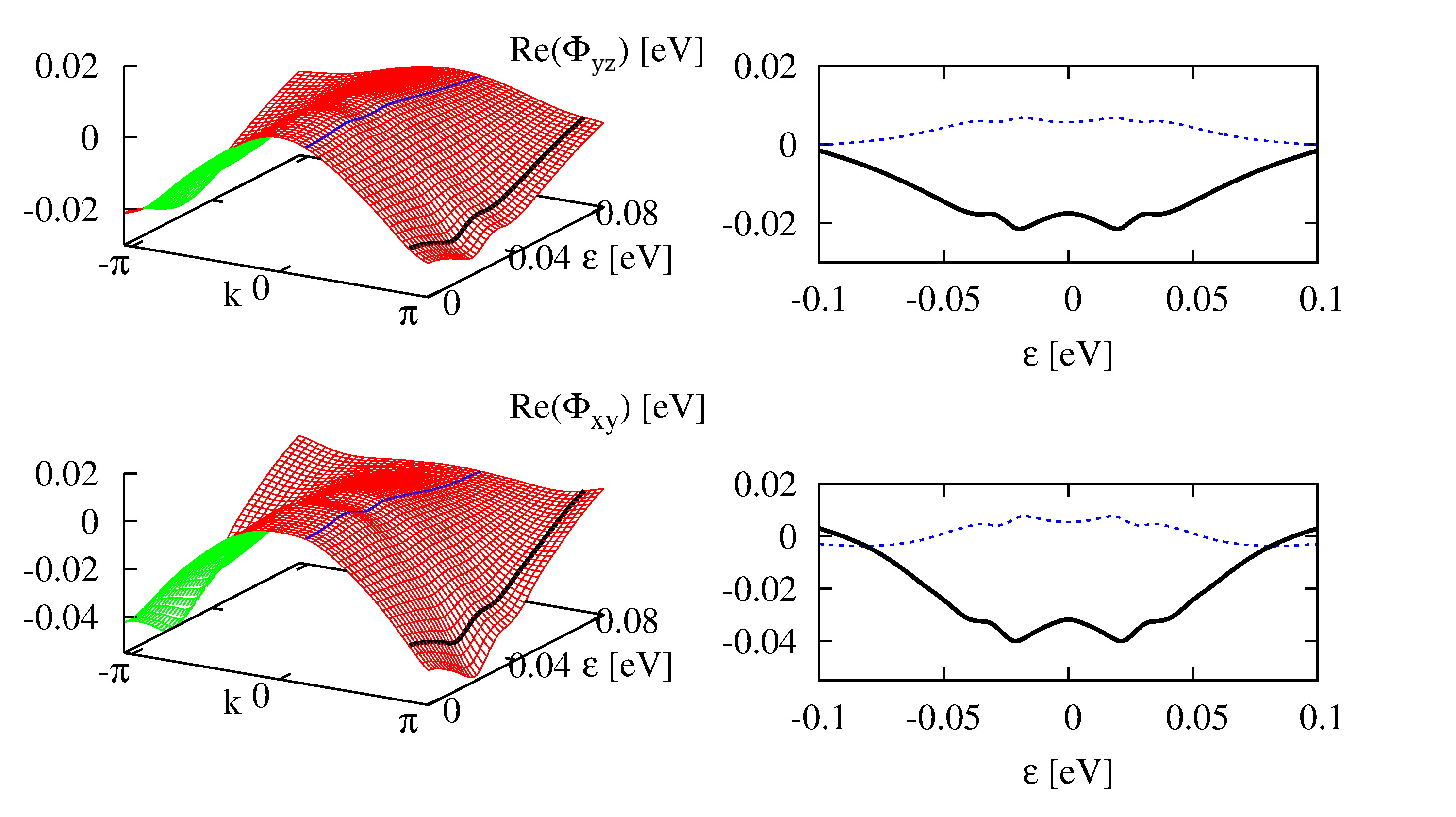} \\
	\includegraphics[angle=0, width=1\columnwidth]{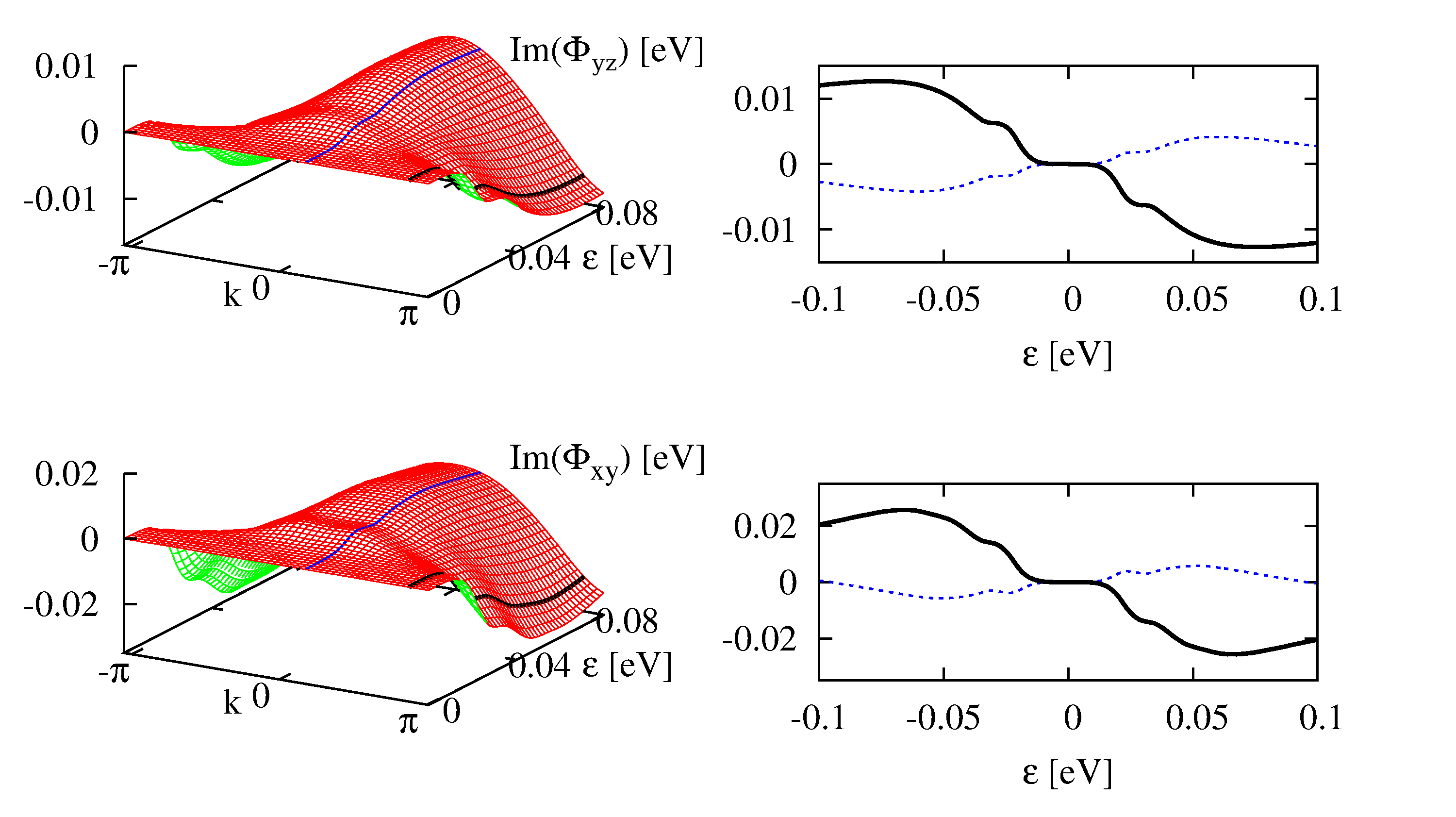} 
   \caption{Energy and momentum dependence of the real and imaginary part of the off-diagonal self-energy $\Phi_n$ at $T=15\,$K along the cut $\{k_x\in[-\pi:\pi),\,k_y=k_z=0\}$.}
   \label{fig4d}
 \end{center}
\end{figure}

\begin{figure}[t]
 \begin{center}
	\includegraphics[angle=0, width=1\columnwidth]{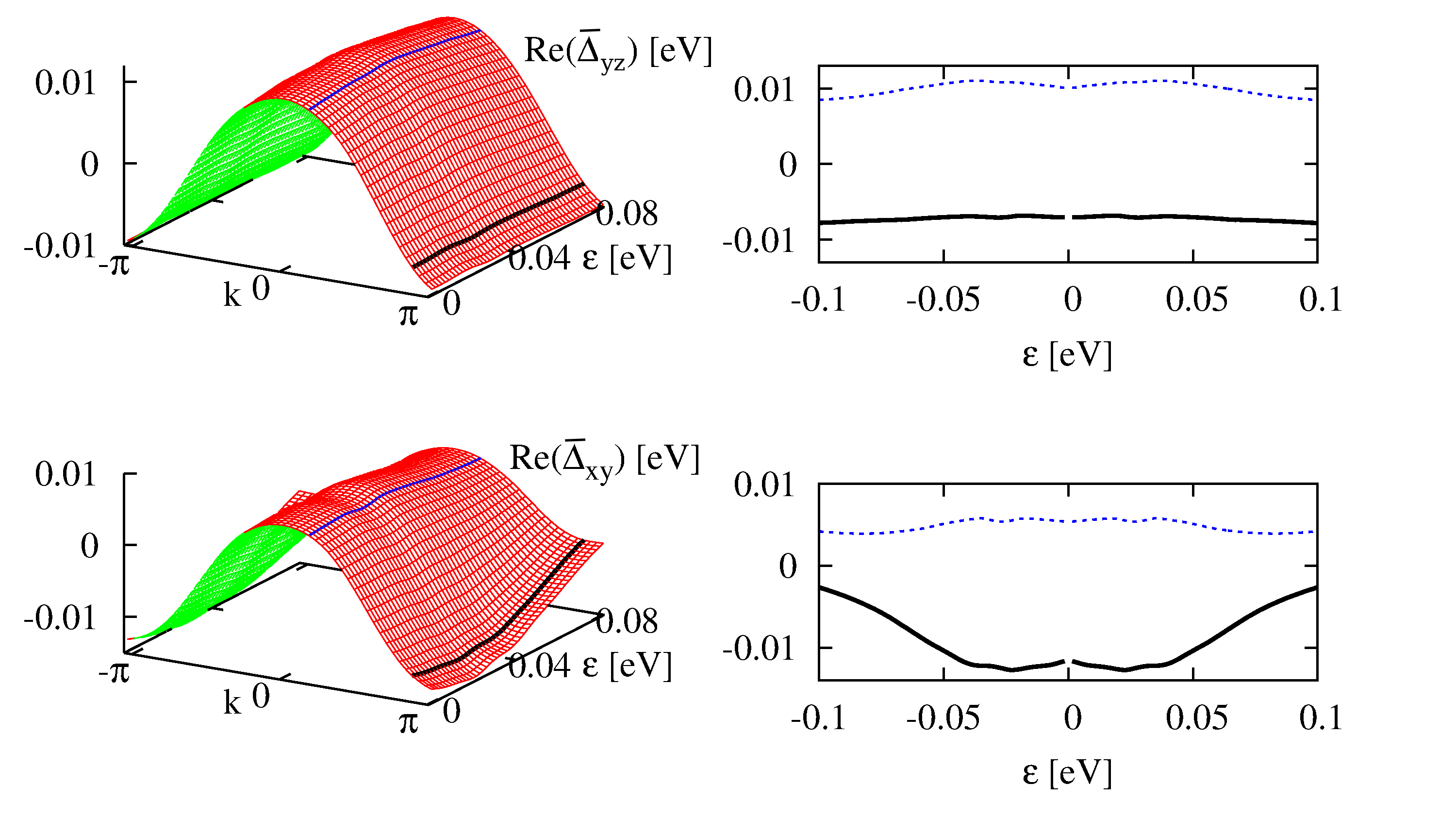} \\
	\includegraphics[angle=0, width=1\columnwidth]{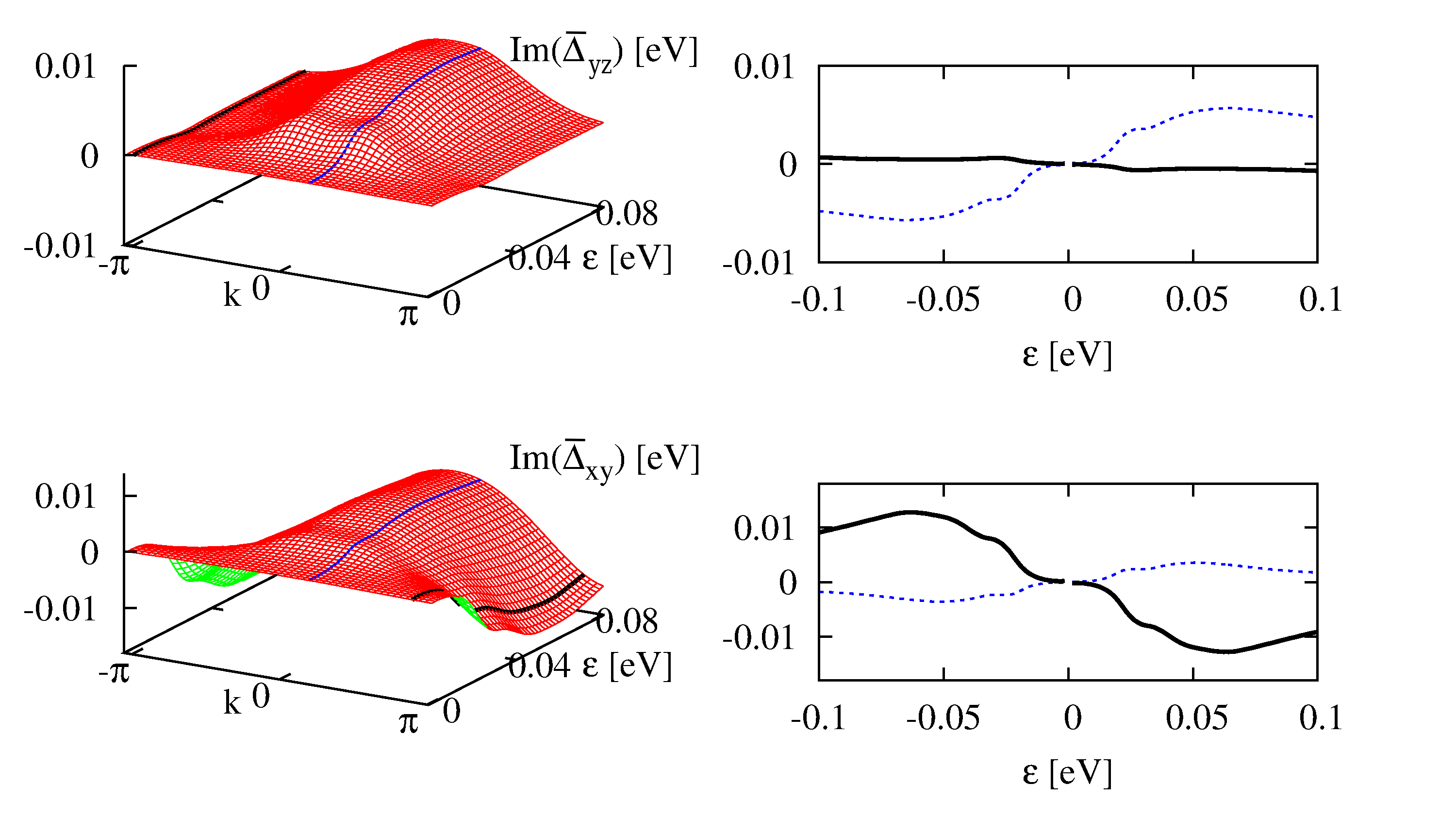} 
   \caption{Energy and momentum dependence of the real and imaginary part of the renormalized order parameter $\bar \Delta_n=(\Delta^{\pm}+\Phi_n^{\rm R})/Z_n$ at $T=15\,$K along the cut $\{k_x\in[-\pi:\pi),\,k_y=k_z=0\}$.}
   \label{fig4f}
 \end{center}
\end{figure}

In Figs.~\ref{fig4d} and \ref{fig4f} the energy dependence of the real and imaginary parts of the off-diagonal self-energy as well as the renormalized gap is shown. 
The gap function, shown in Fig.~\ref{fig4f}, is real in the energy range $<$15 meV, given by the energy of the magnetic resonance. The gap itself is of comparable magnitude. Above this energy range, the gap function acquires a considerable imaginary part, which is of the same order of magnitude as the order parameter itself. 
Note that these effects are very small in the $yz$ channel near $\bm k=(\pm \pi, 0, k_z)$, i.e., at the electron pockets, and in the $xy$ channel near $\bm k=(0,0,k_z)$, i.e. at the large hole pockets.
Thus, strong coupling features in the gap function are almost purely in the $xy$ channel for the electron pockets and in the $xz$ and $yz$ channel for the large hole pockets.
The resonance imprints at energies $\Omega_{\rm res}+ \Delta_{\bm Q}$ and $\Omega_{\rm res}+E_\gamma$ also are indicated in Fig.~\ref{fig4b} ($\Omega_{\rm res}^{T=15\,{\rm K}}\approx 12\,$meV). This effect is most pronounced in $\Phi_n$; however, due to the enhancement of the renormalization function $Z_n$ at these energies (see Fig.~\ref{fig4c}) the renormalized gap $\bar \Delta_n$ weakly varies, as is especially seen in the $yz$ channel in the upper panel of Fig.~\ref{fig4f}.

\section{Local Sum Rule}
\label{appB}
In this Appendix we shortly review a sum rule that holds for the two point correlation function $\langle S^\alpha_{m}(t,\bm R_i)\,S^\beta_{n}(t',\bm R_j) \rangle$, where $S^\alpha_{m}(t,\bm R_i)$ denotes the spatial $\alpha=x,\,y,\,z$ component of the spin at site $\bm R_i$ and in orbital $m$. Then the spin structure factor is defined by 
\begin{eqnarray}
 \nonumber  S^{\alpha,\beta}(\omega,\bm q)&=&\int_{-\infty}^{\infty}dt\,  e^{-\imath \omega t} \\ 
 \nonumber  \times \frac{1}{N} \sum_{ij}\sum_{mn} 
 && \left\langle e^{-\imath \bm q \cdot (\bm R_i - \bm R_j)} S_m^\alpha(t,\bm R_i) S_n^\beta(0,\bm R_j) \right \rangle.
\end{eqnarray}  

Integrating this quantity over frequency yields the static spin structure factor

\begin{eqnarray}
  \nonumber
  S^{\alpha, \beta}(\bm q)= \int_{-\infty}^\infty \frac{d\omega}{2\pi} \, S^{\alpha,\beta}(\omega,\bm q)=\frac{1}{N} \sum_{mn}\left\langle S_m^\alpha(\bm q) S_n^\beta(-\bm q) \right \rangle.
\end{eqnarray}

Further integrating out the momentum and taking the trace gives

\begin{eqnarray}
  \nonumber
  \sum_\alpha \int\frac{d^3\bm q}{(2\pi)^3} \, S^{\alpha\alpha}(\bm q)&=&\frac{1}{N} \sum_{\alpha,i}\sum_{mn}
  \left\langle S_m^\alpha(\bm R_i)\, S_n^\alpha(\bm R_i) \right \rangle \\ 
  \nonumber
  &=& \frac{1}{N}\sum_i \left\langle \bm S(\bm R_i) \cdot \bm S(\bm R_i) \right \rangle\\
  \nonumber
  &=& S(S+1).
\end{eqnarray}

Here we defined the spin $\bm S(\bm R_i)=\sum_m \bm S_m(\bm R_i)$ on each site $i$. The last equality only holds for a system of equal spins. This is realized in the Heisenberg limit, where the hopping of electrons between the different sites is suppressed. However, in more itinerant systems (metallic regime) correlations between the spins are reduced and the integrated weight should be smaller. However it was proven in experiment that the sum rule also holds in the case of iron pnictides. \cite{Inosov2009}

\section{High energy renormalization factor}
\label{appC}

Starting from Eq.~\eqref{eq30} we have introduced a high-energy renormalization factor $Z_{\rm HE}=1-\Sigma'_{\rm HE}(\epsilon)/\epsilon$ in order to treat the renormalization originating from the high-energy incoherent part of the spin fluctuation spectrum. In the procedure explained above we obtained a linear temperature dependence $Z_{\rm HE}=\alpha + \beta\,T$. In this section we provide a motivation for such a temperature dependence. \par 

By introducing a high-energy cutoff $\omega_{\rm c}$ in the bosonic spectrum, the self-energies consist of two contributions, $\Sigma^{\rm R}_n(\epsilon,\bm k)=\Sigma_n^{\rm LE}(\epsilon,\bm k)+\Sigma_n^{\rm HE}(\epsilon,\bm k)$. Whereas the low-energy contribution can be calculated exactly, we have to treat $\Sigma^{\rm HE}$ phenomenologically.
As we have seen in Sec. \ref{sec1b}, the sum rule for the susceptibility implies that $\int_{-\infty}^\infty d \omega \int d^3\bm q S(\omega,\bm q)$ remains constant and temperature independent.  
As we assumed in our model identical weights for the orbital contributions $\chi_n(\omega, {\bm k})$, the sum rule applies to each of them separately.
Introducing the high-energy cutoff we obtain
\begin{eqnarray}
\int d^3\bm q &&\left( \int_{|\omega| < \omega_{\rm c}} d\omega \frac{{\rm Im}\chi_{n}(\omega,\bm q)}{1-\exp(-\omega/T)} 
\right. \nonumber \\ && \qquad \qquad + \left. 
\int_{|\omega| > \omega_{\rm c}} d\omega \frac{{\rm Im} \chi_{n}(\omega,\bm q)}{1-\exp(-\omega/T)}\right) \nonumber \\
&&=I_{{\rm LE}} + I_{{\rm HE}} =I=\mbox{const} .
\end{eqnarray}
 
\begin{figure}[tb]
 \begin{center}
	\includegraphics[width=1.0\columnwidth]{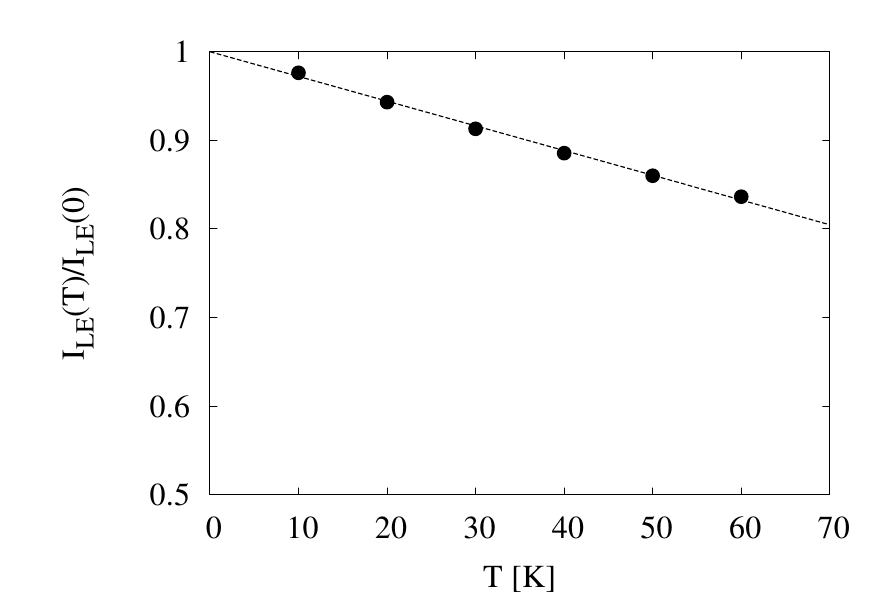}
\caption{Value of the low-energy integrated weight $I_{\rm LE}=\eta_{\rm LE} - \gamma_{\rm LE} T$ as function of temperature $T$, showing a linear dependence.}
   \label{Fig_weight}
 \end{center}
\end{figure}

Whereas the low energy behavior of the susceptibility is motivated by recent experiments \cite{Inosov2009}, the high energy spectrum is unknown. On phenomenological grounds we expect it to be broadened in momentum and to decay at high energies. Furthermore the temperature under consideration is small compared to the energy cutoff, i.e. $\omega_{\rm c}/T >> 1$ and therefore
\begin{eqnarray}
&& \int_{\omega_{\rm c}}^\infty d\omega {\rm Im}\chi_{n}(\omega)\approx I-I_{\rm LE}
\nonumber \\
&=&I-\int d^3\bm q \int_{|\omega| < \omega_{\rm c}}d\omega \frac{{\rm Im}\chi_{n}(\omega,\bm q)}{1-\exp(-\omega/T)}
\end{eqnarray}
By numerically integrating the second term on the right-hand side we find a linear dependence in temperature, $I_{\rm LE}=\eta_{\rm LE} - \gamma_{\rm LE} T$ (see Fig.~\ref{Fig_weight}) implying that 
 \begin{eqnarray}
 \int_{\omega_{\rm c}}^\infty d\omega {\rm Im}\chi_{n}(\omega)\approx I-\eta_{\rm LE} + \gamma_{\rm LE} T .
\end{eqnarray}

Let us now consider the high-energy contribution of the self-energy. We start from an expression for the electronic self-energy \cite{Abrikosov} of electrons coupling to the spin fluctuation mode (in real gauge),

\begin{eqnarray*}
  \hat \Sigma_n^{\rm R}(\epsilon,\bm k)=-2g^2
{\sum_{\bm k'} }'
\sum_{\omega,\zeta}
\frac{{\rm Im}\hat G^{\rm R}_{nn}(\zeta,\bm k'){\rm Im}\chi_n(\omega,\bm k-\bm k')}{
\epsilon -\omega-\zeta + \imath 0} \\
  \times \left[\tanh\left(\frac{\zeta }{2T}\right)+\coth\left(\frac{\omega}{2T}\right)\right]
\end{eqnarray*}

with the abbreviations $\sum'_{\bm k'} \equiv \int \frac{{\rm d}^3{\bm k'}}{(2\pi )^3}$,
$\sum_\omega \equiv \int_{-\infty}^\infty \frac{{\rm d}\omega }{2\pi}$,
and
$\sum_\zeta \equiv \int_{-\infty}^\infty \frac{{\rm d}\zeta}{2\pi}$.
We are interested in the high-energy contribution of the spin susceptibility, for $|\omega |>\omega_{\rm c}$, and therefore restrict the integration over $\omega $ in the expression above to this range.
At high energies the spin-fluctuation contribution will become more and more local, so that it is well approximated by a momentum eigenfunction expansion with a few eigenfunctions $\eta_s(\bm k)$ resulting from close neighbors. At the same time, we relax the approximation that the spin susceptibility only depends on $\bm q=\bm k-\bm k'$.  Thus, we write 

\begin{eqnarray*}
\chi_n(\omega,\bm k-\bm k') \to 
\chi_n(\omega,\bm k,\bm k') = \sum_{s} \eta_s(\bm k) \chi_{n,s}(\omega) \eta_s(\bm k')^\ast
\end{eqnarray*}

with
$\chi_{n,s}(\omega )=\sum'_{\bm k,\bm k'} \eta_s(\bm k)^\ast \chi_n(\omega, \bm k,\bm k') \eta_s(\bm k')$. The eigenfunctions can be classified according to the irreducible representations of the point group, and they are assumed to be orthogonal and normalised, 
$\sum'_{\bm k} \eta_s(\bm k)^\ast \eta_{s'}(\bm k)=\delta_{ss'}$, and built to a complete set, $\sum_s \eta_s(\bm k) \eta_s (\bm k')^\ast=\delta_{\bm k,\bm k'}$. 
Thus, the corresponding components of the self-energy, $\hat \Sigma_{n,s}^{\rm R} (\epsilon ) = \sum'_{\bm k} \eta_s(\bm k)^\ast \hat \Sigma_n^{\rm R} (\epsilon,\bm k)$, are given by

\begin{eqnarray*}
  \hat \Sigma_{n,s}^{\rm HE}(\epsilon)=-2g^2
\sum_{\omega,\zeta }
\frac{{\rm Im}\hat G^{\rm R}_{nn,s}(\zeta ){\rm Im}\chi_{n,s}(\omega)}{
\epsilon -\omega-\zeta  + \imath 0} \\
  \times \left[\tanh\left(\frac{\zeta }{2T}\right)+\coth\left(\frac{\omega}{2T}\right)\right] \cdot \vartheta(|\omega |-\omega_c)
\end{eqnarray*}

with $\hat G^{\rm R}_{nn',s}(\zeta )=\sum'_{\bm k'} \eta_s(\bm k')^\ast \hat G_{nn'}^{\rm R} (\zeta ,\bm k')$.
With the partial density of states $\nu_{n,s}(\zeta )=-\frac{1}{\pi} {\rm Im}G_{nn,s}^{\rm R} (\zeta) $
we obtain

\begin{eqnarray}
 \Sigma_{n,s}^{\rm HE}(\epsilon )&=&g^2 
\int_{-\infty}^\infty \frac{{\rm d}\omega }{2\pi}
\int_{-\infty}^\infty {\rm d}\zeta 
\frac{\nu_{n,s}(\zeta) {\rm Im}\chi_{n,s}(\omega)}{\epsilon -(\omega+ \zeta) + \imath 0} \nonumber \\ 
&& \times \left[\tanh\left(\frac{\zeta}{2T}\right)+
\mbox{sign}(\omega )\right] \vartheta(|\omega|-\omega_c). \qquad
\end{eqnarray}
The second line in this expression ensures that $|\omega+\zeta|$ is always of order $\omega_{\rm c}$, as it mostly contributes for $\zeta > -2T$ for positive $\omega $ and $\zeta< 2T$ for negative $\omega $. As $\omega_{\rm c}\gg T$,
for $|\epsilon |\ll \omega_{\rm c}$ 
this allows us to expand the denominator in the first line in $\epsilon $,
leading to
\begin{widetext}
\begin{eqnarray*}
\Sigma_{n,s}^{\rm HE}(\epsilon)&=& -g^2 
\int_{\omega_{\rm c}}^\infty \frac{d\omega }{2\pi} {\rm Im}\chi_{n,s}(\omega ) 
\int_{-\infty}^\infty {\rm d}\zeta 
\left[ \frac{ \tanh\left(\frac{\zeta}{2T}\right) +1 }{\omega+\zeta} 
\left(
\nu_{n,s-}(\zeta ) + \nu_{n,s+}(\zeta )\frac{\epsilon }{\omega + \zeta} \right)\right]  
\end{eqnarray*}
\end{widetext}
with $\nu_{n,s\pm} (\zeta )=[\nu_{n,s}(\zeta)\pm \nu_{n,s}(-\zeta)]/2$, and where we used ${\rm Im}\chi_{n,s}(-\omega)=-{\rm Im}\chi_{n,s}(\omega )$.
The $\zeta $-integral gives up to second order in a power expansion in temperature
\begin{eqnarray*}
2\int_0^\infty {\rm d}\zeta 
\left[ \frac{ \nu_{n,s-}(\zeta ) }{\omega+\zeta} 
+ \frac{\nu_{n,s+}(\zeta )\, \epsilon }{(\omega + \zeta )^2} \right]  
\\
-\alpha T^2 \left[ \frac{\nu'_{n,s}(0)}{\omega }-\frac{2\nu_{n,s}(0)\epsilon }{\omega^3} \right]
\label{he1}
\end{eqnarray*}
with a numerical constant $\alpha $ of order 1.
We see that the temperature-dependent terms are small compared to the temperature-independent terms. 
The terms independent of $\epsilon $ contribute to the band renormalization.
If all orbitals would contribute equal, it would correspond to
a constant shift of the bands.
The leading contribution is
\begin{eqnarray}
\zeta_{n,s}^{\rm HE}&=& -4g^2
\int_{\omega_{\rm c}}^\infty \frac{d\omega }{2\pi} {\rm Im}\chi_{n,s}(\omega ) 
\int_0^\infty {\rm d}\zeta 
\frac{ \nu_{n,s-}(\zeta ) }{\omega+\zeta} . \quad
\end{eqnarray}
For different orbital contributions to the spin susceptibility, this quantity would lead to orbital-dependent band shifts, and thus to a slightly temperature-dependent renormalization of bands.
The temperature-independent part
can be absorbed in the definition of the tight binding bands and the orbital as well as Brillouin zone average in  the temperature-dependent chemical potential. The remaining contributions are neglected in this paper.

The terms linear in energy contribute to the high-energy renormalization factor $Z_{n,s}^{\rm HE}$. 
We can write in leading order
\begin{eqnarray}
Z_{n,s}^{\rm HE}&=& 1+4g^2
\int_{\omega_{\rm c}}^\infty \frac{d\omega }{2\pi} {\rm Im}\chi_{n,s}(\omega ) 
\int_0^\infty {\rm d}\zeta 
\frac{\nu_{n,s+}(\zeta )}{(\omega + \zeta )^2} . \qquad
\end{eqnarray}

For our calculation we restrict with regard to the diagonal self-energy contributions to the Brillouin zone averages (corresponding to the basis function
$s=0$ with $\eta_0(\bm k)\equiv 1$), and we
define the orbital average $Z_{\rm HE}=\langle Z_{n,0}^{\rm HE} \rangle_n$.
According to that the temperature dependence of the high-energy contribution $Z_{\rm HE}$ is determined by the temperature dependence of $\chi''_{\rm HE}$, leading to 
\begin{eqnarray}
 Z_{\rm HE}= 1+g^2 (\tilde \eta + \tilde \gamma T).
\end{eqnarray}

\section{Kramers-Kronig Relations}
\label{appD}

For the diagonal self-energies in Eq.~\eqref{eq14}
 
we obtain the following Kramers-Kronig relation,
\begin{eqnarray*}
{\rm Re}  [\Sigma_n^{\rm R}(\epsilon,\bm k)-\Sigma_n^{\rm R}(\infty ,\bm k)]&=& 
\frac{1}{\pi} {\cal P} \int_{-\infty}^{\infty} {\rm d}\epsilon'
\frac{{\rm Im}\Sigma_n^{\rm R}(\epsilon',\bm k)}{\epsilon'-\epsilon } ,
\end{eqnarray*}
which implies,
\begin{eqnarray*}
{\rm Re}  [\Sigma_n^{\rm R}(0,\bm k)-\Sigma_n^{\rm R}(\infty ,\bm k)]&=& 
\frac{1}{\pi} {\cal P} \int_{-\infty}^{\infty} {\rm d}\epsilon'\;
\frac{{\rm Im}\Sigma_n^{\rm R}(\epsilon',\bm k)}{\epsilon'} .
\end{eqnarray*}
Here, ${\cal P}$ denotes the principle value integral, and 
$\Sigma_n^{\rm R}(\infty ,\bm k)\equiv 
\lim_{|\epsilon |\to \infty }\Sigma_n^{\rm R}(\epsilon ,\bm k) $.
With the definition
\begin{eqnarray*}
\Sigma^{\pm}_n(\epsilon, \bm k) &=&[\Sigma_n^{\rm R}(\epsilon,\bm k)\pm \Sigma_n^{\rm R}(-\epsilon,\bm k)]/2
\end{eqnarray*}
we can write down
corresponding relations for the renormalization function and band renormalization,
\begin{eqnarray*}
{\rm Re}  Z_n(\epsilon,\bm k)&=& 1-
\frac{2}{\pi} {\cal P} \int_{0}^{\infty} {\rm d}\epsilon' \;
\frac{{\rm Im}\Sigma_n^{+}(\epsilon',\bm k)}{{\epsilon'}^2-\epsilon^2 } \\
{\rm Re}  \zeta_\mu^{n}(\epsilon,\bm k)&=& \zeta_\mu (\bm k)+ {\rm Re} \Sigma_n^{\rm R}(0,\bm k)
\\&+&
\frac{2\epsilon^2}{\pi} {\cal P} \int_{0}^{\infty} \frac{{\rm d}\epsilon'}{\epsilon'} \;
\frac{{\rm Im}\Sigma_n^{-}(\epsilon',\bm k)}{{\epsilon'}^2-\epsilon^2 } .
\end{eqnarray*}
Note that Im$\Sigma_n^{+}(\epsilon,\bm k)<0$ for all $\epsilon$ and $\bm k$ (whereas Im$\Sigma_n^{-}$ can have either sign), and consequently
\begin{eqnarray*}
{\rm Re}  Z_n(0,\bm k)&=& 1-
\frac{2}{\pi} \int_{0}^{\infty} {\rm d}\epsilon' \;
\frac{{\rm Im}\Sigma_n^{+}(\epsilon',\bm k)}{{\epsilon'}^2} 
\ge 1,\\
\lim_{|\epsilon |\to \infty}{\rm Re}  Z_n(\epsilon ,\bm k)&=& 1+
\frac{2}{\pi \epsilon^2} \int_{0}^{\infty} {\rm d}\epsilon' \;
{\rm Im}\Sigma_n^{+}(\epsilon',\bm k)
\le 1,\\
\end{eqnarray*}
Finally, for completeness we also present relations for the band renormalization,
\begin{eqnarray*}
{\rm Re}  \zeta_\mu^{n}(0,\bm k)&=& \zeta_\mu(\bm k) + {\rm Re} \Sigma_n^{\rm R}(0,\bm k)\\
\lim_{|\epsilon |\to \infty}
{\rm Re}  \zeta_\mu^{n}(\epsilon,\bm k)&=& \zeta_\mu (\bm k)+ {\rm Re} \Sigma_n^{\rm R}(0,\bm k)
\\&-&
\frac{2}{\pi} \int_{0}^{\infty} {\rm d}\epsilon' \;
\frac{{\rm Im}\Sigma_n^{-}(\epsilon',\bm k)}{\epsilon'}.
\end{eqnarray*}
In our case, Im$\Sigma^{-}_n$ is predominantly positive, leading to negative corrections at large energies. Finally, we have the sum rule
\begin{eqnarray*}
\int_{-\infty}^{\infty} {\rm d}\epsilon' \; \epsilon' A(\epsilon', \bm k)
&=& \sum_\mu \sum_n |a_\mu^n(\bm k)|^2 \lim_{|\epsilon |\to \infty}{\rm Re}\zeta_\mu^n(\epsilon,\bm k)\\
&=& \sum_\mu \zeta_\mu (\bm k) + \sum_n {\rm Re}\Sigma_n^{\rm R}(\infty,\bm k).
\end{eqnarray*}

\bibliographystyle{unsrt.bst}

\end{document}